\documentclass[prd,
,aps,amsmath,amssymb,nofootinbib,showpacs,showkeys,preprintnumbers,
]{revtex4}
\usepackage{graphicx}
\usepackage{dcolumn}
\usepackage{bm}
\usepackage{amsmath}
\usepackage{amsfonts}
\usepackage{subfigure}
\usepackage{epsfig,graphics}
\usepackage {graphicx}
\usepackage {epsfig}
\usepackage {amsmath}
\usepackage{amsmath}
\usepackage{bm}
\usepackage{graphicx}
\usepackage[T1]{fontenc}		
\usepackage[ansinew]{inputenc}
\usepackage[english]{babel}
\usepackage{amsmath}
\usepackage{amssymb}
\usepackage{slashed}
\usepackage{epsfig}
\usepackage{graphicx}
\usepackage{dcolumn}
\usepackage{bm}
\usepackage{amsmath}
\usepackage{amsfonts}
\usepackage{subfigure}
\usepackage{multirow}
\def\lsi{\raise0.3ex\hbox{$<$\kern-0.75em\raise-1.1ex\hbox{$\sim$}}}
\def\gsi{\raise0.3ex\hbox{$>$\kern-0.75em\raise-1.1ex\hbox{$\sim$}}}

\newcommand{\gsim}{\mathop{\gsi}}
\def\be{\begin{equation}}
\def\ee{\end{equation}}
\def\ba{\begin{eqnarray}}
\def\ea{\end{eqnarray}}
\renewcommand{\d}{\delta}

\newcommand{\m}{\mu}
\newcommand{\n}{\nu}

\renewcommand{\a}{\alpha}
\renewcommand{\b}{\beta}
\renewcommand{\r}{\rho}

\renewcommand{\c}{\chi}
\renewcommand{\l}{\lambda}
\def\xc{\xi_\chi}
\def\xh{\xi_h}
\renewcommand{\L}{\Lambda}

\renewcommand\m{\mu}

\begin{document}
\vspace*{-5mm}
\begin{flushright}

\end{flushright}
\vspace*{5mm}
\title{Higgs-Dilaton Cosmology: From the Early to the Late Universe}
\author{ Juan Garc\'{\i}a-Bellido\footnote{E-mail: juan.garciabellido@uam.es},
  \hspace{0.5mm} Javier Rubio\footnote{E-mail: javier.rubio@uam.es},
  \hspace{0.5mm} Mikhail Shaposhnikov\footnote{E-mail: mikhail.shaposhnikov@epfl.ch},
  \hspace{0.5mm} Daniel Zenh\"ausern\footnote{E-mail: daniel.zenhaeusern@epfl.ch}
  }
\affiliation{ 
 Instituto de F\'{\i}sica Te\'orica CSIC-UAM, 
 Universidad Aut\'onoma de Madrid, Cantoblanco 28049 Madrid, Spain \\
 Institut de Th\'eorie des Ph\'enom\`enes Physiques, 
  \'Ecole Polytechnique F\'ed\'erale de Lausanne,
  CH-1015 Lausanne, Switzerland
 }
\preprint{IFT-UAM/CSIC-11-49}

\begin{abstract}
We consider a minimal scale-invariant extension of the Standard Model of particle physics combined with Unimodular Gravity formulated in \cite{Shaposhnikov:2008xb}. This theory is able to describe not only an inflationary stage, related to the Standard Model Higgs field, but also a late period of Dark Energy domination, associated with an almost massless dilaton.
A number of parameters can be fixed by inflationary physics, allowing to make specific predictions for any subsequent period. In particular, we derive a relation between the tilt of the primordial spectrum of scalar fluctuations, $n_s$, and the present value of the equation of state parameter of dark energy, $\omega_{DE}^0$. We find bounds for the scalar tilt, $n_s<0.97$, the associated running, $-0.0006<d\ln n_s/d\ln k\lesssim-0.00015$, and for the scalar-to-tensor ratio, $0.0009\lesssim r<0.0033$, which will be critically tested by the results of the Planck mission. For the equation of state of dark energy, the model predicts $\omega_{DE}^0>-1$.
The relation between $n_s$ and $\omega_{DE}^0$ allows us to use the current observational bounds on $n_s$ to further constrain the dark energy equation of state to $0< 1+\omega_{DE}^0< 0.02$, which is to be confronted with future dark energy surveys.
\end{abstract}

\maketitle
\vspace*{-2mm}

\section{Introduction}\label{introduction}
At the classical level, the Lagrangian describing the Standard Model of particle physics (SM)
minimally coupled to General Relativity (GR) contains three dimensional parameters: Newton's constant G, the vacuum expectation value (vev)
of the Higgs field or, equivalently, the Higgs boson mass and a possible cosmological constant $\L$.
The masses of quarks, leptons and intermediate vector bosons are induced
by the vev of the Higgs field.
At the quantum level, additional scales, such as $\Lambda_\textrm{QCD}$ and all other scales related to
the running of coupling constants, appear due to dimensional
transmutation.
It is tempting to look for models in which all
these seemingly unrelated scales have a common origin.

In this work we present a detailed analysis of a model realizing this idea, proposed in \cite{Shaposhnikov:2008xb}. We will refer to it as the Higgs-Dilaton model. The model is based on a minimal extension of the SM and GR that contains 
no dimensional parameters in the action and is  therefore scale-invariant at the classical level. Scale invariance is achieved by introducing a new scalar degree of freedom, called Dilaton. The motivation of the model relies on the assumption that the
structure of the theory is not changed at the quantum level. In other
words, the full quantum effective action should still be
scale-invariant and the effective scalar potential should preserve the features of the classical potential. A perturbative quantization procedure maintaining scale invariance was presented in \cite{Shaposhnikov:2008xi} (see also \cite{Englert:1976ep}).
In the Higgs-Dilaton model all scales are induced by the spontaneous breakdown
of scale invariance (SI). As a consequence of the broken symmetry, the physical dilaton is exactly massless. 
Replacing GR by Unimodular Gravity (UG), in which the metric determinant
is fixed to one, ${|g|=1}$, results in the appearence of an arbitrary integration constant in the equations of motion, representing an additional breaking of scale symmetry. As discussed in \cite{Shaposhnikov:2008xb}, in theories with scalar fields non-minimally coupled to gravity, this constant effectively gives rise to a non-trivial potential for the scalar fields. In the case of the Higgs-Dilaton model, the new potential is of the ``run-away'' type in the direction of the dilaton.

While the dynamical breakdown of the scale symmetry by the Higgs field can provide a mechanism for inflation in the early universe \cite{Bezrukov:2007ep}, the light dilaton, practically decoupled from all SM fields, can act as Quintessence (QE), i.e. as dynamical Dark Energy (DE). We find that, under some assumptions, it is possible to relate the observables associated to inflation to those associated to dark energy. Namely, we establish a functional relation between the predicted value for the tilt $n_s$ of the primordial scalar power spectrum and the predicted equation of state parameter $w_{DE}^0$ of dark energy. Further, we find a relation involving the corresponding second order quantities, i.e the running $\alpha_\zeta$ of the spectral tilt and the rate of change $w_{DE}^a$ of the DE equation of state. 

The paper is organized as follows: In section \ref{msi} we introduce and discuss the minimal scale-invariant extension of the Standard Model and General Relativity. In Section \ref{siug}  the idea of Unimodular Gravity is described and applied to the scale-invariant model. We then discuss the cosmology of the resulting Higgs-Dilaton Model. The inflationary period is studied in detail in Section \ref{hdi}. The implications of the model for the late dark energy dominated stage are studied in Section \ref{lu}. Finally, conclusions are presented in Section \ref{conclusions}. For completness, an analysis of slow-roll inflation in the Jordan frame and the differences with respect to the Einstein frame are presented in Appendix \ref{appendix}.

\section{The Higgs-Dilaton Model}\label{model}
In this section we review the Higgs-Dilaton model of \cite{Shaposhnikov:2008xb}, which consists of two moderate extensions of the Standard Model and General Relativity (SM plus GR). In subsection \ref{msi} we show how the introduction of a dilaton allows to extend SM plus GR to a phenomenologically viable scale-invariant theory. After discussing the main properties of the resulting theory (\ref{itd}), we discuss two naturalness issues, the Cosmological Constant Problem and the Gauge Hierarchy Problem, in the context of this model (\ref{natural}). Next, we give some arguments in favor of a particular parameter choice corresponding to the absence of a cosmological constant (\ref{casebeta0}). In subsection \ref{siug} the construction of the model is completed by replacing GR by Unimodular Gravity (UG). The qualitative picture of cosmology in the Higgs-Dilaton model, as found in \cite{Shaposhnikov:2008xb}, is recalled in subsection \ref{qp}.

\subsection{Minimal scale-invariant extension of SM plus GR}\label{msi}

\subsubsection{Introducing the Dilaton}\label{itd}
Let us start by writing down the Lagrangian density that combines GR and the SM\footnote{We use the conventions $\eta_{\mu\nu}=\mathrm{diag}(-1,1,1,1)$ and $R^\alpha_{\phantom{\alpha}\beta\gamma\delta}=\partial_\gamma\Gamma^\alpha_{\beta\delta}+\Gamma^\alpha_{\lambda\gamma}\Gamma^\lambda_{\beta\delta}-(\gamma\leftrightarrow\delta)$.} 
\begin{equation}
\frac{\mathcal{L}}{\sqrt{-g}}=\frac{1}{2}M_P^2 R+{\cal L}_{\textrm{SM}[\l\rightarrow
0]}-\lambda\left(\varphi^\dagger \varphi-v^2\right)^2-\Lambda\;,
\end{equation}
where the first term is the usual Einstein-Hilbert action for GR with $M_P=(8\pi G)^{-1/2}$, the
second term is the SM Lagrangian without the Higgs potential, the
third term is the Higgs potential with the SM Higgs doublet $\varphi$ and its vacuum expectation value $v$, and $\Lambda$ is a cosmological
constant. In this standard theory, to which we will refer as ``SM plus GR'', classical scale invariance is violated by the presence of the
dimensional constants $M$, $v$ and $\Lambda$. Our goal is to let these
scales be dynamical, i.e. replace them by a field. The most obvious
solution, without introducing new degrees of freedom, would be to let
the Higgs field be responsible for all scales. This corresponds
to considering the Lagrangian\footnote{With the conventions used here, a conformally coupled scalar field has $\xi=-1/6$.}
\begin{equation}\label{SIhiggs}
\frac{\mathcal{L}}{\sqrt{-g}}=\xi \varphi^\dagger \varphi R+{\cal
L}_{\textrm{SM}[\l\rightarrow 0]}-\lambda\left(\varphi^\dagger
\varphi\right)^2\;,
\end{equation}
$\xi$ being a new real parameter ("non-minimal coupling").
The associated action is now scale-invariant, i.e. invariant under the global transformations,
\begin{equation}
g_{\mu\nu}(x)\mapsto g_{\mu\nu}(\sigma x)\;,\hspace{10mm}
\Phi(x)\mapsto \sigma^{d_\Phi} \Phi(\sigma x)\label{SItrans}\;,
\end{equation}
where $\Phi(x)$ stands for the different particle physics fields, $d_\Phi$ is their associated scaling dimension and $\sigma$ is an arbitrary real parameter. In a theory that is invariant under all diffeomorphisms (Diff invariant), as is the case for \eqref{SIhiggs}, the symmetry associated to the absence of dimensional parameters can equivalently be written as an internal transformation\footnote{Note that in a theory that is invariant only under a restricted class of diffeomorphisms, such as UG, the absence of dimensional parameters will still guarantee invariance under \eqref{SItrans} but 
not under \eqref{SItransint}.}
\begin{equation}
g_{\mu\nu}(x)\mapsto \sigma^{-2} g_{\mu\nu}(x)\;,\hspace{10mm}
\Phi(x)\mapsto \sigma^{d_\Phi} \Phi(x)\label{SItransint}\;.
\end{equation}

Can the Lagrangian \eqref{SIhiggs} give a satisfactory phenomenology?
Since we are looking for a theory that should eventually be quantized, we want to introduce the requirement that the theory possesses a "Classical Ground State''. The term ``Classical Ground State'' will be used throughout
 this work to refer to solutions of the classical equations of motion, which correspond to constant fields in the particle physics sector 
of the theory and a maximally symmetric geometry, i.e. Minkowski (flat), de Sitter (dS) or Anti de Sitter (AdS) spacetime. The existence 
of such a ground state might
be essential for a consistent quantization of the theory. At the quantum level, the theory should possess a ground state that breaks scale
 invariance and in this way induces masses and dimensional couplings for the excitations (particles). We will require that
 this spontaneous symmetry breaking already appears in the classical theory due to the existence of a symmetry breaking classical ground state\footnote{The authors 
of \cite{Jain:2007ej,Jain:2009ab} propose that scale symmetry could be broken by the pure presence of a time-dependent cosmological background.}.

Let us now look for symmetry-breaking classical ground states in the theory \eqref{SIhiggs}.
If gravity is neglected, i.e. the first term in the Lagrangian is dropped,
the classical ground states correspond to the minima of the scalar potential
$\lambda\left(\varphi^\dagger\varphi\right)^2$.
The only possibility for them to break the scale symmetry,
$\varphi=\varphi_0\neq 0$, is to set $\lambda=0$. In this case the theory possesses an infinite family
of classical ground states satisfying $2 \varphi^\dagger \varphi=h_0^2$, where $h_0$ is an
arbitrary real constant.
If one includes gravity, the set of possible classical ground states becomes richer.
Namely, even if $\lambda\neq 0$
the theory possesses a continuous family of classical ground states satisfying
$2 \varphi^\dagger \varphi=h_0^2$ and $R=4\lambda h_0^2/\xi$, where $h_0$
is an arbitrary real constant. The states with $h_0\neq 0$
break scale invariance spontaneously and induce all scales at the
classical level.
Hence, the goal of having a classical theory in which
all scales have the same origin, spontaneous breakdown of SI, is
achieved. However, the above theory is in conflict with experimental
constraints. In fact, although the non-zero background value of
$\varphi$ gives masses to all other SM particles, the excitations of
the Higgs field itself are massless and, moreover, decoupled from
the SM fields \cite{Salopek:1988qh}. This fact is seen most easily if the
Lagrangian is written in the Einstein-frame by defining the new metric $\tilde
g_{\mu\nu}=M_P^{-2}\xi \varphi^\dagger \varphi g_{\mu\nu}$ and the new canonical Higgs field $\tilde\varphi=M_P\sqrt{1/\xi+6}\ln(\varphi/M_P)$. (This type of variable change will be discussed in detail in section \ref{hdi}.) In the new variables, the SI of the original formulation corresponds to a shift symmetry for
the Higgs field $\tilde\varphi$, which is the massless Goldstone boson associated
to the spontaneous breakdown of SI. A Higgs field with these properties is excluded by Electroweak precision tests \cite{Nakamura:2010zzi}.
Therefore, in order to construct a viable SI theory, it seems
unavoidable to introduce new degrees of freedom. 

The next simplest possibility is to add a new singlet scalar field to the theory. We will refer to 
it as the dilaton $\chi$. The scale-invariant extension for the SM plus GR including the
dilaton reads
\begin{align} \label{SI}
 \frac{\mathcal{L}_{SI}}{\sqrt{-g}}=&\frac{1}{2}\left(\xi_\chi \chi^2 
+2 \xi_h \varphi^\dagger \varphi\right)R+{\cal
L}_{\textrm{SM}[\l\rightarrow 0]} 
-\frac{1}{2}g^{\mu\nu}\partial_\mu\chi\partial_\nu\chi - V(\varphi,\chi)\;,
\end{align}
where the scalar potential is given by\footnote{The parametrization chosen for the scalar potential assumes
${\l\neq 0}$. This only excludes the phenomenologically unacceptable case
 where a quartic term $(\varphi^\dagger \varphi)^2$ is absent.}
\begin{equation}\label{pot}
 V(\varphi,\chi) =
\lambda\left(\varphi^\dagger
\varphi-\frac{\alpha}{2\lambda}\chi^2\right)^2+\beta\chi^4~.
\end{equation}
We will only consider positive values for $\xi_\chi$ and $\xi_h$, such that the coefficient in front
of the scalar curvature is positive, whatever values the scalar fields
take. The positivity of the non-minimal coupling parameters is at the same time the condition for semi-positive definitness of the scalar field kinetic terms. By construction, the action associated to \eqref{SI} is invariant under \eqref{SItrans}, 
respectively \eqref{SItransint}. The theory should possess a symmetry-breaking
classical ground state with $\varphi=\varphi_0\neq 0$ and $\chi=\chi_0\neq 0$. The case $\varphi_0=0$
would correspond to a theory with no electroweak symmetry breaking, while the case $\chi_0=0$ would result in a theory with a massless Higgs field. Both these cases are phenomenologically unacceptable.

Let us again start by neglecting the gravitational part of the action. In its absence,
the ground states correspond to the minima of the potential \eqref{pot}.
It is easy to see that the only possibility to get a ground state satisfying
$\varphi_0\neq0$ and $\chi_0\neq 0$ is to have a potential with a flat direction, i.e. $\alpha>0$ and $\beta=0$, as well
as $\lambda>0$ for stability. The corresponding family of classical ground states is given by $2\varphi^\dagger
\varphi=h_0^2$ and $\chi=\chi_0$ with $h_0^2=\frac{\alpha}{\lambda}\chi_0^2$, where $\chi_0$ is an arbitrary real constant.

Like before, the inclusion of gravity results in the appearance of additional classical ground states for $\beta\neq 0$,  given by
\begin{equation}
h_0^2=\frac{\alpha}{\lambda}\chi_0^2+\frac{\xi_h}{\lambda}R\;,\hspace{10mm} R=\frac{4\beta\lambda\chi_0^2}{\lambda\xi_\chi+\alpha\xi_h}\;.
\label{idgs}
\end{equation}
The solutions with $\chi_0\neq 0$ spontaneously break SI. All scales are induced and 
proportional to $\chi_0$. For instance, one can directly identify the Planck scale as
\begin{equation}
M_P^2=\xi_\chi \chi_0^2+\xi_h h_0^2=\left(\xi_\chi+\xi_h\frac{\alpha}{\lambda}+\frac{4 \beta\xi_h^2}{\lambda\xi_\chi+\alpha\xi_h}\right)\chi_0^2\;.
\end{equation}
Depending on the value of $\beta$, the background corresponds to flat spacetime ($\beta=0$), de Sitter or Anti de Sitter spacetime of
 constant scalar curvature $R$, corresponding to a cosmological constant
\begin{equation}\label{coscon}
\Lambda=\frac{1}{4}M_P^2R=\frac{\beta M_P^4}{(\xi_\chi+\frac{\alpha}{\lambda}\xi_h)^2+4\frac{\beta}{\lambda}\xi_h^2}\;.
\end{equation}
The spectrum of perturbations around a symmetry-breaking solution contains the usual massless spin-2 perturbation in the gravitational sector. The scalar sector 
contains an excitation with mass 
\begin{equation}
m^2=2\alpha M_P^2\frac{(1+6\xi_\chi)+\frac{\alpha}{\lambda}(1+6\xi_h)}{\xi_\chi(1+6\xi_\chi)+\xi_h\frac{\alpha}{\lambda}(1+6\xi_h)}+\mathcal{O}(\beta)\;,
\end{equation}
which will play the role of the physical SM Higgs field, plus a massless Goldstone boson (both perturbations are combinations of the fields $\chi$ and $h$). We use $h$ to
 denote the field $\varphi$ in the unitary gauge. Like in the standard Higgs mechanism, the excitations of the Standard Model fields get
masses proportional to $h_0$. If one extends the SM by introducing right-handed neutrinos
\cite{Asaka:2005an,Asaka:2005pn}, these neutrinos get induced masses proportional to  $\chi_0$\footnote{Gauge invariance does not allow for couplings of $\chi$ to SM fields.}.

In the described model, physics is completely independent
of the value of $\chi_0$, as long as $\chi_0\neq 0$.
This is because only dimensionless ratios
of the different scales can be measured. Therefore, parameters of the model have to be chosen
such that these ratios correspond to the measured ones. For instance,
one should reproduce the hierarchies between the cosmological scale and the
electroweak scale, i.e. $\Lambda/m^4\sim \mathcal{O}(10^{-56})$,
as well as the ratio between the electroweak and
the gravitational scale $m^2/M_P^2\sim \mathcal{O}(10^{-32})$. We
choose the parameter $\beta$ to be responsible for the first ratio and $\alpha$ for the
second ratio. Therefore, these parameters have to take values satisfying $\beta\lll\alpha\lll 1$ and
$\beta, \alpha\lll \xi_\chi, \xi_h$. One then gets approximately  $\Lambda/m^4\simeq \frac{\beta}{\xi_\chi^2}$
and $m^2/M_P^2\simeq\frac{2\alpha}{\xi_\chi}$.
Note that the order of magnitude relation $\sqrt{\Lambda}/m^2\sim m^2/M_P^2$ respectively
$\sqrt{\beta}\sim \alpha$ is the big number coincidence pointed out
by Dirac \cite{Dirac:1938mt}. However, the present model
does not address the question about the
origin of the big differences between theses scales, i.e. the smallness of $\alpha$ and $\beta$, nor does it explain their approximate relation. The non-minimal couplings $\xi_\chi$ and $\xi_h$
will be constrained by cosmological considerations, and $\lambda\lesssim
\mathcal{O}(1)$, as it corresponds to the self coupling of the Higgs field.
Therefore, one can fix the values of $\alpha$ and $\beta$ that give the correct
ratios. In the same fashion, one has to choose values for the SM Yukawa
couplings that produce the observed mass ratios.

As the theory contains a new massless degree of freedom, the dilaton, one has to make sure that it
does not contradict any experimental bounds. A detailed analysis of
the interactions between this massless field and the SM fields is contained in \cite{Blas:2011ac}. Let us cite the relevant findings of that work. It turns out that as a consequence of SI, the massless scalar field completely decouples from all SM fields except for the Higgs field. Since the massless field is the Goldstone boson associated to the broken scale symmetry, there exists a set of field variables in terms of which it couples to the physical Higgs field only derivatively. In addition, for an appropriate choice of field variables, these interactions appear as non-renormalizable operators, suppressed by the scale $M_P/\xi_h$. The analysis of section \ref{hdi} will show that $\xi_h\sim 10^{5}$. The suppression scale of non-renormalizable operators is therefore considerably lower than the Planck scale, but still much larger than all known Particle physics scales.

Other deviations from the SM appear as a consequence of the non-minimal couplings to gravity. In fact, the physical Higgs field, i.e. the field that couples to 
the SM degrees of freedom, is not $h$, but a combination of $h$ and $\chi$. It was shown in \cite{Blas:2011ac} that the resulting deviations from the SM are suppressed by the ratio $m^2/M_P^2$ 
between the physical Higgs mass and the Planck mass, respectively by the small parameter $\alpha$. While the new massless field hardly affects SM phenomenology, we will see that it might play an important role in cosmology. 

At the classical level, the above theory successfully implements the idea that all scales are consequences of the spontaneous breaking of SI. 
All conclusions remain true if SI and the features of the potential can be maintained at the quantum level (in this context, see \cite{Shaposhnikov:2008xi,Shaposhnikov:2008ar,Shaposhnikov:2009nk}). In that case, the presented model is a viable effective field-theory extension of the SM and GR.

\subsubsection{Naturalness issues}\label{natural}
The presented theory contains two important fine tunings related to the very big differences between the Planck scale $M_P$, the electroweak scale $m$ and the cosmological 
scale $\L$. At the quantum level, this can lead to two much-discussed naturalness issues. One of them is part of the Cosmological Constant Problem. In 
standard SM plus GR the effective cosmological constant is the sum of a bare constant and radiative corrections proportional 
to the particle physics mass scales of the theory, e.g. the electroweak scale. Matching the effective cosmological constant with its observed value, tiny compared to, for 
instance, the electroweak scale, requires a tremendous fine-tuning of the bare cosmological constant. In the case of the scale-invariant theory discussed
 here, the situation is somewhat different. Exact SI forbids a term $\sqrt{-g}\Lambda$ in the action.
 Also, if the quantization procedure respects SI, such a term is not generated radiatively. However, as we saw above, 
due to the non-minimal couplings of the scalar fields to gravity, the cosmological constant is in fact associated to the term $\beta\chi^4$. Now, this term 
is not forbidden by scale invariance. Therefore, even if scale invariance can be maintained at the quantum level, the quantum effective potential will contain 
a term $\b_{\rm eff}\chi^4$, where $\b_{\rm eff}$ is a combination of the bare value of $\b$ and other non-dimensional couplings of the theory. These other 
couplings are generally much bigger than the value of $\b_{\rm eff}$ that corresponds to the observed cosmological constant. So, again a strong fine-tuning is needed 
in order to keep $\b_{\rm eff}$ sufficiently small. This tells us that the Cosmological Constant Problem also exists in an exactly scale-invariant theory of the 
type proposed here.

The second naturalness issue is related to the mass of the Higgs boson and is commonly called ``Gauge Hierarchy Problem''. The problem is twofold. The effective
 field theory combining the SM with GR contains two extremely different mass scales, namely, the electroweak scale $v=246~\mathrm{GeV}$ ($v$ being 
the vacuum expectation value of the Higgs field) and the Planck scale $M_P=(8\pi G)^{-1/2}=2.44\cdot 10^{18}~\mathrm{GeV}$. It is considered unnatural
 to have such a huge difference between two scales of the same theory. This is the first part of the Gauge Hierarchy Problem. In the considered type
 of scale-invariant theories, the big difference between the electroweak and the Planck scale remains unexplained.

The other part of the Gauge Hierarchy Problem is related to the stability of the Higgs mass against radiative corrections (for a recent discussion see e.g.
 \cite{Shaposhnikov:2007nj}). Much like the cosmological constant, the mass of the Higgs field gets radiative corrections proportional to the other particle physics mass scales of
 the theory. The logic is the same as in the case of the cosmological constant.
 If there exists a particle physics scale much bigger than the electroweak scale, the measured value of the electroweak scale can only be explained by an important
 fine-tuning of parameters.
In other words, if there exists a new particle physics scale between the electroweak scale $m$ and the Planck scale $M_P$, the ``smallness'' of the Higgs mass constitutes a serious 
theoretical issue. This issue still appears in an exactly scale-invariant theory with spontaneous breaking of the scale symmetry.

If the theory contains no intermediate particle physics scale between $m$ and $M_P$, the situation is different. In that case, whether or not the Higgs mass should be expected to 
contain big radiative corrections of the order $M_P$ depends on the ultraviolet (UV) completion of the theory. At the level of the low-energy effective field theory, the
 UV properties can be
 encoded in the choice of the renormalization scheme.
A renormalization scheme based on the assumption that the UV completion is scale-invariant, and which does not bring in extra particle physics scales, was presented in \cite{Shaposhnikov:2008xi} (see also \cite{Englert:1976ep}). If this scheme
 is applied to the considered minimal scale-invariant extension of SM plus GR, the Higgs mass does 
not obtain corrections proportional to $M_P$ (induced by the vacuum expectation value of the dilaton) and there is no problem of stability of the Higgs mass against radiative corrections. Hence, SI makes for the absence of this part of the Gauge Hierarchy Problem.

\subsubsection{The special case $\beta=0$}\label{casebeta0}
We now want to give some arguments in favor of the case $\beta=0$. This case corresponds to the existence of a flat direction in the Jordan-frame potential \eqref{pot} and hence to the absence of a cosmological constant.

The reasoning of the precedent paragraph tells us that choosing $\b=0$ corresponds to a fine-tuning of the parameters, especially at the quantum level, just like putting $\Lambda=0$ in standard SM plus GR.
From this point of view, such a parameter choice should clearly be disfavored.
Nevertheless, we think that the case $\beta=0$ is specially interesting. One reason is that only if $\beta=0$, SI can be spontaneously broken in the absence of gravity. Put in other words, $\beta=0$ allows flat space-time together with ${(\varphi,\chi)=(\varphi_0,\chi_0)\neq(0,0)}$ to be a classical solution.

Another argument is related to the stability of the ground state. As
discussed above, a scale-invariant theory with spontaneous symmetry
breaking always contains a massless scalar degree of freedom,
Goldstone boson, independently of the value of $\beta$. Now, if
$\beta\neq 0$, the background spacetime of the theory corresponds to
de Sitter (or anti de Sitter) space-time. It is known, however, that a
massless scalar field is unstable in de Sitter spacetime
\cite{Allen:1987tz}.  There are also indications that this is the case for 
the 4-dimensional AdS \cite{Bizon:2011gg}.
Therefore, it is conceivable that a consistent quantization of the
theory might rely on the requirement $\beta=0$ and hence the existence
of flat space-time as a solution (see also \cite{Antoniadis:1985pj,Tsamis:1992sx,
Tsamis:1994ca,Antoniadis:2006wq,Polyakov:2009nq}). 

A third aspect appears in the context of cosmology. The theory with $\beta=0$, not containing a cosmological constant, does not seem to withstand the confrontation with cosmological observations. Just like the case $\beta<0$ (AdS) it can not explain the observed accelerated expansion of the universe without introduction of a new dark energy component. From this point of view, the only viable option seems to be $\beta>0$ (dS). This conclusion is correct if gravity is described by GR. However, as we will see in the upcoming section, the situation is very different if GR in \eqref{SI} is replaced by Unimodular Gravity. In that case, the appearance of an arbitrary integration constant will give rise to a potential for the Goldstone boson of broken scale invariance. As a consequence, for appropriate parameter values and initial conditions, the now pseudo-Goldstone boson can act as a dynamical dark energy component. In this new situation, the case $\beta=0$ will again be peculiar, because it is the only case where dark energy is purely dynamical and has no constant contribution.

Based on these reasons, in what follows, we will single out the case $\beta=0$ and study the associated phenomenology in more detail.

\subsection{Combining scale invariance and Unimodular Gravity}\label{siug}
We now want to add to the idea of SI the idea of Unimodular Gravity (UG) \cite{vanderBij:1981ym,Wilczek:1983as,Zee:1985ug,Buchmuller:1988wx,Unruh:1988in,Weinberg:1988cp,Henneaux:1989zc,Buchmuller:1988yn} and apply it to the Higgs-Dilaton scenario. In UG one reduces the independent components of the metric $g_{\mu\nu}$ by one, imposing that the metric determinant $g\equiv\det(g_{\mu\nu})$ takes some fixed constant value. Conventionally one takes ${|g|=1}$,
hence the name. Fixing the metric determinant to one is not a strong restriction, in the
sense that the family of metrics satisfying this requirement can still
describe all possible geometries.

If we impose the unimodular constraint, the scale-invariant Lagrangian \eqref{SI} becomes
\begin{align} \label{SI-UG}
  {\cal L}_{SI-UG}=&\frac{1}{2}\left(\xi_\chi \chi^2 
+2 \xi_h \varphi^\dagger \varphi\right)\hat R+\hat{\cal
L}_{\textrm{SM}[\l\rightarrow 0]} 
-\frac{1}{2}\hat g^{\mu\nu}\partial_\mu\chi\partial_\nu\chi - V(\varphi,\chi)\;,
\end{align}
where a hat on a quantity, like $\hat{R}$, indicates that it depends on the unimodular metric $\hat g_{\mu\nu}$, which satisfies $\det\hat g_{\mu\nu}=-1$. The potential $V(\varphi,\chi)$ is still given by \eqref{pot}.
As a consequence of the unimodular constraint, the action associated to $ {\cal L}_{SI-UG}$ is no longer invariant under all diffeomorphisms (Diff), but only under transverse diffeomorphisms (TDiff), i.e. coordinate transformations $x^\mu\mapsto x^\mu+\xi^\mu(x)$, with the
condition $\partial_\mu\xi^\mu=0$.
Just as in pure UG (not including non-gravitational fields) the equations of motion derived from the Lagrangian \eqref{SI-UG} contain an arbitrary integration constant $\Lambda_0$, which can be interpreted as an additional initial condition.
It was shown in \cite{Shaposhnikov:2008xb} that the classical solutions obtained from the Lagrangian \eqref{SI-UG} are equivalent to the solutions derived from the equivalent Diff invariant Lagrangian
\begin{align}\label{SI-UGe}
  \frac{{\cal L}_e^{SI-UG}}{\sqrt{-g}}=&\frac{1}{2}\left(\xi_\chi \chi^2 
+2 \xi_h \varphi^\dagger \varphi\right)R+{\cal
L}_{\textrm{SM}[\l\rightarrow 0]} 
-\frac{1}{2}g^{\mu\nu}\partial_\mu\chi\partial_\nu\chi  - V(\varphi,\chi)-\Lambda_0 \;,
\end{align}
where $\Lambda_0$ is the mentioned arbitrary constant. While in the original formulation \eqref{SI-UG} the dimensional constant $\Lambda_0$ only appears in the equations of motion and thereby spontaneously breaks SI, in the equivalent Diff invariant formulation \eqref{SI-UGe} the same constant appears as an explicit symmetry breaking in the action. Nevertheless, this constant should not be understood as a parameter in the action, but rather as an arbitrary initial condition. Given the equivalence of the two formulations, in order to study the phenomenology issued by \eqref{SI-UG}, we will simply study the theory given by \eqref{SI-UGe} for different values of $\Lambda_0$.\footnote{Of course, the choice to analyze the theory in the Diff invariant rather than in the original formulation is purely a matter of convenience.}

We now turn our attention to the physical implications of the
term proportional to $\Lambda_0$. In the first instance, let us only
consider the gravitational and the scalar sector of the theory, i.e.
\begin{equation}
  \frac{{\cal L}}{\sqrt{-g}}=\frac{1}{2}\left(\xi_\chi \chi^2 
+\xi_h h^2\right)R -
\frac{1}{2}(\partial_\mu\chi)^2-
\frac{1}{2}(\partial_\mu h)^2 
- V(h,\chi)-\Lambda_0\;,
\end{equation}
where $h$ is the Higgs field in the unitary gauge.
In order to simplify the physical interpretation we define the
Einstein-frame (E-frame) metric\footnote{The Lagrangian in terms of the original variables is said to be written in the Jordan-frame (J-frame).}
\begin{equation}\label{contran}
 \tilde{g}_{\mu\nu}=M_P^{-2}\left(\xi_\chi \chi^2 
+\xi_h h^2\right)g_{\mu\nu}
\end{equation} 
in terms of which the Lagrangian reads
\begin{equation}\label{scalareinstein2}
 \frac{{\cal
L}}{\sqrt{-\tilde{g}}}=M_P^2\frac{\tilde{R}}{2}-\frac{1}{2}\tilde K
-\tilde U(h,\chi)\;,
\end{equation} 
where $\tilde K$ is a non-canonical but positive definite kinetic term (given below in \eqref{lagrangianE}) and 
$\tilde U(h,\chi)$ is the E-frame potential given by
\begin{equation}\label{potE}
 \tilde U(h,\chi)=\frac{M_P^4}{\left(\xi_\chi \chi^2 
+\xi_h h^2\right)^2}\left(\frac{\lambda}{4}\left(h^2-\frac{\alpha}{\lambda}
\chi^2\right)^2+\beta\chi^4+\Lambda_0\right)\;.
\end{equation} 
Note that the E-frame potential gets singular at
$\chi=h=0$. The reason is that at this point the conformal transformation \eqref{contran} is singular and the change to the E-frame is not allowed. Since for $\chi=h=0$ scale invariance is not broken, we will not be interested in the theory around this point.
Let us discuss the shape of the E-frame potential and the classical ground states of the theory for $\alpha,\l,\xc,\xh>0$
(cf. figure \ref{epotplot}).
If $\Lambda_0=0$, in which case the theory \eqref{SI-UGe} reduces to \eqref{SI}, the potential is minimal along the two valleys
\begin{gather}\label{valleys}
h_0^2=\frac{\alpha}{\lambda}\chi_0^2+\frac{4 \beta\xi_h\chi_0^2}{\lambda\xi_\chi+\alpha\xi_h}\;,
\end{gather}
They correspond to the infinitely degenerate family of classical ground states found in \eqref{idgs}.
As before, if $\beta=0$, the potential vanishes at its minimum, while a non-zero $\beta$
gives rise to a cosmological constant \eqref{coscon}.
In other words, spacetime in the classical ground state is Minkowskian, dS or AdS. These are the results we have already discussed section \ref{msi}. As soon as $\Lambda_0\neq 0$ the valleys get a tilt, which lifts the degeneracy of the classical ground states. For $\Lambda_0<0$ the valleys are tilted towards the origin. The true classical ground state for this case is the trivial one, 
$\chi=h=0$. Hence, we discard this possibility. For $\Lambda_0>0$ the potential is tilted away from the origin, it is of the run-away type. In this case the theory has an asymptotic classical ground state, given by \eqref{valleys} with $\chi_0\rightarrow\infty$. Again, depending on the value of $\beta$ this asymptotic solution corresponds to Minkowski, dS or AdS spacetime with curvature given by \eqref{coscon}.

We see that as a consequence of the non-minimal coupling between the scalar fields and gravity, the arbitrary integration constant $\Lambda_0$ does not play the role of a cosmological constant (as it does in pure UG) but rather gives rise to a peculiar potential for the scalar fields. For $\Lambda_0>0$ the potential is of the run-away type. In the following sections we will see that such a potential can have an interesting cosmological interpretation. In fact, the evolution of the scalar fields along the valley can give rise to dynamical dark energy (quintessence). We will focus on the case $\beta=0$, where dark energy does not contain a constant contribution and is purely due to the term proportional to $\Lambda_0$ (cf. arguments in section \ref{casebeta0}).

While the term proportional to $\Lambda_0$ can play an important role in cosmology, its presence barely affects the particle physics phenomenology of the model. In fact, if the run-away potential is of the order of magnitude of the present dark energy density, the time evolution of the background scalar fields along the valley can be neglected on particle physics time scales. Also, the additional interactions between the Higgs field and the dilaton that are induced by this potential are negligibly small.

\subsection{Higgs-Dilaton Cosmology -- The Qualitative Picture}\label{qp}
\begin{figure}

\begin{center}
\begin{tabular}{ccc}
\includegraphics[scale=0.26]{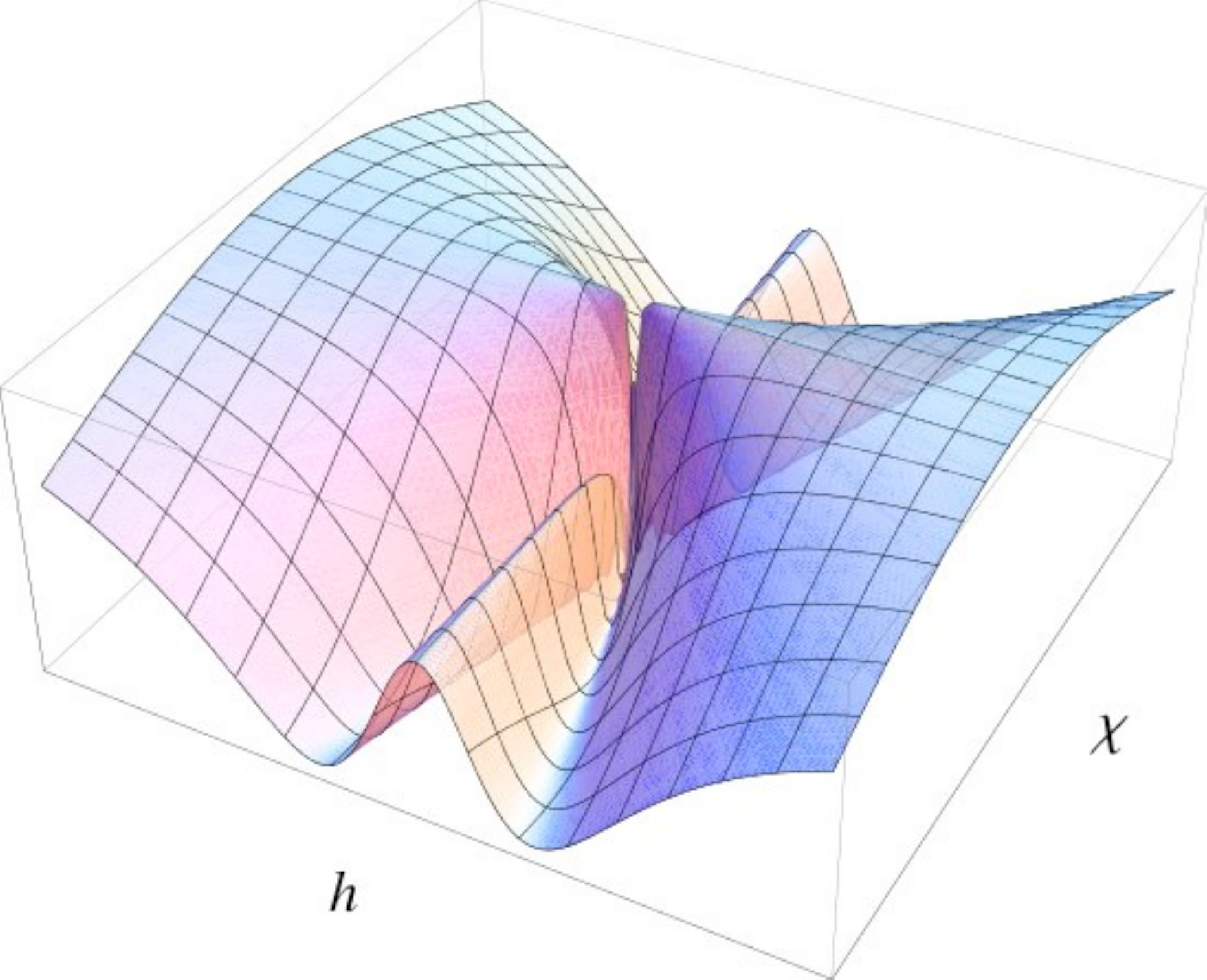}
&
\includegraphics[scale=0.26]{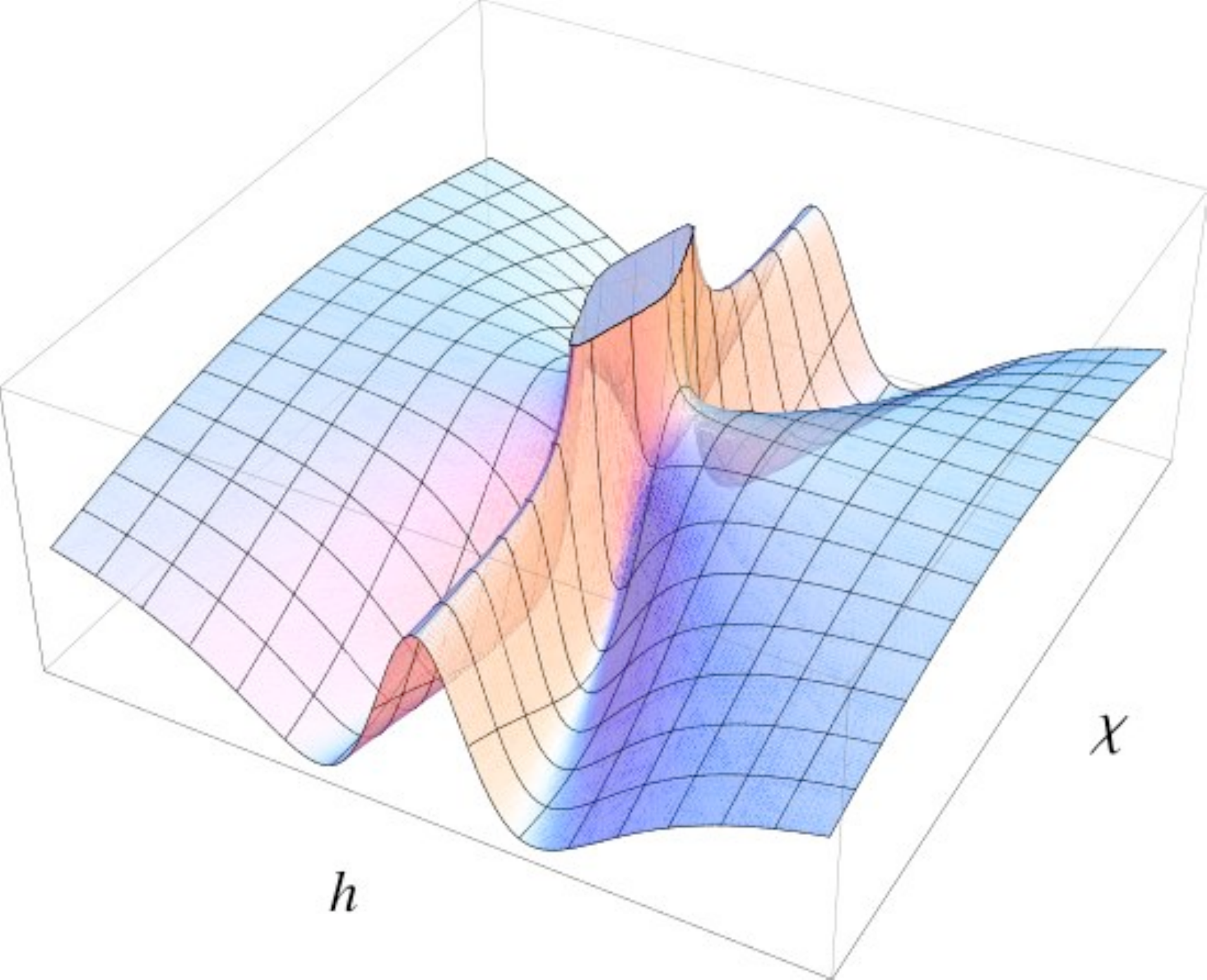}
&
\includegraphics[scale=0.26]{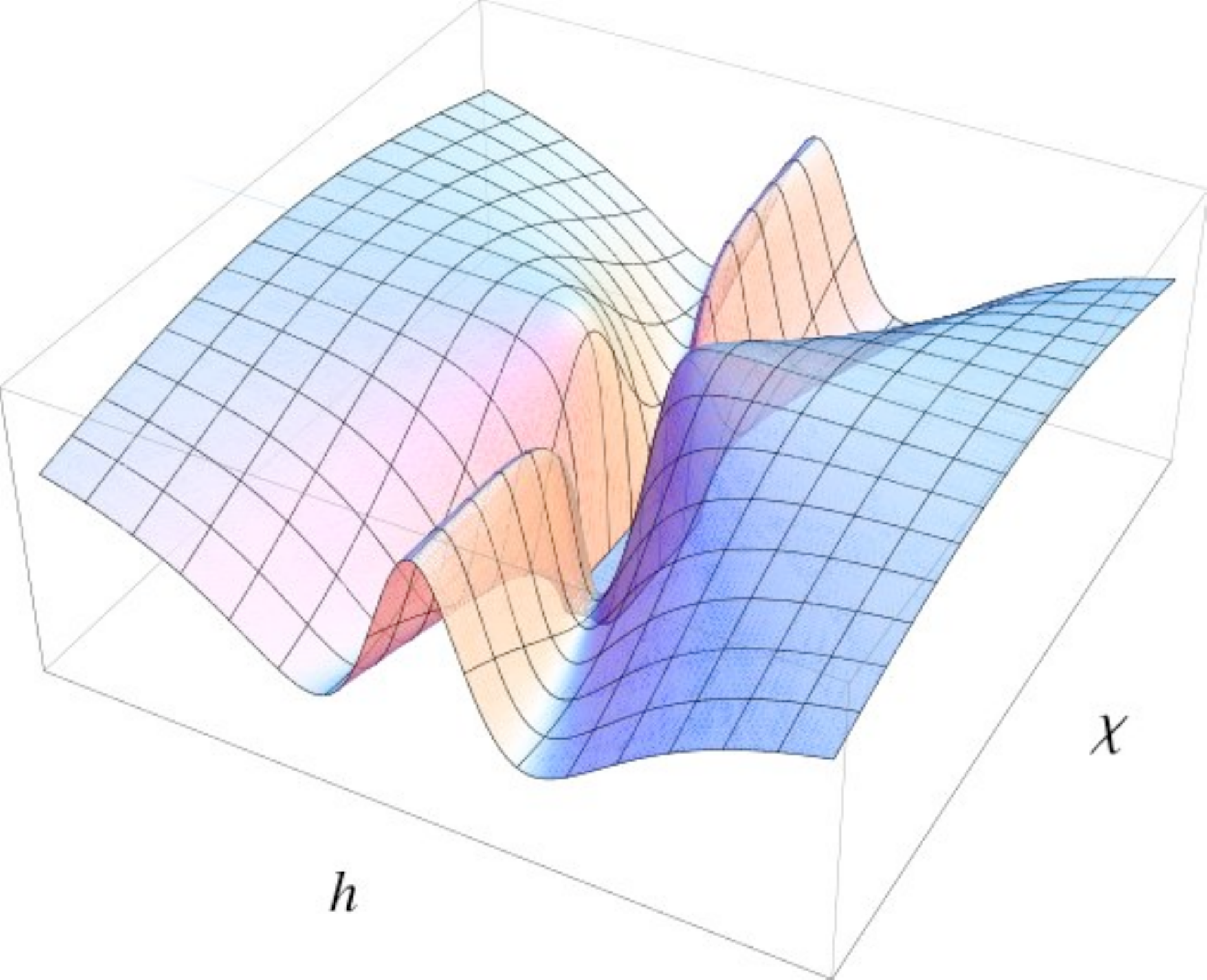}
\\
$\L_0=0$&$\L_0>0$&$\L_0<0$
\end{tabular}
\end{center}
\caption{These plots show the shape of the E-frame potential $\tilde U (h,\chi)$ (equation \eqref{potE0}) for $\Lambda_0=0$, $\Lambda_0>0$ and $\Lambda_0<0$ respectively.}
 \label{epotplot}
 \end{figure}

In this subsection we want to qualitatively describe the cosmological scenario issued by the model \eqref{SI-UGe} (or equivalently \eqref{SI-UG}) presented in the previous two subsections (cf. \cite{Shaposhnikov:2008xb}). We consider the theory in the Einstein frame \eqref{scalareinstein2} and focus on the case $\beta=0$, for which the scalar field potential \eqref{potE} reduces to (cf. figure \ref{epotplot})
\begin{equation}\label{potE0}
 \tilde U(h,\chi)=\frac{M_P^4}{\left(\xi_\chi \chi^2 
+\xi_h h^2\right)^2}\left(\frac{\lambda}{4}\left(h^2-\frac{\alpha}{\lambda}
\chi^2\right)^2+\Lambda_0\right)\;.
\end{equation}

The scalar fields $\chi$ and $h$ are now considered to be homogeneous
background fields evolving in flat Friedmann--Robertson-Walker (FRW)
spacetime. Their evolution is affected by the
non-canonical nature of the kinetic term. However, since the kinetic
term is positive definite, in order to get a qualitative picture, it
is enough to look at the features of the potential. In the absence of
$\Lambda_0$, $\tilde U$ has its minima along the two valleys
$h^2=\frac{\alpha}{\lambda}\chi^2$.
The main effect of $\Lambda_0\neq 0$ is to give a
tilt to the valleys. As discussed in the previous subsection, $\Lambda_0<0$
is phenomenologically unviable. We will only consider the case
$\Lambda_0>0$, in which the valleys are tilted away from the origin.

For an appropriate choice of parameters, the
crude picture of the role of the cosmological scalar fields is the
following: If the scalar fields start off far from the valleys, $\Lambda_0$ can initially be neglected, and the scalar fields roll slowly towards one of the valleys. This roll-down can be responsible
for cosmic inflation. As inflation is mainly driven by the Higgs field, this phase is much like in the case of the Higgs-Inflation model of \cite{Bezrukov:2007ep}.

After the end of inflation preheating takes place. During this phase the scalar field dynamics is dominated by the field $h$ and hence preheating in the present model is expected to be very similar as in the Higgs-Inflation model \cite{Bezrukov:2008ut,GarciaBellido:2008ab} (see also \cite{Bezrukov:2011sz}): The gauge bosons created at the minimum of the potential acquire a large mass while the
Higgs field increases towards a maximal amplitude and starts to decay into all Standard Model leptons and quarks, rapidly depleting the occupation numbers of gauge bosons. The fraction of energy of the Higgs field that goes into SM particles is still very small compared to the energy contained in the oscillations, and therefore the non-perturbative decay is slow.  
As the Universe expands in a matter-like dominated stage with zero pressure, the amplitude of the Higgs field oscillations decreases. Eventually, this amplitude is small enough so that the gauge boson masses become too small to induce a quick decay of the gauge bosons. As a consequence, their occupation numbers start to grow very rapidly via parametric amplification. After about a hundred oscillations, the produced gauge bosons backreact on the Higgs field and the resonant production of particles stops. The Higgs field acquires a large mass via its interaction with the gauge condensate and preheating ends. From there on, the Higgs field as well as the gauge fields decay perturbatively until their energy is transferred to SM particles. 

The phase of preheating is followed by the usual radiation and matter dominated stages, during which the scalar fields are ``frozen'' at some point of the valley. Their energy density is now given by the $\Lambda_0$ term and practically constant. As a consequence, it eventually comes to dominate over radiation and matter and hence provides a dark energy component. In other words, the scalar fields rolling slowly down the potential valley play the role of a thawing quintessence field \cite{Wetterich:1987fm,Ratra:1987rm,Ferreira:1997hj,Caldwell:2005tm}. In this late stage, the fields satisfy $h(t)^2\simeq \frac{\alpha}{\lambda}\chi(t)^2$. On particle physics time-scales the time-variation of the background fields can be neglected. Perturbations around this almost constant symmetry
breaking background can be interpreted as the SM particles plus an
additional almost massless and almost decoupled particle, the dilaton. Note that as long as the background is constant, it is equivalent to quantize perturbations in the
original (Jordan-) frame or in the Einstein-frame (cf. \cite{Coleman:1969sm,Makino:1991sg}).

In the following sections we present a detailed analysis of the inflationary phase (section \ref{hdi}) and the dark-energy dominated phase (section \ref{lu}). A detailed study of preheating in the present model is left for a further work.

Allowing for $\beta\neq 0$ in the potential would not affect the discussion of inflation. It will, however, have an effect on the dark energy phenomenology, on which we will comment in section \ref{lu}.


\section{Higgs-Dilaton inflation}\label{hdi}
As usual, it is assumed that during inflation all the energy of the
system is contained in the inflaton fields and in the gravitational
field. Therefore, during this stage, the SM fields can be neglected.
Let us rewrite the scalar-tensor part of \eqref{SI-UGe} as
\begin{equation}\label{lagrangianJ}
\frac{ \mathcal{L}}{\sqrt{-g}}=\frac{f(\phi)}{2}
R-\frac{1}{2}g^{\mu\nu}\delta_{ab}\partial_\mu \phi^a\partial_\nu
\phi^b-U(\phi)\,,
\end{equation} 
with a non-minimal coupling
\begin{equation}\label{nonmincoup}
f(\phi)\equiv \sum_a \xi_a {\phi^a}^2\,, 
\end{equation}
and the potential
\begin{equation}\label{potentialJ}
U(\phi)=V(\phi)+\Lambda_0=\frac{\lambda}{4}\left(h^2-\frac{\alpha}{\lambda} \chi^2\right)^2+\Lambda_0\,,
\end{equation}
including the SI breaking term $\Lambda_0$.
As discussed
in section \ref{msi}, the parameter $\alpha$ is set to be very tiny 
$\alpha\sim {\cal O}(10^{-30})$, in order to obtain the correct
hierarchy between the electroweak and the Planck scale.
Greek indices $\mu,\nu,...=0,1,2,3$ denote spacetime coordinates
while Latin indices are used to label the two real scalar fields
present in the model: the
dilaton field $\phi^1=\chi$ and the Higgs field in the unitary gauge
$\phi^2=h$. The abstract notation in terms of $\phi^i$ will in the
following allow us to interpret the scalar fields as the coordinates
of a two-dimensional sigma-model manifold. We will be able to
 write expressions and equations that are covariant under variable
changes $\phi\mapsto\phi'(\phi)$.

Whenever the non-minimal coupling is non-zero\footnote{For our choice
of parameters, where $\xi_\chi,\xi_h>0$ this is the case whenever the
scalar fields are away from the origin $(\chi,h)\neq (0,0)$.}
$f(\phi)\neq 0$, one can define the new metric
\begin{equation}\label{conftransf}
 \tilde{g}_{\mu\nu}=\Omega^2 g_{\mu\nu}\,,
\end{equation} 
with $\Omega^2=M_P^{-2}f(\phi)$ to reformulate the Lagrangian 
in the E-frame. Taking into account that the metric
determinant and the Ricci scalar transform as\footnote{The action of the covariant d'Alembertian $\tilde\Box$ on a scalar field $s(x)$ is given by $\tilde\Box s=\frac{1}{\sqrt{-\tilde g}}\partial_\mu\left(\sqrt{-\tilde g}\tilde g^{\mu\nu}\partial_\nu s\right)$.}
\begin{gather}
\sqrt{-g}=\Omega^{-4}\sqrt{-\tilde{g}}\;, \\
R=\Omega^2\left(\tilde{R}+6\tilde{\Box}\ln{\Omega}-6\tilde{g}^{\mu\nu}
\partial_\mu\ln\Omega\,\partial_\nu\ln\Omega\right)\label{Riccitransf}
 \,,
\end{gather}
one obtains 
\begin{equation}\label{lagrangianE}
\frac{
\mathcal{L}}{\sqrt{-\tilde{g}}}=\frac{M_P^2}{2}\tilde{R}-
\frac{1}{2} \tilde K-\tilde U(\phi)\;,
\end{equation}
where the kinetic term is given by
\begin{equation}\label{EK}
\tilde K=\gamma_{ab}\tilde g^{\mu\nu}\partial_{\mu}\phi^a
\partial_{\nu}\phi^b\;,
\end{equation}
and $\gamma_{ab}$ is a generally non-canonical and non-diagonal field space metric, which in terms of the variables
$(\phi^1,\phi^2)=(\chi,h)$ is given by
\begin{equation}
\gamma_{ab}=\frac{1}{\Omega^2}\left(\delta_{ab}+\frac{3}{2}M_P^2\frac{
\Omega^2_ {,a} \Omega^2_{,b}}{\Omega^2}\right)\,.
\end{equation}
Unlike in the single-field case, the non-canonical kinetic term can not
in general be recast in canonical form by redefining the scalar field
variables. In fact, the field-space metric can be brought to canonical form by a
local variable change if and only if its Riemann tensor identically
vanishes. In the present case of a two-dimensional manifold, the
Riemann tensor has only one independent component, and it is enough to
compute the Ricci scalar $R_\gamma$ associated to the field space metric
$\gamma_{ab}$,
\begin{equation}
R_\gamma=(\xi_h-\xi_\chi)\frac{2}{M_P^2}\frac{\xi_\chi^2
(1+6\xi_\chi)\chi^4-\xi_h^2(1+6\xi_h)h^4}
{\left(\xi_h(1+6\xi_h)h^2+\xi_\chi(1+6\xi_\chi)\chi^2\right)^2}\;.
\end{equation} 
For $R_\gamma$ to vanish globally, one would need to have $\xi_\chi=\xi_h$.
As we will see, this case is not allowed by phenomenology.
The E-frame potential is defined as
\begin{equation}\label{potentialE}
\tilde U(\phi)=\tilde V(\phi)+\tilde V_{\Lambda_0}(\phi)\;,
\end{equation}
where we have defined a scale-invariant and a scale-invariance breaking part as
\begin{equation}
\tilde V(\phi)=\frac{V(\phi)}{\Omega^4}\quad\mathrm{and}\quad\tilde V_{\Lambda_0}(\phi)=\frac{\Lambda_0}{\Omega^4}\;.
\end{equation}
We can now write down the
equations of motion derived from the E-frame Lagrangian \eqref{lagrangianE}.
Einstein's equations are
\begin{equation}\label{ein}
 \tilde G_{\mu\nu}=\gamma_{ab}\left(\partial_\mu\phi^a\partial_\nu\phi^b-
\frac{1}{2}\tilde g_{\mu\nu}\tilde g^{\rho\sigma}
\partial_\rho\phi^a\partial_\sigma\phi^b\right)
+\tilde U\,\tilde g_{\mu\nu}\;,
\end{equation} 
where $\tilde G_{\mu\nu}$ is the Einstein tensor computed from the
metric $\tilde g_{\mu\nu}$.
The equations for the scalar fields are
\begin{equation}\label{kg}
 \tilde\Box\phi^c+\tilde g^{\mu\nu}\Gamma^c_{ab}
\partial_\mu\phi^a\partial_\nu\phi^b=\tilde U^{;c}\;,
\end{equation} 
where $\Gamma^c_{ab}$ is the Christoffel symbol computed from the
field-space metric $\gamma_{ab}$,
\begin{equation}
 \Gamma^c_{ab}=\frac{1}{2}\gamma^{cd}\left(\gamma_{da,b}+\gamma_{db,a}-\gamma_{ab,d}
\right)\;,
\end{equation} 
and where we use the notation $\tilde U^{;c}=\gamma^{cd}\tilde U_{,d}$.
Notice that equations \eqref{ein} and \eqref{kg} are covariant under
redefinitions of the scalar field variables $\phi^i\mapsto\phi'{}^i(\phi)$.

We choose to do our analysis in the
Einstein-, rather than in the Jordan frame. The reason for this choice
is that in the literature predictions for measurable
quantities are usually computed in the Einstein frame, where gravity has
the standard GR form. At the classical level there is, apart
from such practical arguments, nothing that would privilege one or
the other frame. After all, the choice of the frame simply corresponds to a choice of
variables.

\subsection{Exploiting scale invariance}
By construction, all terms in the Lagrangian \eqref{lagrangianJ}, except the one proportional to $\Lambda_0$, are invariant under the scale transformations \eqref{SItransint}. We will see that if the $\Lambda_0$-term is to be associated with dark energy, it must be negligibly small during inflation. The approximate scale-invariance of the theory will considerably simplify the analysis of the inflationary period.

\subsubsection{The Noether current of scale invariance}
Let us start by computing the Noether current associated to the scale transformations \eqref{SItransint}, which for an infinitesimal value of the parameter $\sigma$ become
\begin{equation}
g_{\m\n}\mapsto g_{\m\n}+\sigma\Delta g_{\m\n}\;,\hspace{10mm} \phi^i\mapsto\phi^i+\sigma\Delta\phi^i\;. 
\end{equation} 
The explicit expressions for $\Delta g_{\m\n}$ and $\Delta\phi^i$ depend on the choice of the field variables. For the original variables one has $\Delta g_{\m\n}=-2 g_{\m\n}$, $\Delta \c=\c$ and $\Delta h=h$. The associated current is given by (see e.g. \cite{Peskin:1995ev})
\begin{equation}\label{J}
\sqrt{-g}J^\m=\frac{\partial\mathcal{L}}{\partial\left[\partial_\m g_{\a\b}\right]}\Delta g_{\a\b}+
\frac{\partial\mathcal{L}}{\partial\left[\partial_\m\phi^i\right]}\Delta\phi^i\,.
\end{equation}
and satisfies
\begin{equation}
D_\m J^\m=-\frac{1}{\sqrt{-g}}\frac{\partial\left[\Lambda_0\sqrt{-g}\right]}{\partial g_{\mu\nu}}\Delta g_{\mu\nu}=4\Lambda_0\;,
\end{equation}
where $D_\mu$ denotes the covariant derivative constructed with the metric $g_{\mu\nu}$.

In the E-frame, scale transformations do not act on the metric, ${\Delta\tilde g_{\m\n}=0}$, and are simply given by
\begin{equation}
\phi^i\mapsto\phi^i+\sigma\Delta\phi^i\;.
\end{equation}
In this case, the expression for the current is
\begin{equation}\label{ce}
\sqrt{-\tilde g}\tilde J^\m=\frac{\partial\mathcal{L}}{\partial\left[\partial_\m\phi^i\right]}\Delta\phi^i\;,
\end{equation}
while the conservation law becomes
\begin{equation}\label{cce}
\tilde D_\m \tilde J^\m=-\frac{\partial \tilde V_{\Lambda_0}}{\partial\phi^i}\Delta\phi^i=\frac{4\Lambda_0}{\Omega^4}\;,
\end{equation}
where the covariant derivative $\tilde D_\mu$ is constructed with the metric $\tilde g_{\mu\nu}$. Whenever $\Lambda_0$ vanishes, scale invariance becomes exact and the associated current $J^\mu$, respectively $\tilde J^\mu$ is conserved.

\subsubsection{New variables}\label{newvar}
The approximate conservation law can lead us to a very convenient choice for the scalar field variables in the E-frame formulation. In fact, one can always choose a set of variables $({\phi'}^1,{\phi'}^2)=(\rho,\theta)$ such that a scale transformation only acts on a single variable $\rho$. Moreover, $\rho$ can always be defined such that the scale transformation acts on it like a shift $\rho\mapsto\rho+\sigma M_P$. This transformation is a symmetry of the Lagrangian if $\Lambda_0=0$. This means that $\rho$ can appear in the Lagrangian only through $\tilde V_{\Lambda_0}$ and through derivatives in the kinetic term. Moreover, the variable $\theta$ can be defined such that the field-space metric $\gamma'_{ab}$, which is independent of $\rho$, is diagonal. If the Lagrangian \eqref{lagrangianE} is expressed in terms of variables that satisfy these requirements, the current $\tilde J^\mu$ takes the form
\begin{equation}
\tilde J^\mu=M_P\tilde g^{\mu\nu}\gamma'_{\rho \rho}(\theta)\partial_\nu\rho\;.\label{Jnew}
\end{equation}
Now, in order to find the relation between the variables $({\phi'}^1,{\phi'}^2)=(\rho,\theta)$ and the original variables $({\phi}^1,{\phi}^2)=(\chi,h)$, let us express the current $\tilde J^\mu$ in terms of the original variables,
\begin{equation}\label{Jold}
\tilde J^\mu=\tilde g^{\mu\nu}\frac{M_P^2}{2(\xi_\chi\chi^2+\xi_h h^2)}\partial_\nu\left((1+6\xi_\chi)\chi^2+(1+6\xi_h)h^2\right)\;.
\end{equation}
Comparing this expression to \eqref{Jnew}, it is clear that $\rho$ must be a function of the combination of the fields $\chi$ and $h$ acted upon by the partial derivative, i.e. $\rho=\rho\left[\left(1+6\xi_\chi\right)\chi^2+\left(1+6\xi_h\right)h^2\right]$. For the scale transformation to correspond to a shift of $\rho$ by $\sigma M_P$, this function has to be chosen as
\begin{equation}\label{rhodef}
\rho=\frac{M_P}{2}\ln\left(\frac{(1+6\xi_\chi)\chi^2+(1+6\xi_h)h^2}{M_P^2}\right)\;.
\end{equation}
The variable $\theta$, as it does not transform under scale transformations, has to be a function of the ratio between $h$ and $\chi$, i.e. $\theta=\theta\left[\frac{h}{\chi}\right]$. There is some freedom in the choice of this function. One can notice that the argument of the logarithm in \eqref{rhodef} corresponds to the radius of an ellipse in the $(\chi,h)$-plane. We will therefore define $\theta$ as the angular coordinate of the ellipse, i.e.
\begin{equation}\label{thetadef}
\theta=\arctan\left(\sqrt{\frac{1+6\xi_h}{1+6\xi_\chi}}\frac{h}{\chi}\right)\,.
\end{equation}
Let us note that in terms of the variables $(\rho,\theta)$, since one has $\Delta\rho=M_P$ and $\Delta \theta=0$, the current conservation law \eqref{cce} corresponds to the equation of motion for $\rho$. Further, one can see from \eqref{cce} that the dependence of $\tilde V_{\Lambda_0}$ on $\rho$ is such that $\tilde V_{\Lambda_0}\propto \exp\left(-4\rho/M_P\right)$.

In terms of the new variables, the E-frame kinetic term $\tilde K$ and the potential $\tilde U=\tilde V+\tilde V_{\Lambda_0}$ (cf. \eqref{lagrangianE}) are given by
\begin{equation}\label{Knew}
\tilde K=\left(\frac{1+6\xi_h}{\xi_h}\right)\frac{ 1}{\sin^2\theta+\varsigma\cos^2\theta} \left( \partial \rho\right)^2+\frac{M^2\, \varsigma}{\xi_\chi}   \frac{\tan^2\theta+\mu}{\cos^2\theta\left(\tan^2\theta+\varsigma \right)^2} \left(\partial \theta\right)^2
\end{equation}
and
\begin{align}\label{Vlnew}
&\tilde V(\theta)=\frac{\lambda M_P^4}{4\xi_h^2}\left(\frac{\sin^2\theta-\frac{\alpha}{\lambda}\frac{1+6\xi_h}{1+6\xi_\chi}\cos^2\theta}{\sin^2\theta+\varsigma\cos^2\theta}\right)^2\;,
\\
&\tilde V_{\Lambda_0}(\rho,\theta)=\Lambda_0\left(\frac{1+6\xi_h}{\xi_h}\right)^2 \frac{e^{-4\rho/M_P}}{\left(\sin^2\theta+\varsigma\cos^2\theta\right)^2}\,,\label{VLnew}
\end{align}
where we have defined the parameters
\begin{equation}\label{constants}
 \mu\equiv \frac{\xi_\chi}{\xi_h} \,,\hspace{10mm} \varsigma\equiv \frac{\left(1+6\xi_h\right)\xi_\chi}{\left(1+6\xi_\chi\right)\xi_h}\,.
\end{equation}
We will see in section \ref{cmbcop} that for a successful description of inflation the parameters have to be such that $\xi_\chi\sim\mathcal{O}(10^{-3})$ and $\xi_h\sim\mathcal{O}(10^{5})$ and hence $\mu\ll 1$. In this case, one can neglect $\mu$ in the kinetic term \eqref{Knew}. In this approximation, the action can be further simplified by introducing the variables
\begin{equation}
\tilde\rho=\gamma^{-1}\rho \quad\quad \mathrm{and}\quad\quad \tilde\theta=\frac{M_P}{a}\tanh^{-1}\left(\sqrt{1-\varsigma}\cos\theta\right)\,,
\end{equation}
with the parameters
\begin{equation}
a=\sqrt{\frac{\xi_\chi(1-\varsigma)}{\varsigma}} \,,\hspace{10mm} \gamma=\sqrt{\frac{\xi_\chi}{1+6\xi_\chi}}\,.\end{equation}
In terms of these variables and for $\mu\ll 1$, the kinetic term takes the simple form
\begin{equation}\label{Knew2}
\tilde K\simeq e^{2b(\tilde\theta)}(\partial\tilde\rho)^2+(\partial\tilde\theta)^2\;,
\end{equation}
with
$$b(\tilde\theta)= \frac{1}{2}\ln \left(\varsigma\cosh^2\left(a\tilde\theta/M_P\right)\right)\, ,$$
which has been studied in the literature previously \cite{GarciaBellido:1995fz,GarciaBellido:1995qq,DiMarco:2002eb}. The potentials are given by
\begin{align}\label{Vlnew2}
&\tilde V(\tilde\theta)=\frac{\lambda M_P^4}{4\xi_h^2(1-\varsigma)^2}\left(1-\varsigma\cosh^2(a\tilde\theta/M_P)-\frac{\alpha}{\lambda}\frac{1+6\xi_h}{1+6\xi_\chi}\sinh^2(a\tilde\theta/M_P)\right)^2\;,
\\
&\tilde V_{\Lambda_0}(\tilde\rho,\tilde\theta)=\frac{\Lambda_0}{\gamma^{2}}\,\varsigma^2\cosh^4(a\tilde\theta/M_P)e^{-4\gamma\tilde\rho/M_P}\,.\label{VLnew2}
\end{align}

\subsubsection{Departure from scale invariance}
Let us now look at the E-frame equations of motion \eqref{ein} and \eqref{kg} in order to see in which region of field space the effect of a non-zero $\Lambda_0$ will be important. $\Lambda_0$ enters the equations through $\tilde U=\tilde V+\tilde V_{\Lambda_0}$ and through $\tilde U^{;c}=\tilde V^{;c}+\tilde V_{\Lambda_0}^{;c}$. We therefore define the two new parameters
\begin{align}\label{ups1}
&\upsilon_1=\frac{\tilde V_{\Lambda_0}}{\tilde V}\;,\\\label{ups2}
&\upsilon_2=\frac{\sqrt{\tilde V_{\Lambda_0}{}^{;a}\tilde V_{\Lambda_0}{}_{;a}}}{\sqrt{\tilde V^{;b}\tilde V_{;b}}}\;,
\end{align}
that characterize the departure from scale invariance. The parameter $\upsilon_1$ compares the importance of the scale invariance breaking part of the potential $\tilde V_{\Lambda_0}$ to the scale-invariant part $\tilde V$ in the Einstein equations \eqref{ein}. In the region of field space where $\upsilon_1\ll 1$, $\tilde V_{\Lambda_0}$ can be neglected in Einstein's equations.
The parameter $\upsilon_2$ provides a coordinate invariant measure of the importance of $\tilde V_{\Lambda_0}{}^{;a}$ compared to $\tilde V^{;a}$ in the scalar field equations \eqref{kg}. In fact, locally one can always choose a coordinate system in which $|\tilde V_{\Lambda_0}{}^{;1}/\tilde V^{;1}|=|\tilde V_{\Lambda_0}{}^{;2}/\tilde V^{;2}|=\upsilon_2$. Hence, in the region of field space where $\upsilon_2\ll 1$, $\tilde V_{\Lambda_0}{}^{;a}$ can be neglected in the scalar field equations. In the region where both $\upsilon_1\ll 1$ and $\upsilon_2\ll 1$ hold, the effect of $\Lambda_0$ is negligible and the equations of motion become practically scale-invariant. We will refer to this region as the scale-invariant region. It will turn out that for phenomenologically viable values of the parameters, the whole period of observable inflation takes place in this region.

\subsection{Evolution of the homogeneous background}
Let us now consider homogeneous scalar fields $\phi^i=\phi^i(t)$ in spatially
flat FLRW space-time characterized by the line element
\begin{equation}
ds^2=\tilde g_{\mu\nu}dx^\mu dx^\nu=-dt^2+a^2(t)d\vec{x}^2\,.
\end{equation}

Before writing down the equations of motion, we introduce some notation for vectors lying in the tangent and cotangent bundles of the field manifold, that will allow us to write many of the upcoming expressions in a very compact way. The notation corresponds to the one of \cite{Peterson:2010np}. We denote vectors in boldface, i.e. $\bm A=(A^1,A^2)$. The inner product of two vectors $\bm A$ and $\bm B$ is given by
\begin{equation}
\bm{A\cdot B}\equiv\bm A^\dagger \bm B=\gamma_{ij}A^i B^j\,,
\end{equation}
and the norm of a vector $\bm A$ is
\begin{equation}
|\bm A|\equiv\sqrt{\bm{A\cdot A}}\,,
\end{equation}
where a dagger ${}^\dagger$ on a naturally contravariant or covariant vector denotes its dual, e.g. $\bm{\dot\phi}^\dagger\equiv (\gamma_{ij}\dot\phi^j)$ and $\bm\nabla^\dagger\equiv(\gamma^{ij}\nabla_j)$. Here and in the following we use $\nabla_j$ to denote the covariant derivative constructed from the field-space metric $(\gamma_{ij})$.

For homogeneous fields in flat FLRW space-time, equations \eqref{ein} and \eqref{kg} reduce to the Friedmann equations and the equations of motion for the scalar fields,
\begin{align}\label{fre}
 &H^2=\frac{1}{3M_P^2}\left(\frac{1}{2}\big|\bm{\dot\phi}\big|^2+\tilde U\right)\;,
\\\label{fr2e}
&2\dot{
 H}+3 H^2=-\frac{1}{M_P^2}\left(\frac{1}{2}\big|\bm{\dot\phi}\big|^2-\tilde
U\right)\,,
\\\label{sfe}
&\frac{D\bm{\dot\phi}}{dt}+3 H\bm{\dot\phi}=-\bm\nabla^\dagger\tilde U\;,
\end{align}
where a dot stands for a derivative with respect to $t$, $H=\dot{a}/ a$ and the action of $D$ on a contravariant vector $X^i$ is defined as $DX^i=d X^i+\Gamma^c_{ab}
X^a d\phi^b$. To this we can add the equation for the current
\eqref{cce}, which for homogeneous fields reduces to
\begin{equation}\label{cchom}
\frac{d}{d t}\left(a^3 \bm{\dot\phi}\bm\cdot\Delta\bm\phi\right)=4\tilde V_{\Lambda_0}\,,
\end{equation}
where $\Delta\bm\phi=(\Delta\phi^1,\Delta\phi^2)$.
This relation is of course not independent of the equations of motion.
However, it will prove useful in the following. 

For the discussion of inflation, it is helpful to change the time parameter from $t$ to the e-fold time parameter $N=\ln a(t)$. The field equations can then be written as
\begin{align}
&H^2=\frac{\tilde U}{3M_P^2-\frac{1}{2}\big|\bm{\phi'}\big|^2}\,,\label{fr1n}\\
&\frac{H'}{H}=-\frac{1}{2}\frac{\big|\bm{\phi'}\big|^2}{M_P^2}\,,\label{fr2n}\\
&\frac{\dfrac{D\bm{\phi'}}{dN}}{3-\frac{1}{2}\big|\bm{\phi'}\big|^2/M_P^2}+\bm{\phi'}=-M_P^2\bm\nabla^\dagger\ln\tilde U\,,\label{sfn}
\end{align}
where a prime denotes a derivative with respect to the e-fold parameter $N$.

\subsubsection{Slow-roll inflation and background trajectories}
In the present model, inflation can occur due to a phase of slow-roll
of the scalar fields over the flat part of the potential towards one of the potential valleys (cf. figure \ref{epotplot}). 
Let us define the slow-roll parameter $\epsilon$ and the slow-roll vector $\bm\eta$ as \cite{Lyth:1993eu,Peterson:2010np}\footnote{The definition of the vector $\bm\eta$ used here differs from the definition in \cite{Peterson:2010np} by the factor $|\bm{\phi'}\big|$.}
\begin{align}
&\epsilon\equiv-\frac{H'}{H}=\frac{1}{2}\frac{\big|\bm{\phi'}\big|^2}{M_P^2}\,,\label{srp1ex}\\
&\bm\eta\equiv\frac{D\bm{\phi'}}{dN}\Big/|\bm{\phi'}\big|\,,\label{srp2ex}
\end{align}
in terms of which equations \eqref{fr1n} and \eqref{sfn} read
\begin{align}
&H^2=\frac{1}{M_P^2}\frac{\tilde U}{3-\epsilon}\,,\label{frpar}\\
&\frac{\bm\eta}{3-\epsilon}\big|\bm{\phi'}\big|+\bm{\phi'}=-M_P^2\bm\nabla^\dagger\ln\tilde U\,.\label{sfpar}
\end{align}
The exact condition for inflation, i.e. for $\ddot a>0$, is given by $\epsilon<1$.

The slow-roll regime is characterized by the fact that the equations
\eqref{frpar} and \eqref{sfpar} are well-approximated by the ''slow-roll equations''
\begin{align}
&H^2=\frac{\tilde U}{3M_P^2}\,,\label{frsr}\\
&\bm{\phi'}=-M_P^2\bm\nabla^\dagger\ln\tilde U\,.\label{sfsr}
\end{align}
The conditions for the validity of the slow-roll approximation are\footnote{Note that $\epsilon=\epsilon_1$ and $|\eta|=\frac{1}{2}|\epsilon_2|$, where $\epsilon_1\equiv\frac{H'}{H}$ and $\epsilon_2\equiv\frac{D\ln\epsilon}{dN}$ are the multifield generalizations of the standard first two horizon-flow parameters defined in \cite{Schwarz:2001vv}.}
\begin{equation}\label{srcond}
\epsilon\ll 1\quad\mathrm{and}\quad \eta\equiv |\bm\eta|\ll 1\;.
\end{equation}
Still following ref. \cite{Peterson:2010np} we introduce the kinematical orthonormal basis vectors
$\bm e_\|=\frac{\bm{\phi'}}{\big|\bm{\phi'}\big|}$, pointing in the direction of the field trajectory, and $\bm e_\perp$, pointing in the direction of $(\bm I-\bm e_\|\bm e_\|^\dagger)\bm\eta$, where $\bm I$ is the $2\times 2$ unit matrix. This allows us to write the second slow-roll parameter as
\begin{equation}
\eta=\sqrt{\eta_\|^2+\eta_\perp^2}\;,
\end{equation}
where the speed-up rate $\eta_\|$ and the turn-rate $\eta_\perp$ are defined as
\begin{align}\label{sur}
&\eta_\|=\bm e_\|\bm\cdot\bm\eta=-\frac{3-\epsilon}{\big|\bm\phi'\big|}\left(\big|\bm\phi'\big|+M_P^2\,\bm e_\|\bm\cdot\bm\nabla \ln \tilde U\right)\,,\\\label{tr}
&\eta_\perp=\bm e_\perp\bm\cdot\bm\eta=-M_P^2\,\frac{3-\epsilon}{\big|\bm\phi'\big|}\,\bm e_\perp\bm\cdot\bm\nabla \ln \tilde U\,.
\end{align}
In the case of one-field inflation, $\eta_\perp$ is equal to zero and $\eta=|\eta_\||$. In the upcoming section we will see that for the Higgs-Dilaton model, as long as the fields are in the scale-invariant region, $\eta_\perp$ goes to zero very quickly, and $\eta$ can be computed like in a single field model.

Within the slow-roll approximation, i.e. making use of equations \eqref{frsr} and \eqref{sfsr}, one can compute approximations in terms of the potential for the slow-roll parameters $\epsilon$ and $\eta$ 
\begin{align}
&\epsilon^{(\mathrm{SR})}= \frac{1}{2}M_P^2\big|\bm\nabla\ln\tilde U\big|^2\,,\label{srp1app}\\
&\eta^{(\mathrm{SR})}=\sqrt{(\bm e_\|^{(\mathrm{SR})})^\dagger \bm M^2\bm e_\|^{(\mathrm{SR})}}\,,\label{srp2app}
\end{align}
such as for the speed-up rate $\eta_\|$ and the turn-rate $\eta_\perp$
\begin{align}
&\eta_\|^{(\mathrm{SR})}=-(\bm e_\|^{(\mathrm{SR})})^\dagger \bm M\bm e_\|^{(\mathrm{SR})}\,,\\
&\eta_\perp^{(\mathrm{SR})}=-(\bm e_\|^{(\mathrm{SR})})^\dagger \bm M\bm e_\perp^{(\mathrm{SR})}\,,
\end{align}
where the matrix $\bm M$ is defined as
\begin{equation}\label{massmatrix}
\bm M\equiv M_P^2\bm{\nabla^\dagger\nabla}\ln\tilde U
\end{equation}
and the kinematical unit vectors in the slow-roll approximation are given by
\begin{equation}\label{unitapp}
\bm e_\|^{(\mathrm{SR})}=-\frac{\bm\nabla^\dagger\ln\tilde U}{\big|\bm\nabla\ln\tilde U\big|}\,,
\quad\mathrm{and}\quad
\bm e_\perp^{(\mathrm{SR})}=-\frac{\bm M+\eta_\|^{(\mathrm{SR})}\bm I}{\sqrt{(\eta^{(\mathrm{SR})})^2-(\eta_\|^{(\mathrm{SR})})^2}}\bm e_\|^{(\mathrm{SR})}\,.
\end{equation}
Hence, instead of the exact slow-roll conditions \eqref{srcond} one can use the approximate slow-roll conditions
\begin{equation}\label{srcondapp}
\epsilon^{(\mathrm{SR})}\ll1\quad\mathrm{and}\quad \eta^{(\mathrm{SR})}\ll 1\,,
\end{equation}
which should be understood as consistency conditions for the slow-roll approximation. Once the system is in the slow-roll regime, i.e. the exact conditions \eqref{srcond} are satisfied, the approximate conditions \eqref{srcondapp} guarantee that the system remains in the slow-roll regime and describes a phase of inflation. We will approximate the time where inflation ends, as the moment where $\epsilon^{(\mathrm{SR})}=1$.
\begin{figure}
\begin{center}
	\includegraphics[scale=0.2]{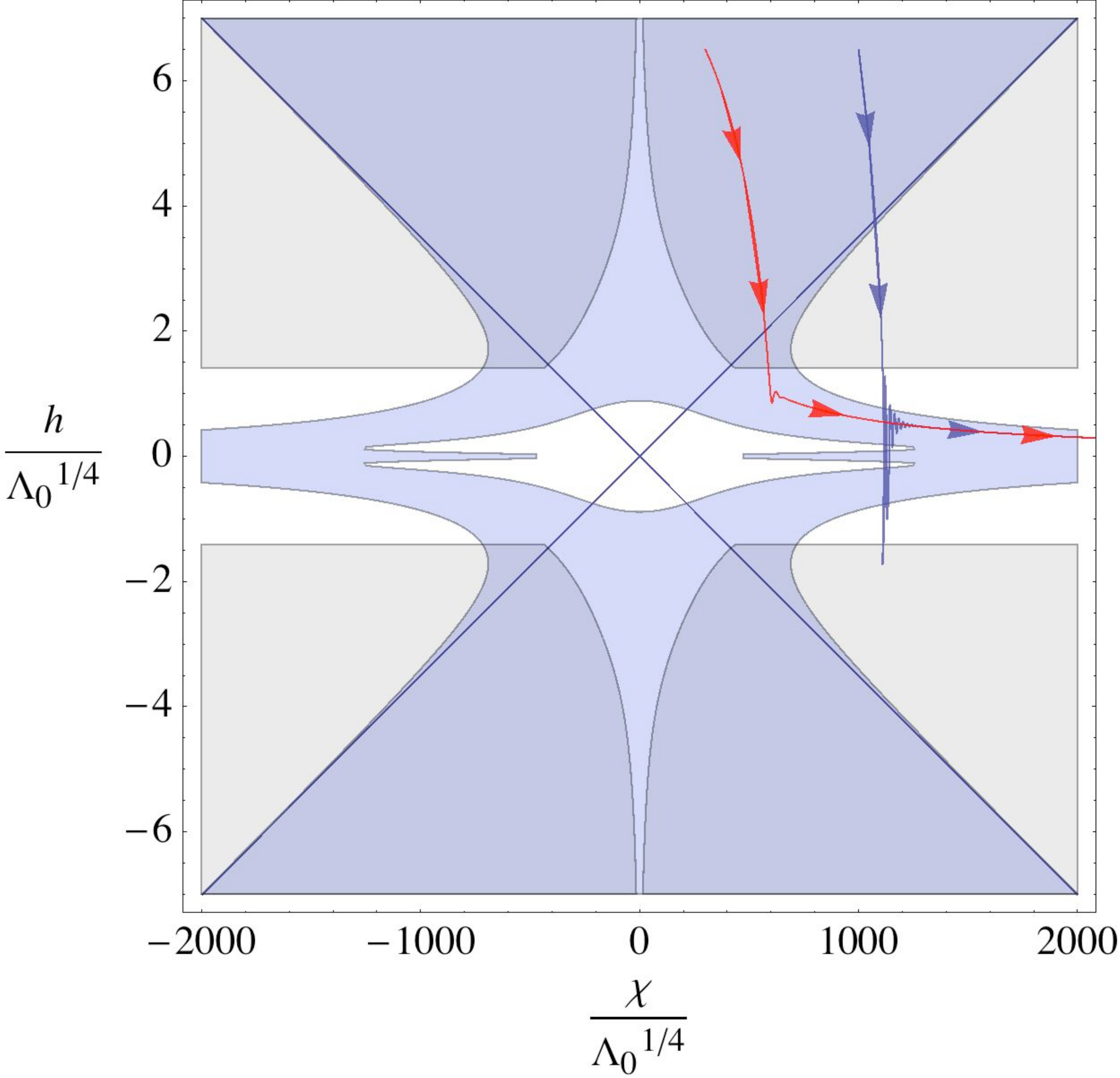}
\end{center}
\caption{The blue region is the slow-roll region for
 $\xi_\chi\ll 1$, $\xi_h\gg 1$ and $\alpha=0$ given by $\epsilon^{(\mathrm{SR})}<1$. The inclusion of the second slow-roll condition $\eta^{(\mathrm{SR})}<1$ does not change the essential properties of this region. The general features of the slow-roll region are the same whenever $\xi_\chi<\frac{1}{2}$ and
 $\xi_h>\frac{1}{2}$. For $\xi_\chi<\frac{1}{2}$ and $\xi_h<\frac{1}{2}$
 the central fast-rolling region vanishes. For $\xi_\chi>\frac{1}{2}$
the slow-roll region does not extend to infinity along the $\chi$-axis in
 which case the scalar fields can not act as dark energy in the late
 stage of evolution. The shaded region corresponds to the scale-invariant region delimited by
 $\upsilon_1<1$ and $\upsilon_2<1$. This is the region where the influence of $\Lambda_0\neq 0$ is small.
 The presence of the slow roll-region along the
 $\chi$-axis such as the central fast-roll region are effects of
 $\Lambda_0>0$. For $\Lambda_0=0$ the slow-roll region is simply given by
 the triangles delimited by the two diagonal lines. Note that in this case the units of the axis have to be chosen differently. The red line
 represents a trajectory of type a, never leaving the SR-region. The
 blue line is a trajectory of the type b, which leaves the SR-region
 and oscillates strongly before rolling down the valley. These trajectories were found by numerically solving the exact equations \eqref{sfe}.}
 \label{SRregion}
 \end{figure}

Let us now discuss the regions in the $(\chi,h)$-plane for which the
approximate slow-roll conditions hold (cf. Fig. \ref{SRregion}). The slow-roll region extends
to infinity along the potential valleys if $\xi_\chi<\frac{1}{2}$. As will be shown in section
\ref{lu}, only if this condition
holds, the scalar fields can constitute a dark energy component in the
late stage. Further, during inflation it is safe to neglect the term in the
potential proportional to $\alpha$. In fact, for $\alpha=0$ the
potential possesses only one valley which goes along the $\chi$-axis.
For $\alpha\neq 0$ this valley splits into two valleys that lie at
the angles $\theta=\pm\arctan(\alpha)$ with respect to the
$\chi$-axis. For $\alpha\lll 1$ these angles are very small. We
will see that inflation in our model occurs far from these valleys
where the effect of a non-zero $\alpha$ is irrelevant. Hence, we will
put $\alpha=0$ for the rest of this section.
The plot of the slow-roll region for $\xi_\chi<\frac{1}{2}$ and
$\alpha=0$ is presented in Fig. \ref{SRregion}.

Next, we want to analyze the different trajectories the fields can take if
the initial conditions are chosen in the slow-roll region. We will only consider trajectories
starting in the first quadrant $\chi,h>0$. Trajectories starting in other quadrants are exactly analog. The
shape of the potential \eqref{potentialE} makes that all trajectories tend to
approach one of the potential valleys. There are trajectories
(type a) that on their way to the valley never leave the slow-roll
region. Numerical computations show that such trajectories undergo
only very
few slow oscillations before asymptotically approaching the valley.
One can not expect a successful
reheating period from this type of behaviour \cite{Bezrukov:2008ut,GarciaBellido:2008ab}.
The good trajectories
(type b) are those that at some point leave the slow-roll region. 
After the exit of the slow-roll region, which at the same time
marks the end of inflation, these trajectories undergo a fast-roll
towards the valley and therefore oscillate strongly around its
minimum. Typical examples for both types of trajectories are given in
Figure \ref{SRregion}.

Looking at Fig. \ref{SRregion} we observe that all good trajectories (type b, blue line) go
through the scale-invariant region before leaving the slow-roll
region. Therefore, for these trajectories, the end of inflation always takes place within the scale-invariant region. We will see in section \ref{iccons} that requiring the scalar fields to act as a dark-energy
component in the late phase will give a bound on the initial conditions. Qualitatively, this bound tells that the scalar fields during inflation have to be very far from the origin. Therefore, not only the end but the whole period of observable inflation (i.e. the final $\sim60$ e-folds) takes place in the scale-invariant region. This fact considerably simplifies the analysis. In particular, during inflation the scale-invariance breaking part $\tilde V_{\Lambda_0}$ of the potential \eqref{potentialE} can be neglected, i.e. $\tilde U\simeq\tilde V$. As a consequence, if one uses the set of variables $(\rho,\theta)$ introduced in section \ref{newvar}, the potential only depends on $\theta$. Consequently, also the slow-roll parameters $\epsilon^{(\mathrm{SR})}$ and $\eta^{(\mathrm{SR})}$ are functions of $\theta$ only. Further, one finds that the lowest order approximation of the turn rate $\eta_\perp$ vanishes, i.e. $\eta_\perp^{(\mathrm{SR})}=0$.

In the scale-invariant region and written in terms of the variables $(\rho,\theta)$, the slow-roll equations for the scalar fields \eqref{sfsr} read
\begin{align}
&\rho'=0\,,\label{rhosisr}\\
& \theta'=-\frac{4\xc}{1+6\xc}\cot\theta\left(1+\frac{6\xc\xh}{\xc\cos^2\theta+\xh\sin^2\theta}\right)\,.\label{thetasisr}
\end{align}
From the first equation one finds that the background trajectories in the scale-invariant slow-roll region correspond to a constant value for $\rho$, $\rho=\rho_0$, respectively to ellipses in the $(\chi,h)$-plane described by
 \begin{equation}\label{ellipse}
\left(1+6\xc\right)\c^2+\left(1+6\xh\right)h^2=M_P^2\,e^{2\rho_0/M_P}\,.
\end{equation}
The second scalar field equation \eqref{thetasisr}, as a consequence of scale invariance, does not depend on $\rho_0$. This equation can be integrated in order to get the
number of e-folds before the end of inflation as a function of the
angle $\theta$,
\begin{equation}\label{efoldstheta}
 N(\theta,\theta_\textnormal{end})=\frac{1}{4\xc}\left[\ln\left(\frac{\cos\theta_\textnormal{end}}{
\cos\theta}\right)+3\xc\ln\left(\frac{
\xc\cos^2\theta_\textnormal{end}+\xh\sin^2\theta_\textnormal{end}
+6\xc\xh}{
\xc\cos^2\theta+\xh\sin^2\theta
+6\xc\xh} \right)\right]\;,
\end{equation} 
where $\theta_\textnormal{end}$ is the value of $\theta$ at the end of
inflation, which can be found from the condition
\begin{equation}\label{thetaend}
\epsilon^{(\mathrm{SR})}(\theta_\textnormal{end})=\frac{8\xc^2(1+6\xh)}{1+6\xc}\frac{\cot^2\theta_\textnormal{end}}
{\xc\cos^2\theta_\textnormal{end}+\xh\sin^2\theta_\textnormal{end}}
=1\,.
\end{equation} 
After inserting values for $\xc$, $\xh$ and requiring a minimal number
of inflationary e-folds $N(\theta_\textnormal{min},\theta_\textnormal{end})=N_\textnormal{min}$, equations
\eqref{efoldstheta} and \eqref{thetaend} can be solved to obtain a lower bound
$\theta_\textnormal{min}<\theta_\textnormal{initial}$ on the initial
conditions for inflation. In what follows, we will derive bounds on the
parameters $\xc$ and $\xh$, which are related to the spectra of primordial
perturbations.

\subsubsection{One-field attractor in the scale-invariant region}
In the previous section we have seen that if the slow-roll conditions hold and if the system is in the scale-invariant region, i.e. $\upsilon_1\ll1$ and $\upsilon_2\ll1$, the trajectories are given by $\rho=\rho_0$.
In other words, there exists a set of variables in term of which only one of the two fields evolves during inflation.

In this section we are going to show that in the scale-invariant region $\rho'=0$ is an attractor independently of slow-roll and also after inflation. We will further show that during inflation in the scale-invariant region, the turn-rate $\eta_\perp$ goes to zero very rapidly, also beyond the lowest order slow-roll approximation. This fact will have important consequences for the evolution of perturbations of the scalar fields.

Let us look at the equation for the scale-current \eqref{cchom}. In the scale-invariant region, the term on the righ-hand side can be neglected. In terms of the variables $(\rho,\theta)$ and using the e-fold time parameter, one obtains
\begin{equation}
\rho'=\frac{\mathrm{cst.}}{H\gamma_{\rho\rho}}e^{-3N}\;,
\end{equation}
where the constant depends on initial conditions. $\gamma_{\rho\rho}$ can be read from \eqref{Knew} and represents a bounded function. During inflation the factor $H^{-1}$ is generally nearly constant but grows at most like $e^N$ at the end of inflation. For a matter domination like stage one obtains $H^{-1}\propto e^{\frac{1}{2}N}$ and for a radiation domination like stage $H^{-1}\propto e^{\frac{2}{3}N}$. One can read from equations \eqref{fr1n} and \eqref{fr2n} that whenever the potential is positive, which is always the case in the scale-invariant region of our model, $H^{-1}$ grows more slowly than $e^{3N}$. We conclude that as soon as the system can be approximated by the scale-invariant equations, $\rho'=0$ is an attractor.

Let us now turn our attention to the turn-rate $\eta_\perp$. In the scale-invariant region, equation \eqref{tr} reduces to
\begin{equation}\label{tr2}
\eta_\perp=-M_P^2\,\frac{3-\epsilon}{\big|\bm\phi'\big|}\,\bm e_\perp\bm\cdot\bm\nabla \ln \tilde V=-M_P^2|\bm\nabla \ln \tilde V|\,\frac{3-\epsilon}{\big|\bm\phi'\big|}\,\bm e_\perp\bm\cdot\bm e_{(\bm\nabla \tilde V)}\,,
\end{equation}
where $\bm e_{(\bm\nabla \tilde V)}$ is a unit vector pointing in the direction of $\bm\nabla\tilde V$. From the current conservation law \eqref{cchom}, still neglecting $\tilde V_{\Lambda_0}$ and noticing that $\Delta\bm\phi$ is perpendicular to $\bm\nabla\tilde V$, we find
\begin{equation}\label{ccev}
\bm e_\|\bm\cdot\bm e_{(\bm\nabla \tilde V)_\perp}=\frac{\mathrm{cst.}}{|\Delta\bm\phi||\bm\phi'|H}e^{-3N}\,,
\end{equation}
where $\bm e_{(\bm\nabla \tilde V)_\perp}$ is a unit vector perpendicular to $\bm e_{(\bm\nabla \tilde V)}$. At every point in field-space one can find an orthogonal matrix $\bm O$ such that $\bm e_\perp=\bm O\bm e_\|$ and $\bm e_{(\bm\nabla \tilde V)}=\bm O\bm e_{(\bm\nabla \tilde V)_\perp}$. Hence, one has
\begin{equation}
\left|\bm e_\perp\bm\cdot\bm e_{(\bm\nabla \tilde V)}\right|=\left|\left(\bm O\bm e_\perp\right)\bm\cdot\left(\bm O\bm e_{(\bm\nabla \tilde V)}\right)\right|=\left|\bm e_\|\bm\cdot\bm e_{(\bm\nabla \tilde V)_\perp}\right|\;.
\end{equation}
Inserting this relation and \eqref{ccev} in \eqref{tr2} and making use of \eqref{srp1ex} and \eqref{srp1app}, we find the following result for the turn rate
\begin{equation}
\left|\eta_\perp\right|=\mathrm{cst.}\cdot\frac{(3-\epsilon)\sqrt{\epsilon^{(\mathrm{SR})}}}{\epsilon H|\Delta\bm\phi|}e^{-3N}\,,
\end{equation}
where we have redefined the constant related to initial conditions. During slow-roll inflation, $\epsilon$ is growing, but remains smaller than unity, $H$ is decreasing very slowly and $\epsilon^{(\mathrm{SR})}$ is small. $|\Delta\bm\phi|$ is a bounded function, which in terms of the variables $(\rho,\theta)$ is given by $|\Delta\bm\phi|=M_P\sqrt{\gamma_{\rho\rho}}$. We conclude that during slow-roll inflation and for approximate scale invariance, the turn rate $\eta_\perp$ goes to zero exponentially fast. Further, one can show that if scale invariance is slightly violated, $\eta_\perp$ is proportional to $\upsilon_2$.

\subsection{Linear perturbations}\label{perturb}
The theory of cosmological perturbations as stemming from quantum fluctuations during inflation was developed in \cite{Sakharov:1965,Lukash:1980,Mukhanov:1981xt,
Chibisov:1982nx,Lyth:1984gv} (see also \cite{Mukhanov:1990me} and references therein). 
Including scalar and tensor perturbations and fixing the Newtonian
transverse traceless gauge, the line element can be written as
\begin{equation}
ds^2=-\left(1+2\Phi\right)
dt^2+a(t)^2\left(\left(1-2\Psi\right)\delta_{ij}+h_{ij}^{TT}
\right)dx^i dx^j\;.
\end{equation} 
$\Phi$ and $\Psi$ are the Bardeen potentials \cite{Bardeen:1980kt}.
For comparison with CMB observations, we will be interested in the power spectrum of the comoving curvature perturbation, which is defined as \cite{Bardeen:1980kt,Kodama:1985bj}
\begin{equation}
 \zeta\equiv\Psi-\frac{H}{\dot H}\left(\dot\Psi+H\Phi\right)\;.
\end{equation} 
Through the perturbed Einstein equations, $\zeta$ is related to the linear perturbations of the scalar fields, $\bm\delta\bm\phi$, like (see e.g. \cite{Ringeval:2007am})
\begin{equation}
\zeta=\frac{1}{|\bm\phi'|}\left(\Psi\bm\phi'+\bm{\delta\phi}\right)\bm\cdot\bm e_\|\;,
\end{equation}
where the quantity in parenthesis is the multifield version of the Mukhanov-Sasaki variable \cite{Sasaki:1986hm,Mukhanov:1988jd}
The evolution equation for $\zeta$ is given by \cite{GarciaBellido:1995qq,DiMarco:2002eb,Peterson:2010np}
\begin{equation}\label{zih}
\zeta'=\frac{2}{(aH)^2|\bm\phi'|^2}\Delta\Psi-2\eta_\perp\, \frac{\delta\phi_\perp}{|\bm\phi'|}\,,
\end{equation}
where $\Delta=\delta^{ij}\partial_i\partial_j$. The quantity $\delta\phi_\perp\equiv\bm{\delta\phi}\bm\cdot\bm e_\perp$ is the component of the field perturbations perpendicular to the background field trajectory sometimes called relative isocurvature (or entropy) perturbation (see e.g. \cite{Gordon:2000hv}).
In the long wavelength limit, $k\ll aH$, the first term in \eqref{zih} can be neglected and the evolution equation becomes
\begin{equation}\label{zoh}
\zeta'=-2\eta_\perp\, \frac{\delta\phi_\perp}{|\bm\phi'|}\,,
\end{equation}
This is a well-known result, showing that for multi-field inflation,
$\zeta$ is not in general conserved outside the Hubble horizon \cite{GarciaBellido:1995qq,DiMarco:2002eb}. There
are two cases in which the source term on the right-hand side of the
equation vanishes. One is if the perturbation vector $\bm{\delta\phi}$
is tangent to the field trajectory, i.e. $\delta\phi_\perp=0$. This
corresponds to the complete absence of relative isocurvature perturbations during inflation and is not satisfied in our scenario. The second possibility is to have the potential gradient $\bm\nabla U$ parallel to the background field trajectory, resulting in the vanishing of the turn rate $\eta_\perp=0$.
In our model, as we have seen in the previous sections, inflation takes place in the scale-invariant region, where $\eta_\perp=0$ is an attractor. Hence, as a consequence of scale-invariance, the comoving curvature perturbation is practically conserved outside the horizon, just like in single-field inflation.\footnote{Using the results of \cite{Peterson:2010np}, one finds that $\delta\phi_\perp/|\bm\phi'|$ is of the same order of magnitude as $\zeta$ and does not grow considerably during inflation.} As mentioned before, corrections due to deviations from scale invariance are suppressed by the parameter $\upsilon_2$. This parameter being extremely small (cf. section \ref{iccons}), we will neglect such corrections.

We will from now on use the fact that inflation takes place in the scale-invariant region and suppose that initial conditions are such that the attractor $\eta_\perp=0$ has been reached before the observable scales
cross the horizon.  In that case, $\zeta$ is constant outside the
Hubble horizon and the result for the primordial
spectrum of $\zeta$ is like in the case of one-field inflation. To lowest order in the slow-roll
parameters it can be expressed as \cite{Mukhanov:1990me,Liddle:1993fq,Chiba:2008rp}
\begin{equation}\label{Pzeta}
 \mathcal{P}_\zeta(k)\simeq\frac{1}{2
M_P^2\epsilon^*}\left(\frac{H^{*}}{2\pi}
\right)^2\simeq \frac{1}{24\pi^2{\epsilon^{\mathrm{(SR)}}}^*}\frac{\tilde V^*}{M_P^4}\;,
\end{equation} 
where quantities with an asterisk are evaluated at the moment of horizon crossing, i.e. when $aH=k$.
The scalar spectral index is given by
\begin{equation}\label{ssi}
 n_s(k)-1\equiv\frac{d\ln\mathcal{P}_\zeta}{d\ln
k}\simeq-2(\epsilon^*+\eta_\|^*)\simeq -2({\epsilon^{\mathrm{(SR)}}}^*+{\eta_\|^{\mathrm{(SR)}}}^*)\;,
\end{equation} 
while the running of the spectral index can be expressed as \cite{Leach:2002ar}
\begin{equation}\label{ssa}
\alpha_\zeta(k)\equiv\frac{dn_s}{d\ln
k}\simeq -4\epsilon^*\eta_\|^*-2\eta_\|^*\xi^*\simeq-4{\epsilon^{\mathrm{(SR)}}}^*{\eta_\|^{\mathrm{(SR)}}}^*-2\left({\eta_\|^*\xi^*}\right)^{\mathrm{(SR)}}\;,
\end{equation}
where $\xi$ is the third Hubble-flow parameter $\xi\equiv\frac{d^2\ln \epsilon}{dN^2}$. In the slow-roll approximation the combination $\eta_\|\xi$ can be expressed in terms of the potential as \cite{Peterson:2010np}
$$\left({\eta_\|\xi}\right)^{\mathrm{(SR)}}=\bm e_\|^{(\mathrm{SR})\dagger}\bm X\bm e_\|^{(\mathrm{SR})}\,.$$
with the matrix $\bm X\equiv M_P^2\bm\nabla\ln\tilde U\bm\nabla^\dagger\bm M$, where
$\tilde U$ can be replaced by $\tilde V$ due to scale invariance.\footnote{Let us mention again that, since inflation takes place in the scale-invariant region, one can choose variables $(\rho,\theta)$ for which the approximate slow-roll parameters $\epsilon^{\mathrm{(SR)}}$, $\eta_\|^{\mathrm{(SR)}}$ and the combination $\left({\eta_\|\xi}\right)^{\mathrm{(SR)}}$ depend only on $\theta$. In models of one-field inflation with a canonical kinetic term an alternative common definition of approximate slow-roll parameters is given by $\epsilon_s\equiv\frac{1}{2}M_P^2\left(\frac{U'}{U}\right)^2$, $\eta_s\equiv M_P^2\frac{U''}{U}$ and $\xi_s\equiv M_P^4\frac{U''' U'}{U^2}$. In the case of one or several fields with non-canonical kinetic terms, these definitions generalize to $\epsilon_s\equiv \frac{1}{2}M_P^2\big|\bm\nabla\ln U\big|^2$, $\eta_s\equiv(\bm e_\|^{(\mathrm{SR})})^\dagger \bm M_s\bm e_\|^{(\mathrm{SR})}$ and $\xi_s\equiv(\bm e_\|^{(\mathrm{SR})})^\dagger \bm X_s\bm e_\|^{(\mathrm{SR})}$ where the matrices $\bm M_s$ and $\bm X_s$ are defined as $\bm M_s\equiv M_P^2\frac{1}{U}\bm\nabla^\dagger\bm\nabla U$ and $\bm X_s\equiv M_P^4\frac{1}{U^2}\bm\nabla U\bm\nabla^\dagger\bm\nabla^\dagger\bm\nabla U$
and the unit vectors are given in \eqref{unitapp}. These parameters are related to the approximate slow-roll parameters used in the present work as
$\epsilon^{(\mathrm{SR})}=\epsilon_s$, $\eta_\|^{(\mathrm{SR})}=2\epsilon_s-\eta_s$ and $\left({\eta_\|\xi}\right)^{\mathrm{(SR)}}=8\epsilon_s^2-6\epsilon_s\eta_s+\xi_s$.}

Now, even if the relative entropy perturbations $\delta\phi_\perp$ do not affect the evolution of
$\zeta$, since they are completely decoupled, they can in general be present at the end of inflation. 
The total entropy perturbations, however, are given by $\mathcal{S}\propto \eta_\perp\delta\phi_\perp/|\bm\phi'|$ \cite{Gordon:2000hv}. And, as we have shown above, scale invariance during inflation leads to $\eta_\perp\simeq 0$ and hence to a strong suppression of $\mathcal{S}$.
As has been shown in Ref.~\cite{Finelli:2000ya} (see also \cite{Tsujikawa:2002nf} for a general perspective), the large scale suppression of entropy perturbations during inflation avoids the resonant growth of these fluctuations also during (p)reheating. Working with this assumption, we
will be able to relate the primordial spectra to CMB observations.

The primordial spectrum of the tensor perturbations is given, to 
the lowest order in the slow-roll approximation, by
\cite{Mukhanov:1990me,Liddle:1993fq}\footnote{This result is based on the
slow-roll approximation and
involves no further assumptions.}
\begin{equation}\label{Ph}
 \mathcal{P}_g(k)\simeq\frac{8}{M_P^2}\left(\frac{H^{*}}{2\pi}
\right)^2\simeq \frac{2}{3\pi^2}\frac{\tilde V^*}{M_P^4}
\; ,
\end{equation}  
which results in a tensorial spectral index
\begin{equation}
n_g(k)\equiv\frac{d\ln\mathcal{P}_g}{d\ln
k}\simeq-2\epsilon^*\simeq-2{\epsilon^{\mathrm{(SR)}}}^*\;.
\end{equation} 
The ratio of the tensor and the scalar spectra to first order
in slow-roll is then given by
\begin{equation}\label{str}
 r\equiv\frac{\mathcal{P}_g}{\mathcal{P}_\zeta}\simeq
16\epsilon^{*}\simeq 16{\epsilon^{\mathrm{(SR)}}}^*\,,
\end{equation} 
and we have the consistency condition like in one-field inflation
\begin{equation}\label{scter}
r=-8n_g\,,
\end{equation}
valid to the lowest non-trivial order in the slow-roll approximation.

\subsection{CMB constraints on parameters and predictions of the model}\label{cmbcop}
In this section we are going to explicitly compute the primordial spectra and confront them with CMB observations. As discussed in the previous sections, the whole period of observable inflation takes place in the scale-invariant region. We can therefore use equations
\eqref{efoldstheta} and \eqref{thetaend} for the background. Moreover, due to scale-invariance, $\zeta$ is conserved for large wavelengths during inflation. Further, we make the assumption that after inflation entropy perturbations die away before having an observable effect. This allows
us to directly compare the primordial spectra \eqref{Pzeta} and \eqref{Ph} to observations of the CMB.

We will show in the following, that the running of the scalar spectral index and the amplitude of tensor perturbations 
are related and very small, cf. \eqref{relar}. Therefore, we can consider those observational bounds (WMAP7 + BAO + $H_0$) for the scalar tilt and the amplitude of the scalar power spectrum, which are based on the standard $\L$CDM model and the assumptions that the primordial spectrum obeys a power-law and that tensor modes can be neglected (see Ref. \cite{Komatsu:2010fb})
\begin{eqnarray}\label{ampobs}
\mathcal{P}_\zeta(k_0)&=&(2.43\pm 0.27)\times 10^{-9}\,,\\
n_s(k_0)&=&0.968\pm0.036\label{tiltobs}\,,
\end{eqnarray}
where $k_0/a_0=0.002\,\mathrm{Mpc}^{-1}$ and the indicated errors correspond to the $99\%$ confidence levels.

Let us start by computing the spectral quantities $\mathcal{P}_\zeta(k_0)$,
$n_s(k_0)$,  $\alpha(k_0)$ and $r(k_0)$ evaluated at the pivot scale $k_0$, in terms of the parameters $\xc$, $\xh$ and $\l$.
This is done in four steps:
\begin{enumerate}
\item[i.] Equation \eqref{thetaend} is solved for
$\theta_{end}=\theta_{end}(\xc,\xh)$.

\item[ii.] 
$\theta_{end}$ is inserted into \eqref{efoldstheta}
from which one determines $\theta^*=\theta^*(\xc,\xh,N^*)$.

\item[iii.]
Expressions \eqref{Pzeta},
\eqref{ssi}, \eqref{ssa} and \eqref{scter} are evaluated at $\theta^*$ to find the spectral quantities as
functions of $\xc$, $\xh$, $\l$ and $N^*$.

\item[iv.] 
$N^*$, the number of e-folds between the moment
where $k_0$ exits the horizon and the end of inflation, is expressed as a function of the parameters $\xc$, $\xh$ and $\l$. 
\end{enumerate}
In order to determine $N^*$ (step (iv)) we need to know the post-inflationary
evolution of the universe including the details of the reheating 
process. If there are uncertainties related to the
post-inflationary history, these can be accounted for by varying the value
of $N^*$. One can compute $N^*$ approximately, by making a few assumptions about the post-inflationary evolution.
First, during the reheating phase the scale factor is expected to evolve
like in a matter dominated universe. The reason is that during this stage the present model behaves much like the Higgs-Inflation model (cf. \cite{Bezrukov:2008ut,GarciaBellido:2008ab}). In this sense matter-like scaling of the universe during reheating is not
really an assumption but rather a property of the considered model. Next, we make the usual assumptions that reheating is 
followed by the standard radiation and matter dominated
stages. Further assuming that the transitions between the different phases are
instantaneous, one can derive the following relation (cf. \cite{Liddle:1993fq})
\begin{equation}
N^*\simeq-\ln\frac{k_0}{a_0 H_0}-\ln\left(\frac{\varrho_{0}^{cr}/\Omega^\gamma_0}{\tilde V(\theta^*)}\right)^{1/4}
+\ln\left(\frac{\tilde V(\theta^*)}{\tilde V(\theta_{end})}\right)^{1/4}
-\frac{1}{3}\ln\left(\frac{\tilde V(\theta_{end})}{\varrho_{rh}}\right)^{1/4}\;.
\end{equation}
Here, $a_0$, $H_0$, $\varrho_{0}^{cr}$ and $\Omega^r_0$ stand for the current values
of the scale factor, the Hubble parameter, the critical density and the abundance of
radiation, respectively. $\varrho_{rh}$ denotes the radiation
energy density at the end of reheating, i.e. at the onset of the hot big bang.
After inserting the observational value $\Omega^r_0 h^2\simeq 4.2\cdot 10^{-5}$ (for $T^\gamma_0\simeq2.73 K$ \cite{Mather:1998gm}), where $h$ is the dimensionless Hubble parameter, the above formula can be written as
\begin{equation}\label{Nrh}
N^*\simeq59-\ln\frac{k_0\mathrm{Mpc}}{0.002 a_0}-\ln\frac{10^{16}
\,\mathrm{GeV}}{\tilde V(\theta^*)^{1/4}}
+\ln\left(\frac{\tilde V(\theta^*)}{\tilde V(\theta_{end})}\right)^{1/4}
-\frac{1}{3}\ln\left(\frac{\tilde V(\theta_{end})}{\varrho_{rh}}\right)^{1/4}\;.
\end{equation}
Notice that the dependence on $H_0$ has cancelled out\footnote{In the analog formula of
\cite{Liddle:1993fq} this fact remains somewhat hidden.}.
A detailed determination of $\varrho_{rh}$ goes beyond the scope
of this work and is postponed for a future publication. We can, however, consider two limiting cases. An upper limit on $\varrho_{rh}$ is simply given by
\begin{equation}\label{rhorhmax}
\varrho_{rh}^{max}=\tilde V(\theta_{end})\,,
\end{equation}
corresponding to instantaneous reheating at the end of inflation. A lower limit can be found in the same way as for the Higgs-Inflation model in \cite{Bezrukov:2008ut,GarciaBellido:2008ab}. Namely, one can look for the value of $\theta$, below which the particle interactions become those of the Standard Model, apart from the suppressed interactions discussed in section \ref{itd}, and therefore guarantee immediate reheating. Inspecting the kinetic term \eqref{Knew} and the potential \eqref{Vlnew}, we find that this happens as soon as $\tan^2{\theta}<min(\mu,1)$.\footnote{Note that the value of $\varsigma$ lies always between $1$ and the value of $\mu$.} We will therefore set the lower limit
\begin{equation}\label{rhorhmin}
\varrho_{rh}^{min}=\tilde V(\theta_{rh}^{min})\,,\quad\quad\mathrm{with}\quad\quad \tan^2\theta_{rh}^{min}=min(\mu,1)\,.
\end{equation}

The value of $\lambda$ to be used when computing the spectral parameters corresponds to
the Higgs self-coupling evaluated at the scale of inflation \cite{Bezrukov:2008ut,GarciaBellido:2008ab}. It contains the uncertainty related to the Higgs mass $m_H^2$. We expect the running of $\l$ to be
similar as in the case of the Higgs-Inflation model \cite{Barvinsky:2008ia,DeSimone:2008ei,Bezrukov:2008ej,Bezrukov:2009db,Bezrukov:2010jz,Ferrara:2010yw,Ferrara:2010in,Kallosh:2000ve}. As was shown in \cite{Bezrukov:2008ut,GarciaBellido:2008ab},
for $m_H^2\simeq 130-180\,\textnormal{GeV}$, the coupling $\l$ evaluated at the scale of inflation lies in the range $\l\simeq 0.1-1$.

\begin{figure}
\begin{center}
	\includegraphics[scale=1]{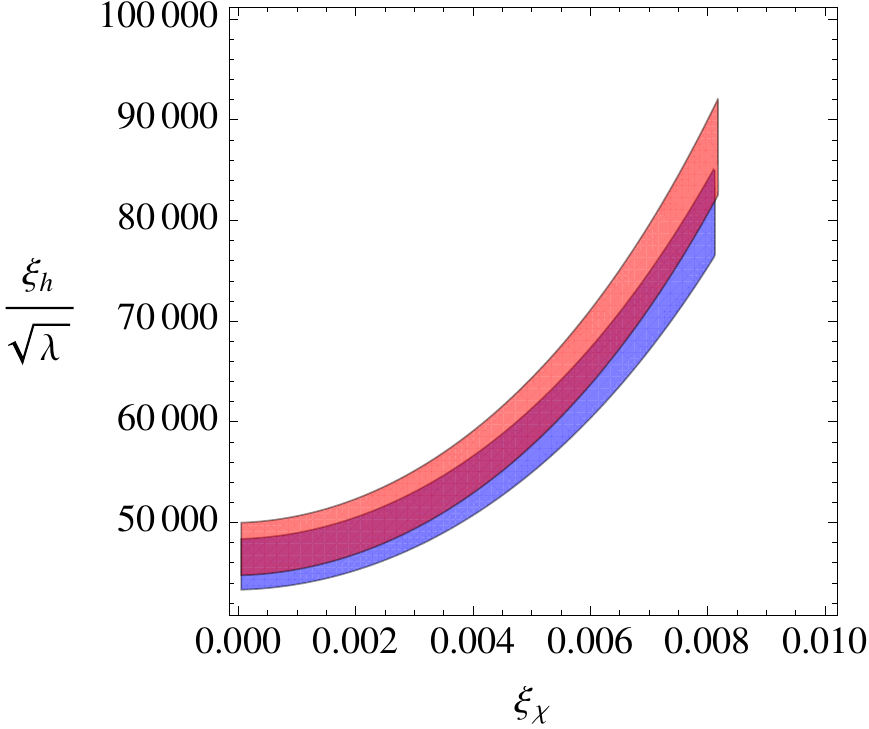}
\end{center}
\caption{This plot shows the parameter regions for which the amplitude $P_\zeta(k_0)$ and the tilt $n_s(k_0)$ of the scalar spectrum lie in the observationally allowed region (WMAP7 + BAO + $H_0$ at $99\%$ confidence level), for $\lambda=1$. (The variation of the result induced by variation of $\lambda$ in the interval $0.1<\lambda<1$ is negligible.) The red region is obtained for $\varrho_{rh}=\varrho_{rh}^{max}$ (instantaneous reheating), while the blue region corresponds to $\varrho_{rh}=\varrho_{rh}^{min}$ (long reheating). The fact that the bands are cut on the right comes from the constraint on the scalar tilt $n_s(k_0)$, cf. \eqref{tiltcal}, while the band-shape is due to the constraint on the amplitude $P_\zeta(k_0)$,
 cf. \eqref{ampcal}.}
\label{parrange}
\end{figure}

\begin{figure}
\begin{center}
	\includegraphics[scale=1]{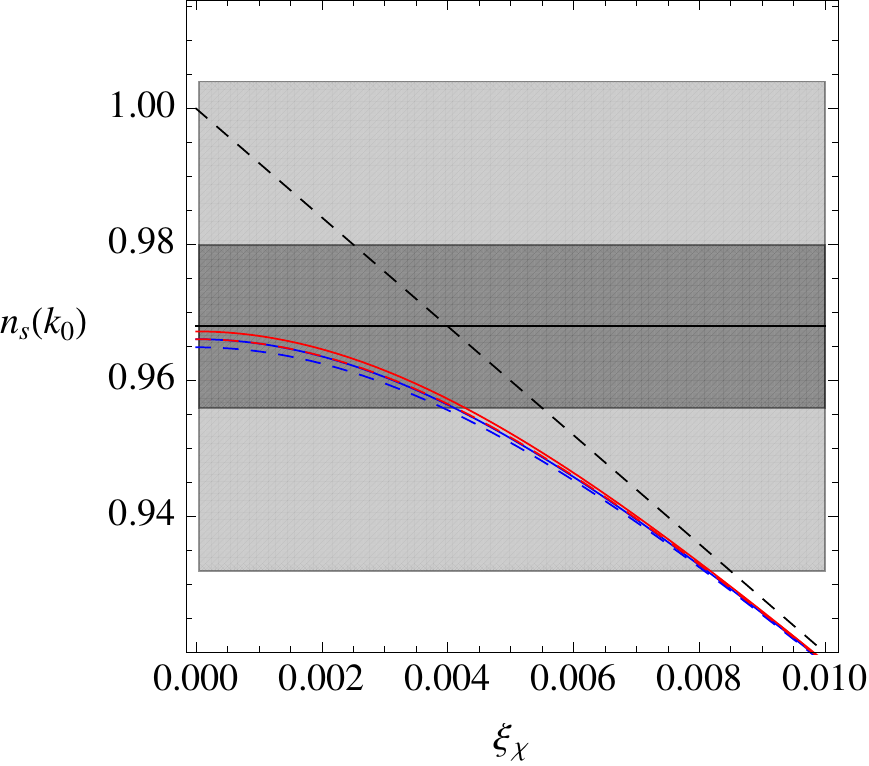}
\end{center}
\caption{
The spectral tilt as a function of the non-minimal coupling parameter $\xi_\chi$. The other parameters are set to $\xi_h=65000$ and $\lambda=1$. Note, however, that changing the ratio $\xi_h/\sqrt{\lambda}$ in the observationally allowed range affects the result only by a negligible amount. The solid curves correspond to the numerical results, the blue one is obtained for $\varrho_{rh}=\varrho_{rh}^{min}$ (long reheating) and the red one for $\varrho_{rh}=\varrho_{rh}^{max}$ (instantaneous reheating). The blue and red dashed curves are obtained from the analytical approximation \eqref{tiltcal} for $\bar N^*_{min}$  and  $\bar{N}^*_{max}$. The black dashed curve represents the asymptotic solution \eqref{nsassol}, which is a good approximation if  $\frac{1}{4N^*}<\xi_\chi\ll1$ . The horizontal line and the shaded 
regions correspond to the observational mean value and the $1\sigma$ and $3\sigma$ confidence intervals, cf. \eqref{tiltobs}.
}
\label{nsplot}
\end{figure}

\begin{figure}
\begin{center}
\begin{tabular}{cc}
          \includegraphics[scale=0.9]{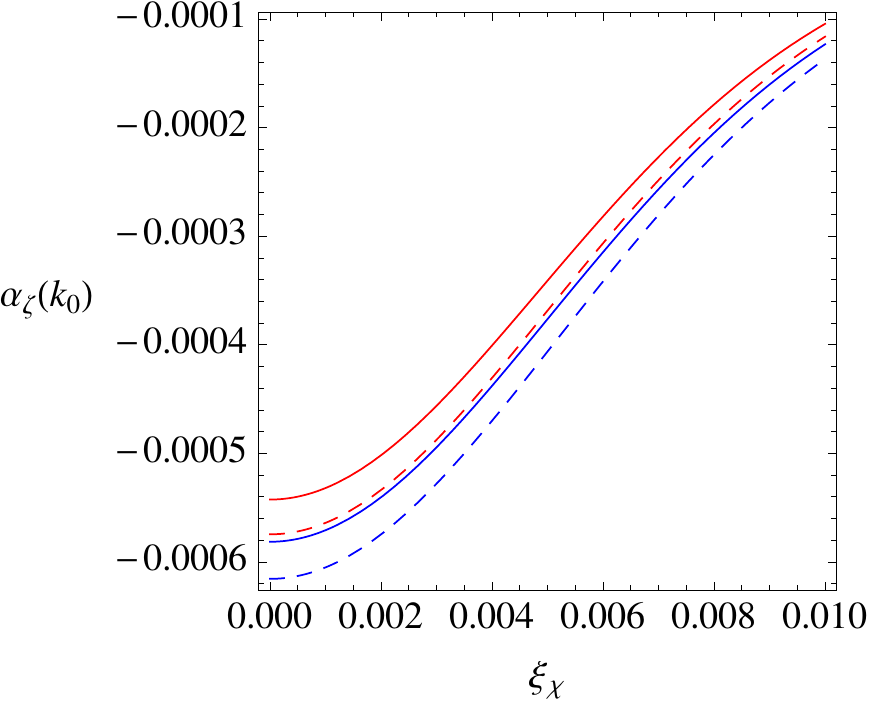} &\includegraphics[scale=0.9]{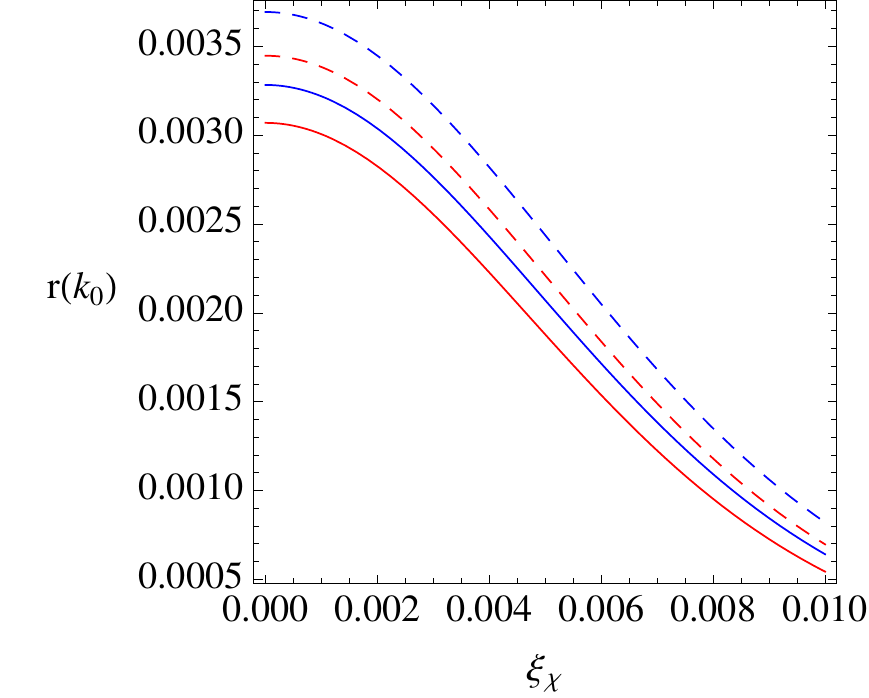}
 \end{tabular}
\end{center}
\caption{The running of the scalar spectral tilt and the tensor-to-scalar ratio (right) as a function of the coupling $\xi_\chi$. The other parameters are set to $\xi_h=65000$ and $\lambda=1$. Note, however, that changing the ratio $\xi_h/\sqrt{\lambda}$ in the observationally allowed range affects the result only by a negligible amount. Blue solid curves show the numerical results for $\varrho_{rh}=\varrho_{rh}^{min}$ (long reheating), while the red solid curves are the numerical results for the case $\varrho_{rh}=\varrho_{rh}^{max}$ (instantaneous reheating). The dashed curves are obtained from the approximate expressions \eqref{rcal} and \eqref{runcal}, inserting $\bar N^*_{min}$  and  $\bar{N}^*_{max}$}.
\label{alpharplot}
\end{figure}

Step (ii), i.e. the exact inversion of \eqref{efoldstheta} can not be done analytically. Hence, before computing approximate analytical results, we execute the four steps numerically. This allows us to plot the region in parameter space, for which $\mathcal{P}_\zeta(k_0)$ and $n_s(k_0)$ lie within the observational constraints \eqref{ampobs} and \eqref{tiltobs}, cf. figure \ref{parrange}.
It turns out that in the observationally allowed range and for $\l\simeq 0.1-1$, the spectral quantites depend on $\xi_h$ and $\lambda$ almost only through the combination $\xi_h/\sqrt{\lambda}$. This fact will become explicit in the approximate analytical results to be derived below.
The red region in figure \ref{parrange} is obtained under the assumption of instantaneous reheating ($\varrho_{rh}=\varrho_{rh}^{max}$) while the blue region corresponds to the case of long reheating, i.e. $\varrho_{rh}=\varrho_{rh}^{min}$. We obtain the bounds
\begin{equation}
\begin{array}{rcccll}
0&<&\xc&\lesssim&0.008\;,&\\
&&&&&\quad\mathrm{for}\quad \varrho_{rh}=\varrho_{rh}^{min}\;,\\
43000&\lesssim&\dfrac{\xh}{\sqrt{\lambda}}&\lesssim&85000\;,&
\end{array}\label{minbound}
\end{equation}
and
\begin{equation}
\begin{array}{rcccll}
0&<&\xc&\lesssim&0.008\;,&\\
&&&&&\quad\mathrm{for}\quad \varrho_{rh}=\varrho_{rh}^{max}\;.\\
44500&\lesssim&\dfrac{\xh}{\sqrt{\lambda}}&\lesssim&92000\;,&
\end{array}\label{maxbound}
\end{equation}
In this region of parameter space, the quantities $n_s(k_0)$, $\alpha(k_0)$ and $r(k_0)$ vary only with $\xi_\chi$. The numerical results for these quantities are given in figures \ref{nsplot} and \ref{alpharplot}, respectively.

We have found that the parameters need to satisfy $\xc\ll 1$ and $\xh\gg 1$.
With this knowledge, we again carry out the above four steps and derive approximate analytical results. From \eqref{thetaend},
we obtain
\begin{equation}
\theta_{end}=2\times 3^{\frac{1}{4}}\sqrt{\xc}\left(1+\mathcal{O}\left[\xc,\frac{1}{\xh}\right]\right)\label{endcal}\;.
\end{equation}
In order to approximately solve \eqref{efoldstheta} for $\theta^*$, we can neglect the second term on the right-hand side. The inversion then gives
\begin{equation}\label{incaltemp}
\theta^*\simeq\arccos\left(\cos(\theta_{end}) e^{-4\xc N^*}\right)\;.
\end{equation}
Here the sign for approximate equality ``$\simeq$'' refers to the approximation made when inverting \eqref{efoldstheta}. This approximation will constitute
 the main source of error in the approximate expressions for the spectal quantities. One can get a more accurate approximation by reinserting the first approximation into the right-hand side of \eqref{efoldstheta} in order to compute the second order approximation of an iterative solution. However, as the expressions get considerably more complicated, we stick to the first order approximation, which already comes very close to the numerical results.
Inserting $\theta_{end}$ from \eqref{endcal} into \eqref{incaltemp}, one obtains
\begin{equation}\label{incal}
\theta^*\simeq \arccos\left(e^{-4\xc N^*}\right)\left(1+\mathcal{O}\left[\xc,\frac{1}{\xh}\right]\right)\;.
\end{equation}
We can now evaluate the spectral parameters at the approximate value for $\theta^*$. Inserting \eqref{incal} into \eqref{Pzeta}, \eqref{ssi}, \eqref{ssa} and \eqref{str} and recalling \eqref{scter} we obtain\footnote{Note that these approximate results can equivalently be derived from the approximate action given by \eqref{Knew2} and \eqref{Vlnew2} with $\alpha=0$.}
\begin{eqnarray}
P_\zeta(k_0)&\simeq&\frac{\l \sinh^2\left(4\xc
N^*\right)}{1152\pi^2\xc^2\xh^2}\left(1+\mathcal{O}\left[\xc,\frac{1}{\xh},\frac{1}{N^*}\right]\right)
\label{ampcal}\\
n_s\left(k_0\right)-1&\simeq &-8\xc\coth\left(4\xc
N^*\right)\left(1+\mathcal{O}\left[\xc,\frac{1}{\xh},\frac{1}{N^*}\right]\right)\label{tiltcal}\;,\\
\alpha_\zeta(k_0)&\simeq &-32\xc^2 \sinh^{-2}\left(4\xc N^*\right)\left(1+\mathcal{O}\left[\xc,\frac{1}{\xh},\frac{1}{N^*}\right]\right) \label{runcal}\;,\\
r(k_0)=-8n_g(k_0)&\simeq &192\xc^2 \sinh^{-2}\left(4\xc N^*\right)\left(1+\mathcal{O}\left[\xc,\frac{1}{\xh},\frac{1}{N^*}\right]\right)\label{rcal}\;.
\end{eqnarray}
One can see that in this approximation, $\alpha_\zeta(k_0)$, $r(k_0)$ and $n_g(k_0)$ are related as
\begin{equation}\label{relar}
\alpha_\zeta(k_0)\simeq-\frac{1}{6}\,r(k_0)=\frac{4}{3}\,n_g(k_0)\;,
\end{equation}
which can be understood as an approximate consistency condition for the Higgs-Dilaton model.

\begin{figure}
\begin{center}
	\includegraphics[scale=0.9]{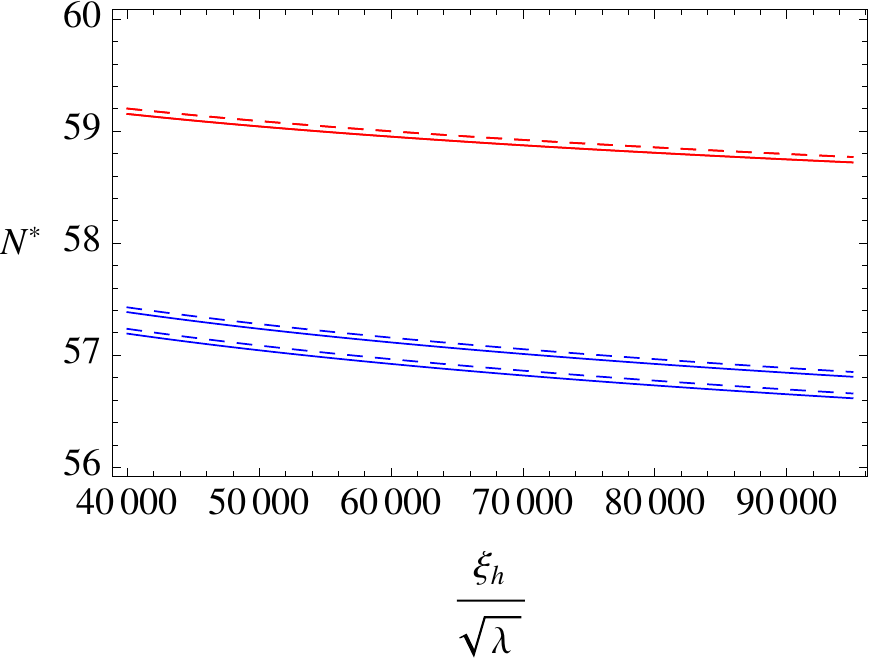}
\end{center}
\caption{
The number of efolds $N^*$ between horizon crossing of the scale $k_0$ and the end of inflation as a function of $\xi_h/\sqrt{\lambda}$, for $\xi_\chi=0.001$. Note that changing $\xi_\chi$ in the observationally allowed range affects the result by a negligible amount. The solid curves correspond to the numerical results, the blue ones correspond to $N_{min}^*$ (long reheating) and the red ones to $N_{max}^*$ (instantaneous reheating). The upper one of the solid blue curves is obtained for $\lambda=1$ and the lower one for $\lambda=0.1$, showing the slight dependence of $N_{min}^*$ on $\lambda$. The dependence of $N_{max}^*$ on $\lambda$ is negligible. The blue and red dashed curves are obtained from the analytical approximations\eqref{Nmincal} and \eqref{Nmaxcal}.}
\label{Nstarplot}
\end{figure}

In order to find $N^*$ in terms of the parameters of the theory, we insert \eqref{incal} into \eqref{Nrh}, from which we derive the approximate results\footnote{The numerical factor is given to the first decimal.}
\begin{eqnarray}
N^*_{min}&\simeq &\left(64.3-\frac{1}{12}\ln\l-\frac{2}{3}\ln{\frac{\xh}{\sqrt{\l}}}\right)\left(1+\mathcal{O}\left[\xc,\frac{1}{\xh},\frac{1}{N^*}\right]\right)\,,\label{Nmincal}\\
N^*_{max}&\simeq&\left(64.5-\frac{1}{2}\ln{\frac{\xh}{\sqrt{\l}}}\right)\left(1+\mathcal{O}\left[\xc,\frac{1}{\xh},\frac{1}{N^*}\right]\right)\,,\label{Nmaxcal}
\end{eqnarray}
where the subscripts ``min'' and ``max'' stand for the cases $\varrho_{rh}=\varrho_{rh}^{min}$ and
$\varrho_{rh}=\varrho_{rh}^{max}$ respectively. These approximate results together with the numerical results for $N_{min}^*$ and $N_{max}^*$ are plotted in figure \ref{Nstarplot}.

Equations \eqref{ampcal}-\eqref{rcal} together with \eqref{Nmincal} and \eqref{Nmaxcal} constitute our approximate analytical results for the spectral parameters in terms of the parameters $\xi_\chi$, $\xi_h$ and $\lambda$, in the two limiting cases of instantaneous and late reheating. One can see that the results depend on $\xi_h$ and $\lambda$ mainly through the combination $\xi_h/\sqrt{\lambda}$. Independent variation of $\lambda$ affects the spectral parameters only through $N^*_{min}$. Further, we remark that for $0.1<\lambda<1$ and $\xi_h/\sqrt{\lambda}$ in the observationally allowed range \eqref{minbound} or \eqref{maxbound}, both $N^*_{min}$ and $N^*_{max}$ vary only very little with the parameters (cf. figure \ref{Nstarplot}). For the precision required here, it is enough to know $N^*$ at the precision of a whole number. Therefore, inserting parameters of the allowed order of magnitude, we can set the approximate values to
\begin{eqnarray}
\bar N^*_{min}&=&57\,,\label{Nminapp}\\
\bar N^*_{max}&=&59\,.\label{Nmaxapp}
\end{eqnarray}
In the two limiting cases, the spectral parameters can be evaluated at these values for $N^*$.
Neglecting the small variation of $N^*$ with the parameters of the theory, we observe that the amplitude $P_\zeta$ of the scalar spectrum depends on the combination $\xi_h/\sqrt{\lambda}$ and on $\xi_\chi$. The expression\eqref{ampcal} allows to understand the shape of the allowed parameter region in figure \ref{parrange}. The other spectral parameters, unlike the amplitude, are practically independent of $\xi_h$ and $\lambda$ and depend on the single parameter $\xi_\chi$. Hence, we can plot $n_s$, $\alpha_\zeta$ and $r$ as functions of $\xi_\chi$ (cf. figures \ref{nsplot} and \ref{alpharplot}). Compared to the numerical results, the approximate formula for the spectral tilt has an accuracy of the order of $10^{-3}$ while the accuracy of the approximate results for $\alpha_\zeta$ and $r$ is of the order of $5\cdot 10^{-4}$. Given the uncertainties in the observational values, these accuracies are largely sufficient.

As long as the quantity $4\xc N^*$ is smaller than one (i.e. $\xi_\chi\lesssim 0.004$), the series expansions of the hyperbolic functions in \eqref{ampcal}-\eqref{rcal} converge rapidly and the expressions can be further approximated by
\begin{eqnarray}
P_\zeta(k_0)&\simeq&\frac{\l {N^*}^2}{72\pi^2 \xh^2}\left(1+\frac{1}{3}(4\xc N^*)^2+...\right)\;,\\
n_s\left(k_0\right)-1&\simeq&-\frac{2}{N^*}\left(1+\frac{1}{3}(4\xc N^*)^2+...\right)\;,\\
\alpha_\zeta(k_0)&\simeq&-\frac{2}{{N^*}^2}\left(1-\frac{1}{3}(4\xc N^*)^2+...\right)\;,\\
r(k_0)&\simeq&\frac{12}{{N^*}^2}\left(1-\frac{1}{3}(4\xc N^*)^2+...\right)\;.
\end{eqnarray}
From these expressions, we can see that in the limit
$\xc\rightarrow 0$, the predictions of the Higgs-Dilaton inflation model reduce to those found for
the Higgs-Inflation model \cite{Bezrukov:2008ut}. Hence, one can think of
$\xc$ as the deviation of our predictions from those of the Higgs-Inflation model. In the Higgs-Dilaton scenario, the results for $n_s$, $\alpha_\zeta$ and $r$ in the limit $\xi_\chi\rightarrow 0$ constitute a prediction of bounds on these quantities. We find (cf. figures \ref{nsplot} and \ref{alpharplot})
\begin{eqnarray}
n_s(k_0)&<&0.97\simeq 1-\frac{2}{N^*}\,,\\
\alpha_\zeta(k_0)&>&-0.0006\simeq-\frac{2}{{N^*}^2}\,,\\
r(k_0)&<&0.0033\simeq\frac{12}{{N^*}^2}\,.
\end{eqnarray}
These bounds are non-trivial predictions of our model. Further, given that $n_s$, $\alpha_\zeta$ and $r$ are functions of $\xi_\chi$ only, the bound $\xi_\chi\lesssim 0.008$ deduced from the observational lower bound on $n_s$ translates to (cf. figure \ref{alpharplot})
\begin{eqnarray}
\alpha_\zeta(k_0)&\lesssim&-0.00015\,,\\
r(k_0)&\gsim&0.0009\,.
\end{eqnarray}
The upper bound on $n_s$ is well in accord with the observational constraints. Results of the Planck mission are expected to reduce the errors by a factor of a few and will hence provide an important test of the Higgs-Dilaton model \cite{Planck:2006uk}.
While the present observational limits on $\alpha_\zeta$ and $r$ are too weak to compete with the bounds derived above \cite{Komatsu:2010fb}, the results of Planck might also improve this situation.

We will see in section \ref{daqe} that, if the scalar fields constitute a dark-energy component at late times, its equation of state parameter $w_{DE}^0$ is also a function of the parameter $\xc$ only. As a consequence, the observational lower bound on $n_s$ will induce a bound on $w_{DE}^0$. Thereby, the Higgs-Dilaton model provides a non-trivial connection between observables related to the early and the late universe.

Besides the parameter region in which both $\xi_\chi$ and $4\xi_\chi N^*$ are small, the observational bounds do not completely exclude the region where $\xi_\chi$ is small, but $4\xi_\chi N^*$ is somewhat bigger than unity. In this region, as can be deduced from the approximate results \eqref{tiltcal}-\eqref{rcal}, the predicted values for $\alpha_\zeta$ and $r$ go to zero exponentially with growing $4\xi_\chi N^*$, while the spectral tilt becomes asymptotically linear in $\xi_\chi$ (cf. figure \ref{nsplot}), i.e.
\begin{equation}\label{nsassol}
n_s(k_0)-1\simeq -8\xi_\chi\,,\quad\quad\textnormal{for}\quad\frac{1}{4N^*}<\xi_\chi\ll1\,.
\end{equation}
This fact will allow us to speculate about a somewhat deeper connection between $n_s$ and $w_{DE}$ in section \ref{daqe}.

From the approximate results \eqref{tiltcal}-\eqref{rcal} one can see that $\alpha_\zeta$ and $r$ are suppressed with respect to $n_s-1$. In the region where $4\xi_\chi N^*<1$, the suppression factor is $1/N^*$, while in the region where $4\xi_\chi N^*>1$, the suppression factor is smaller than $\xi_\chi$. This fact justifies the use of the observational bounds \eqref{ampobs} and \eqref{tiltobs}, that are based on the assumption of a power-law spectrum for scalar perturbations and the absence of tensor modes.

The above results provide limits on the reheating temperature $T_{rh}$, defined as the initial temperature of the homogeneous radiation dominated universe. $T_{rh}$ is related to $\varrho_{rh}$ through
\begin{equation}\label{rhorh}
\varrho_{rh}=\frac{\pi^2}{30}g_{\rm eff}(T_{rh})T_{rh}^4\,,
\end{equation}
where $g_{\rm eff}(T_{rh})$ is the effective number of relativistic degrees of freedom present in the thermal bath at the temperature $T_{rh}$. Counting all degrees of freedom of the Standard Model plus the dilaton, one has $g_{\rm eff}(T_{rh})=107.75$. To lowest non trivial order in $\xi_\chi$ and $1/\xi_h$, the limits on $\varrho_{rh}$, i.e. $V(\theta_{rh}^{min})<\varrho_{rh}<V(\theta_{end})$ (cf. \eqref{rhorhmax} and \eqref{rhorhmin}) are found to be
\begin{eqnarray}
V(\theta_{rh}^{min})&\simeq&\frac{\lambda}{144\xi_h^4}M_P^4\,,\\
V(\theta_{end})&\simeq&\frac{\lambda}{\xi_h^2}X M_P^4\,,
\end{eqnarray}
where $X=7-4\sqrt{3}\simeq 0.7$. The bounds on $\varrho_{rh}$, together with the obtained bounds on $\xi_h/\sqrt{\lambda}$ \eqref{minbound} and \eqref{maxbound}, translate into the following bounds on the reheating temperature $T_{rh}$
\begin{equation}\label{Trhbounds}
3.8\,\sqrt{\frac{65000}{\xi_h}}\cdot10^{12}\,\mathrm{GeV}\lesssim T_{rh}\lesssim2.5\cdot10^{15}\,\mathrm{GeV}\,.
\end{equation}

Finally, let us note that the findings of this section allow us to constrain the region of initial conditions for the scalar fields which lead to successful inflation. Based 
on the assumption that the last $N^*$ e-folds of inflation take place in the scale-invariant region, we have found the field value $\theta^*$ 
close to which the observable scales exit the Hubble horizon during inflation.
 The initial conditions for inflation have to be such that $\theta_{in}\geq\theta^*$. In terms of the original variables, this condition reads
\begin{equation}\label{incond1}
\frac{h_{in}}{\chi_{in}}\geq\sqrt{\frac{1+6\xc}{1+6\xh}}\tan\theta^*\;.
\end{equation}
$\theta^*$ was found to be $\theta^*\simeq \arccos\left(e^{-4\xc N^*}\right)$ \eqref{incal}.
For typical parameter values $\xc= 0.005$, $\xh= 65000$ and $N^*=58$
one obtains $\theta^*\simeq 1.25$ and $\frac{h_{in}}{\chi_{in}}\gsim 0.005$.
Considerations related to dark energy (section \ref{iccons}) will yield an additional constraint on the initial conditions. The region of acceptable initial conditions satisfying both constraints
 is shown in Fig. \ref{icplot}.

\section{Implications for the late universe}\label{lu}
We have shown in the previous section that a number of parameters of the theory can be constrained by two independent inflationary observables: the amplitude and the tilt of the primordial spectrum of scalar perturbations $\mathcal{P}_\zeta$, cf. Figure \ref{parrange}. The theory is therefore completely specified 
at the inflationary stage, and any subsequent period should be consistent with that choice of parameters. In this section we focus on the late dark energy dominated stage previously described in section \ref{qp}, during which the dilaton field is rolling down along one of the potential valleys. In subsection \ref{daqe} we show how 
this results in the dilaton playing the role of a quintessence field. We then derive consistency conditions among the inflationary observables and those associated to the dark energy dominated stage, which could allow us to either confirm or exclude the model in the coming years. In subsection \ref{iccons}, we derive a constraint that has to be imposed on the initial conditions of the scalar fields in order to have a successful description of dark energy. This will prove a posteriori that the whole period of observable inflation must take place in the scale-invariant region.

\subsection{The dilaton as quintessence field}\label{daqe}
After the phase of reheating, the system enters the radiation dominated
stage, at the beginning of which the total energy density is given by
$\varrho_{rh}$ (cf. \eqref{rhorh}). At that moment, the scalar fields have almost settled down in one of the
potential valleys, i.e. $h(t)^2\simeq\frac{\alpha}{\lambda}\chi(t)^2$ or, in terms of the variables $(\tilde\rho,\tilde\theta)$,  $\tanh^2(a\,\tilde\theta(t)/M_P)\simeq\frac{1-\varsigma}{1+\frac{\alpha}{\lambda}\frac{1+6\xi_h}{1+6\xi_\chi}}=1-\varsigma+\mathcal{O}(\alpha)$. We will work with the assumption that the equality is exact and that the fields evolve exactly along the valley (cf. \cite{Shaposhnikov:2008xb})\footnote{The validity of this approximation can be checked numerically. Let us further note that the trajectory going exactly along the valley is an asymptotic, but not an exact solution of the equations of motion.}. In this case, at the level of homogeneous fields, we are left with a single degree of freedom $\tilde\rho(t)$. As discussed in section \ref{model}, also at the level of perturbations the field $\tilde\rho$ (or equivalently $\rho$) is almost decoupled from the SM fields. Hence, we will from now on treat $\tilde\rho$ as a field minimally coupled to gravity and not interacting with matter and radiation. Given that $\mu\ll 1$ and $\alpha\ll1$, its dynamics is described by the E-frame Lagrangian \eqref{lagrangianE} with $\tilde K$ and $\tilde U=\tilde V+\tilde V_{\Lambda_0}$ given by \eqref{Knew2}, \eqref{Vlnew2} and \eqref{VLnew2}, where one inserts the constraint $\tanh^2(a\,\tilde\theta(t)/M)\simeq 1-\varsigma$, i.e.
\begin{equation}
\frac{\mathcal{L}}{\sqrt{-\tilde g}}\simeq\frac{M_P^2}{2}\tilde R-\frac{1}{2}(\partial\tilde\rho)^2-\tilde V_{QE}(\tilde\rho)\;,
\end{equation}
with
\begin{equation}
 \tilde
V_{QE}(\tilde\rho)=\frac{\Lambda_0}{\gamma^{4}} e^{-4\gamma \tilde\rho/M_P} \,.
\end{equation}
As already pointed out in \cite{Shaposhnikov:2008xb}, it is remarkable that the exponential potential, which was proposed for Quintessence (QE) long time ago \cite{Wetterich:1987fk,Wetterich:1987fm,Ratra:1987rm}, appears automatically in the present model. It hence turns out that the dilaton field $\tilde \rho$ can play the role of QE.

Let us now discuss in more detail the influence of the field $\tilde\rho$ on standard homogeneous cosmology. The equation of motion for the homogeneous field $\tilde\rho=\tilde\rho(t)$ in spatially flat FLRW spacetime is given by
\begin{equation}\label{qeeq}
 \ddot {\tilde\rho}+3H\dot{\tilde\rho}+\frac{d V_{QE}}{d\tilde\rho}=0\;.
\end{equation} 
Defining energy density $\varrho_{{QE}}$, pressure $p_{{QE}}$ and equation of state parameter $w_{{QE}}$ of the scalar field $\tilde\rho$ as
\begin{eqnarray}
\varrho_{{QE}}&\equiv&\frac{1}{2}\dot{\tilde\rho}^2+V_{QE}\,,\\
p_{{QE}}&\equiv&\frac{1}{2}\dot{\tilde\rho}^2-V_{QE}\,,\\
w_{{QE}}&\equiv&\frac{p_{{QE}}}{\varrho_{{QE}}}\,,
\end{eqnarray}
its equation of motion \eqref{qeeq} can equivalently be written as
\begin{equation}
\dot\varrho_{{QE}}=-3H\varrho_{{QE}}\left(1+w_{{QE}}\right)\,.
\end{equation}
On the other hand, in the presence of a barotropic fluid of energy density $\varrho_b$, for instance relativistic or non-relativistic matter, the Hubble parameter is given by the first Friedmann equation as
\begin{equation}\label{freq}
 H^2=\frac{1}{3M_P^2}\left(\varrho_{b}+\varrho_{QE}\right)\,, 
\end{equation} 
which in terms of the relative abundances $\Omega=\varrho/3M_P^2 H^2$ can be written as the cosmic sum rule ${\Omega_{b}+\Omega_{{QE}}=1}$.
The cosmological model described by
equations \eqref{qeeq} and \eqref{freq} with a scalar field evolving
in an exponential potential has been widely studied in the literature
(for a recent review see \cite{Copeland:2006wr}). We want to recap
the main results established in the literature and show how they apply
to our model.

For the qualitative analysis of the system we rewrite equations \eqref{qeeq} and
\eqref{freq} in terms of the observable quantities $\Omega_{{QE}}$
and ${\delta_{{QE}}\equiv 1+w_{{QE}}}$ as (cf. e.g. \cite{Scherrer:2007pu})
\begin{align}
&\delta_{QE}'=-3\delta_{QE}(2-\delta_{QE})+4\gamma(2-\delta_{QE})\sqrt{
3\delta_{QE}\Omega_{QE}}\; , \label{beta}\\
&\Omega'_{QE}=3(\delta_b
-\delta_{QE})\Omega_{QE}(1-\Omega_{QE})\label{omega}\;,
\end{align}
where prime, as before, denotes the derivative with respect to the number of
e-folds $N=\ln a$. Further, $\delta_b\equiv 1+w_b$, where $w_b$
is the equation of state parameter of the barotropic fluid. For
radiation one has $\delta_b=4/3$, while for non-relativistic matter
$\delta_b=1$. An additional dark energy component with constant equation of state would have $\delta_b<2/3$.
In the scale-invariant model analyzed here, a component of this type is present
as soon as the action \eqref{SI-UG} contains a term $\beta \chi^4$ with $\beta>0$ (cf. also section \ref{msi}).
For the reasons mentioned in section \ref{casebeta0}, we will mainly focus on the case $\beta=0$, in which $\tilde\rho$ will be alone responsible for dark energy.\footnote{Let us mention, that even in the case $\beta<0$, appropriate initial conditions lead to a dark energy dominated phase.}
It has been shown in \cite{Wetterich:1987fm,Copeland:1997et,Ferreira:1997hj} that as long as $0\leq\delta_b\leq2$ and depending on the value of the parameter $\gamma$, the system approaches one of
two qualitatively very different attractor solutions.

For $4\gamma>\sqrt{3\delta_b}$, the variables evolve towards the
stable fixed point $\Omega_{QE}=3\delta_b/16\gamma^2$ and
$\delta_{QE}=\delta_b$. This means that the scalar field
inherits the equation of state parameter of the
barotropic fluid. Hence, the energy density of the scalar field
scales like the energy density of the fluid.
Unless the scalar field gives the dominating contribution to the
energy density from the very beginning, it will never become dominating.
Therefore, these so-called ``scaling solutions'' can not be responsible
for the late-time acceleration of the universe.
In this case, the accelerated expansion must be due to another mechanism,
e.g. a barotropic dark energy component with $\delta_b<2/3$.
In other words, a scaling field can at best provide a small contribution to dark energy.

For $4\gamma<\sqrt{3\delta_b}$, the situation is very different. The stable fixed point is given by
${\Omega_{QE}=1}$ and ${\delta_{QE}=16\gamma^2/3}$. Hence, in this case,
the asymptotic solution describes a scalar field dominated universe,
which is accelerating if $4\gamma<\sqrt{2}$, i.e. $\xc\lesssim\frac{1}{2}$.
This means that the scalar field with exponential potential and $4\gamma<\sqrt{3\delta_b}$ can
describe the late-time acceleration of the universe, provided that
the system has not quite reached the fixed-point by today\footnote{As
has been shown in \cite{LopesFranca:2002ek}, this statement holds even
if $4\gamma>\sqrt{2}$.}.

In the previous section, we have found that our model can successfully
describe inflation if $\xc\lesssim 8\times 10^{-3}$. This yields the bound
$4\gamma\simeq 4\sqrt{\xc}\lesssim 0.36$. Hence, for this parameter choice,
he system evolves towards the second type of fixed point, corresponding to a scalar-field dominated universe in accelerated expansion. 
This allows us to draw a non-trivial conclusion. Namely, if the parameters of the model are fixed by the requirements of inflation, and for $\beta=0$, the late time behavior of the system necessarily corresponds to an accelerating universe, dominated by $\tilde\rho$.

Current observations \cite{Komatsu:2010fb} show that the present abundance of dark energy is 
${\Omega^0_{DE}\simeq 0.74}$. For $\beta=0$, dark energy is entirely due to $\tilde\rho$ and we can identify $\Omega_{{QE}}^0=\Omega^0_{DE}$ and $w_{{QE}}^0=w^0_{DE}$. The observed value shows that dark energy is not clearly dominating the present universe, which means that the system must not have reached its fixed point yet.

We now want to qualitatively discuss the scenario in which the
field $\tilde\rho$ is irrelevant during the radiation and matter
dominated stages, but has become important recently and is
now responsible for the present acceleration of the
universe. During the radiation and
matter dominated stages one must have $\Omega_{QE}\ll 1$.
As long as this is the case, the second term on the right-hand side
of \eqref{beta} is small compared to the first one. Hence,
$\delta_{QE}$ is driven towards a very small value 
$\delta_{QE}\ll 1$ and $w_{QE}\simeq -1$. This
shows that even if initially $\varrho_{QE}$ were dominated by
kinetic energy, the kinetic part would soon die away and $\Omega_{QE}$
become potential dominated
\footnote{In principle
one could imagine a scenario in which after reheating
$\Omega_{QE}$ is non-negligible, as
long as $\delta_{QE}\simeq 1$. Since the kinetic part of
$\varrho_{QE}$ decreases as $a^{-6}$ it would soon fall below
$\varrho_\text{radiation}$ and radiation would start dominating
provided that the potential energy of ${\tilde\rho}$ is small
enough. However, we do not expect this to happen in our model,
because the field ${\tilde\rho}$ is almost constant during reheating.}.
As a consequence, the value of ${\tilde\rho}$ is almost constant during
the radiation and matter dominated epochs and remains practically
equal to its value at the end of reheating. However, since $\varrho_{QE}$
decreases more slowly than the energy densities of radiation and
matter, $\Omega_{QE}$ becomes relevant at some point.
At this point, ${\tilde\rho}$ starts rolling down the potential and
$\delta_{QE}$ starts growing towards its attractor value. The
initial value of $\Omega_{QE}$ has to be small enough such that
$\Omega_{QE}$ remains negligible throughout radiation domination
and only becomes important in the late matter dominated stage. The
described scenario in which the quintessence field remains constant
for a long time and then starts rolling down the potential goes under
the name of  ``thawing quintessence'' \cite{Caldwell:2005tm}.
Two recent studies treating the case of an exponential potential
can be found in \cite{Scherrer:2007pu} and \cite{Sen:2009yh}.

\begin{figure}
\begin{center}
	\includegraphics[scale=0.9]{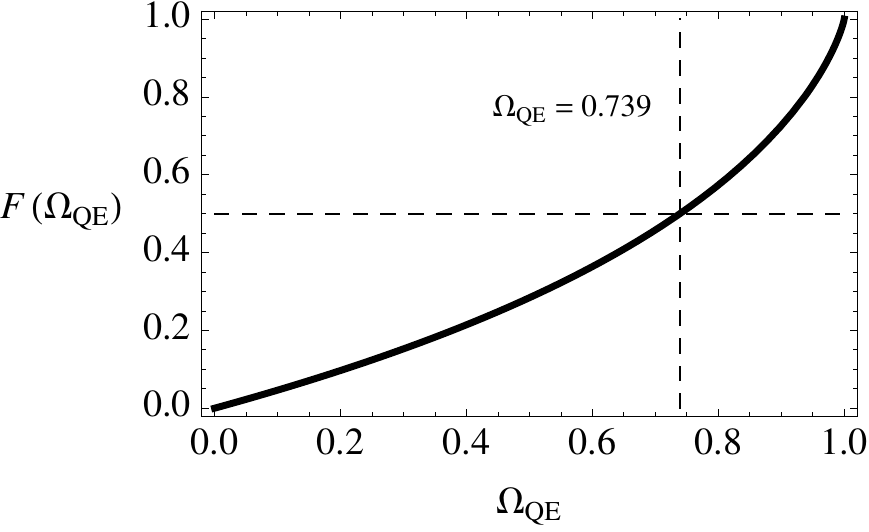}
\end{center}
\caption{
The function $F(\Omega_{QE})$. Note that it becomes exactly 1/2 for $\Omega_{QE}=0.739$, which is very close to the observed abundance of dark energy $\Omega_{DE}^0=0.725\pm0.048$ (\textit{WMAP7+BAO+}$H_0$ at $99\%$ confidence level \cite{Komatsu:2010fb}). At this value the functional relation between the spectral tilt $n_s$ of CMB anisotropies and the equation of state parameter $w_{QE}^0$ is particularly simple.}
\label{FOplot}
\end{figure}

\begin{figure}
\begin{center}
        \includegraphics[scale=1]{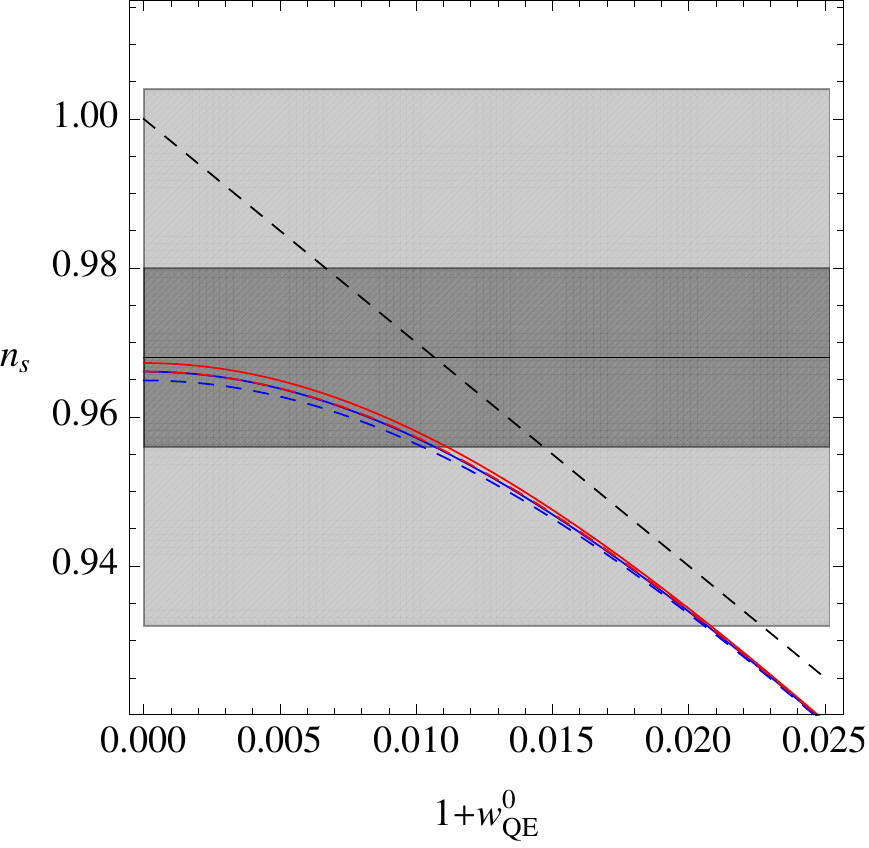}
\end{center}
\caption{This plot shows the approximate functional relationship between $n_s$ and $w^0_{DE}$. The plain curves are numerical results. The red 
plain curve is obtained for $\varrho_{rh}=\varrho_{rh}^{max}$ (instantaneous reheating), while the blue plain curve represents the case $\varrho_{rh}=\varrho_{rh}^{min}$ (long reheating).
The dashed curves are obtained from the approximate relation \eqref{relation}. The red one for $N^*=\bar N^*_{max}=59$ (instantaneous reheating) 
and the blue one for $N^*=\bar N^*_{min}=57$ (long reheating).
The black dashed straight line corresponds to the relation \eqref{relation2}, approximately valid in the region $\frac{2}{3N^*}<1+w_{DE}^0\ll 1$. Note that it does not exactly correspond to the straight line in figure \ref{nsplot}, which would not correspond to a straight line in the present plot.
The shaded region represents the experimental mean value and the associated $1\sigma$ and $2\sigma$ confidence intervals.}\label{nsplotomega}
\end{figure}

In the approximation where $\delta_{{QE}}=1+w_{QE}\ll 1$, the system of equations \eqref{beta} and \eqref{omega} can be integrated, and one finds the interesting relation \cite{Scherrer:2007pu}
\begin{equation}\label{scherrer}
1+w_{QE}\simeq\frac{16 \gamma^2}{3}F(\Omega_{QE})\;,
\end{equation} 
where
\begin{equation}
F(\Omega_{QE})=\left[\frac{1}{
\sqrt { \Omega_{QE}} } -\frac{1}{2}
\left(\frac{1}{\Omega_{QE}}-1\right)\ln
\frac{1+\sqrt{\Omega_{QE}}}{1-\sqrt{\Omega_{QE}}}\right]^2\;.
\end{equation}
The function $F(\Omega_{QE})$ is monotonically increasing from $0$ to $1$ (cf. figure \ref{FOplot}).
For the observed value  $\Omega_{QE}^0=\Omega_{DE}^0\simeq0.74$, one gets
$F(\Omega_{QE}^0=0.74)\simeq 0.5$.
Inserting this into \eqref{scherrer} and identifying $w_{{QE}}^0=w^0_{DE}$, we obtain the following result for the present equation of state parameter of dark energy (provided by quintessence field $\tilde\rho$)
\begin{equation}\label{omegade}
1+w^0_{DE}\simeq\frac{8}{3}\frac{\xc}{1+6\xc}\;.
\end{equation}
We can now plug into this relation the upper bound $\xc\lesssim0.008$ (cf. eq. \eqref{maxbound}), derived from the observational bound on $n_s$, as well as the theoretical lower bound $\xi_\chi>0$, such that (cf. Fig. \ref{nsplotomega})
\begin{equation}\label{eos}
 0\leq1+w^0_{DE}\lesssim 0.02\;.
 \end{equation}
Thus, we have found that the
parameter bound from inflation implies a strong bound on the
equation of state parameter of dark energy. This is a rather non-trivial result.
The current observational constraint $-0.52<1+w^0_{DE}<0.32$ (WMAP7 + BAO + $H_0$ at $99\%$ confidence level \cite{Komatsu:2010fb}) is much too weak to compete with this theoretical prediction. From this point of view, the energy density $\varrho_{QE}$
is practically indistinguishable from a cosmological constant. Nevertheless, the observational
bound is expected to improve considerably in the near future. While the Dark Energy Survey collaboration aims at measurement of $w^0_{DE}$ with an accuracy of $\sim 5\%$ \cite{Annis:2005ba}, the expected accuracy from the Euclid consortium is $\sim 2\%$ \cite{Refregier:2010ss}.
These measurements, together with the projected improvement on the determination of $n_s$ from the Planck mission \cite{Planck:2006uk}, should provide an important consistency check of the Higgs-Dilaton model in the near future.

In fact, the theoretical predictions of our model (with $\beta=0$) can be further refined.
Namely, since both the scalar spectral index $n_s$ and the equation of state
parameter $w^0_{DE}$ depend mainly on $\xc$, it is possible to establish a functional relation between these two very different observables. 
Combining \eqref{omegade} with the approximate relation \eqref{tiltcal} allows us to express the scalar tilt
 $n_s$ as a function of $\delta^0_{DE}$ and the number of e-folds $N^*$ as
\begin{equation}\label{relation}
n_s-1\simeq-\frac{12\d^0_{DE}}{4-9\d^0_{DE}}\coth\left(\frac{6N^*\d^0_{DE}}{4-9\d^0_{DE}}\right)\;.
\end{equation}
We plot this relation and the corresponding numerical result in figure \ref{nsplotomega}.
The plot is equivalent to the plot of Fig. \ref{nsplot}, except that the independent variable is changed from $\xc$ to 
$\delta_{DE}=1+w^0_{DE}$ with the help of \eqref{omegade}. As before, we see that the result is rather insensitive to variations of $N^*$ in the range $\bar N^*_{min}<N^*<\bar N^*_{max}$, cf. \eqref{Nminapp} and \eqref{Nmaxapp}. One can also derive a relation involving the respective second order quantities
\begin{equation}\label{relationsec}
\alpha_\zeta\simeq\frac{48w^a_{DE}}{(-4+9\d^0_{DE})^3}\left((-4+9\d^0_{DE})\coth\left(\frac{6N^*\d^0_{DE}}{4-9\d^0_{DE}}\right)+6N^*\d^0_{DE}\sinh^{-2}\left(\frac{6N^*\d^0_{DE}}{4-9\d^0_{DE}}\right)\right)\,,
\end{equation}
connecting the running $\alpha_\zeta$ of the scalar spectral index to the equation of state parameter $w^0_{DE}$ of dark energy and its rate of change $w^a_{DE}$, defined through
$$w_{DE}(a)=w^0_{DE}+w^a_{DE}\ln(a/a_0)\,.$$
While the first order consistency condition \eqref{relation} should become testable in the near future, a the test of the second order relation \eqref{relationsec} will be more challenging. 

We have mentioned previously that the parameter region where $n_s$ is well-approximated by the asymptotic linear in $\xi_\chi$ is not excluded by observations. In terms of $\delta_{{DE}}$ this region is given by $\delta_{{DE}}\ll1$ and $1<\frac{6N^*\d^0_{DE}}{4-9\d^0_{DE}}\simeq \frac{3}{2}N^*\d^0_{DE}$, in which the relation \eqref{relation} becomes approximately
\begin{equation}\label{relation2}
-3(1+w^0_{DE})\simeq (n_s-1)\,,\quad\quad\textnormal{for}\quad\frac{2}{3N^*}<\delta_{DE}\ll1\,,
\end{equation}
which can equivalently be written as a relation between ``first orders'' in the early and the late universe
\begin{equation}\label{omegans}
\frac{d\ln\varrho^0_{DE}}{d\ln a} \simeq \frac{d\ln P_\zeta}{d\ln k}.
\end{equation}
Whether this is a fundamental consequence of SI or just a coincidence remains yet unclear. Note that if relation (\ref{omegans}) should hold, it would not only imply that the
deviation $\d^0_{DE}$ of dark energy from a cosmological constant is proportional to the deviation $n_s$ of the primordial spectrum from the scale-invariant one. In fact, if we could take it at face value, it would {\em imply} a concrete value for the present abundance of dark energy: $F(\Omega_{DE})=1/2\Rightarrow\Omega_{DE}=0.739$, surprisingly close to the observed value. In the same region of parameter space, also the respective second order quantities are proportional to each other,
again for $\Omega_{DE}=0.74$, 
\begin{equation}\label{omegatilt}
3 w^a_{DE} \simeq \alpha_\zeta \;,
\end{equation}
or equivalently
\begin{equation}\label{omegatilt2}
\frac{d^2\ln\varrho^0_{DE}}{(d\ln a)^2} \simeq \frac{d^2\ln P_\zeta}{(d\ln k)^2}\,.
\end{equation}

Let us stress again that the links between the observables $n_s$ and $\alpha_\zeta$, related to inflation, and $w^0_{DE}$ and $w_{DE}^a$, related to dark energy, are non-trivial predictions of the present model. They relate two a priori totally independent periods and allow us to use the measurable observables from CMB anisotropies to make predictions for the widely unknown DE sector. On the other hand, one should bear in mind that these results rely on several important assumptions. In particular, the functional relations are based on the requirement that the J-frame potential has a flat direction ($\beta=0$).

\subsection{Dark energy constraints on initial conditions}\label{iccons}
Let us now show how the obtained results justify the assumption we made about
inflation taking place in the scale-invariant region.
From \eqref{eos} we infer that $\varrho_{QE}$ is dominated by the potential energy contribution and hence $\varrho_{QE}\simeq V_{QE}({\tilde\rho})$. This allows us to deduce from the observational value $\Omega^0_{QE}={\Omega^0_{DE}\simeq 0.74}$ today's value of ${\tilde\rho}$,\begin{equation}\label{fieldbound}
{\tilde\rho}_0\simeq -\frac{1}{4\gamma}M_P\ln\left(\gamma^4\frac{\L_{\rm eff}}{\Lambda_0}\right)\;,
\end{equation}
where we have defined an effective cosmological constant as
\begin{equation}
\L_{\rm eff}\equiv 3M_P^2H_0^2\Omega^0_{DE}\simeq10^{-120}M_P^4\;.
\end{equation}
Now, numerical simulations show that the field $\tilde\rho$ has been almost constant from
the end of inflation till today. Therefore, the value of $\tilde\rho_0$ provides
an order of magnitude estimate for the value of $\tilde\rho$ at the end of inflation.
During the analysis of inflation we have made the assumption that the whole period of observable inflation, i.e. the last $\sim 60$ e-folds, took place in the scale-invariant region, where $\upsilon_1,\upsilon_2\ll 1$ and hence $\rho\simeq cst.$.
We can now check this assumption by computing $\upsilon_1$ and $\upsilon_2$, cf. \eqref{ups1} and \eqref{ups2}, at $\r\simeq \r^*\simeq \r_{end}\simeq \gamma\tilde\rho_0$. Using \eqref{fieldbound}
and working in the usual approximation $\xc\ll 1$ and $\xh\gg1$, we obtain
\begin{equation}
\upsilon_1\simeq\frac{144\xc^2\xh^2}{\lambda}\frac{\L_{\rm eff}}{M_P^4}\frac{1}{\sin^{4}\theta}\;,\hspace{10mm}
\upsilon_2\simeq \frac{24 \xc\xh^2}{\lambda}\frac{\L_{\rm eff}}{M_P^4}
\frac{1}{\sin^{2}\theta\cos\theta}\;.\label{ups2bound}
\end{equation}
From \eqref{endcal} and \eqref{incal} we have $\theta_{end}\simeq 2*3^{\frac{1}{4}}\sqrt{\xc}$ and $\theta^*\simeq \arccos\left(e^{-4\xc N^*}\right)$. Evaluating $\upsilon_1$ and $\upsilon_2$ for values $\xc$, $\xh$ and $N^*$ of the orders of magnitude found in section \ref{cmbcop} and $0.1<\lambda<1$,
we find that for the whole interval $\theta_{end}<\theta<\theta^*$, $\upsilon_1,\upsilon_2 \lll1$, and hence that the deviation from exact scale invariance is negligible. This justifies a posteriori the neglecting of $\Lambda_0$ during inflation. Let us note that this conclusion is not altered if one takes into account the slight change of the scalar fields between the end of inflation and today. The change of $\tilde\rho$ during the reheating oscillations and during the thawing quintessence stage are of the percent level.

\begin{figure}
\begin{center}
	\includegraphics[scale=1]{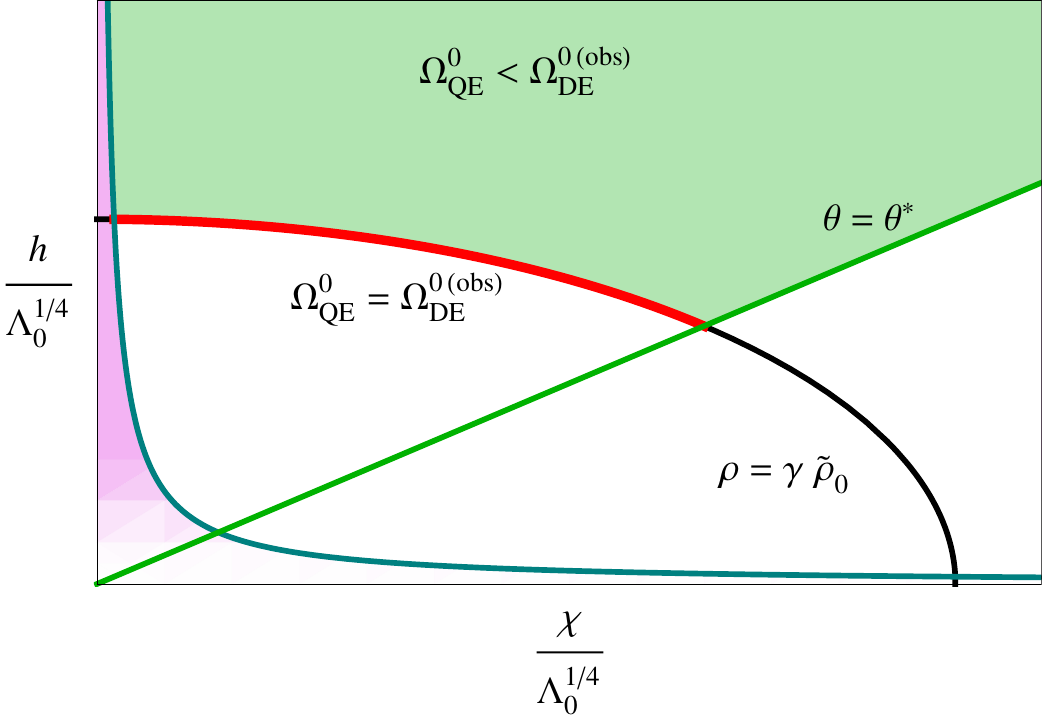}
\end{center}
\caption{This plot shows the different regions of initial conditions giving rise to qualitatively different evolutions. For a successful 
description of inflation initial conditions have to lie above the line $\theta=\theta^*\simeq \arccos\left(e^{-4\xc N^*}\right)$. For $\L_0>0$, the scalar fields contribute to dark energy
in the late stage. Initial conditions have to lie above the arc of an ellipse given by $\r\simeq\gamma\tilde\rho_0\simeq -\frac{M_P}{4}\ln\left(\gamma^4\frac{\L_{\rm eff}}{\Lambda_0}\right)$, for this contribution not to exceed the observed value 
of $\Omega_\textrm{DE}^0$. Hence, the yellow region corresponds to initial conditions giving rise to successful inflation and a contribution to dark energy
 not exceeding $\Omega_{DE}^{0\,\,(\rm obs)}=0.74$. The red segment of the ellipse corresponds to initial conditions for which the scalar fields yield the total observed dark energy.
The hyperbola is given by $\upsilon_2=1$. Initial conditions below the hyperbola lie in the non-scale-invariant region, where $\L_0$ is important. Trajectories 
starting here tend to move away from the origin before entering the scale-invariant region and following a scale-invariant trajectory. Therefore, such initial 
conditions can also be acceptable as long as the corresponding trajectories enter the scale-invariant region at or above the line given by $\r$. Note that while we only describe the quadrant $\c/\L_0^{1/4},
h/\L_0^{1/4}>0$, the reasoning would be completely analog in the other quadrants.}
\label{icplot}
\end{figure}

In section \ref{cmbcop} we have seen that for a successful description of inflation, the initial conditions for the scalar fields need to satisfy
 ${\theta_{in}>\theta^*}$, respectively $\frac{h_{in}}{\chi_{in}}\geq\sqrt{\frac{1+6\xc}{1+6\xh}}\tan\theta^*$. We recall that for typical values $\xc=0.005$, $\xh=65000$ and $N^*=58$ one
 obtains $\frac{h_{in}}{\chi_{in}}\gsim 0.005$.
The observational value for $\Omega^0_{DE}\simeq 0.74$ (respectively \eqref{fieldbound}) together with the knowledge that the field $\rho$ remains almost constant from horizon crossing during inflation until today, allows us to further restrict the region of allowed initial conditions (cf. Fig. \ref{icplot}). 
Namely, if $\tilde\rho$ is alone responsible for dark energy, and as long as the initial conditions lie in the scale-invariant region ($\upsilon_1,\upsilon_2\ll 1$), the relation \eqref{fieldbound} yields approximately the initial value for the field $\rho$, i.e.
\begin{equation}\label{boundrho}
\r_{in}\simeq \r^*\simeq\r_{end}\simeq \gamma\tilde\rho_0\simeq-\frac{M_P}{4ti}\ln\left(\gamma^4\frac{\L_{\rm eff}}{\Lambda_0}\right)\;.
\end{equation}
In terms of the original variables, this corresponds to a relation between  $\c_{in}$ and $h_{in}$ given by ($\xi_\chi\ll 1$ and $\xi_h\gg 1$)
\begin{equation}\label{boundchih}
\frac{\c^2_{in}}{\Lambda_0^{1/2}}+6\xh\frac{h_{in}^2}{\Lambda_0^{1/2}}\simeq\frac{1}{\xc}\frac{M_P^2}{\Lambda_{\rm eff}^{1/2}}\sim10^{60}\;.
\end{equation}
Together with the bound $h_{in}/\chi_{in}\gsim 10^{-3}$, this shows that initial conditions have to approximately satisfy $h_{in}/\L^{1/4}_0\gsim10^{30}$. Hence, the initial value of $h$ has to be much larger than the scale $\L_0^{1/4}$.
For $\tilde\rho$ to exactly produce the observed abundance of dark energy, the initial values have to be chosen precisely on a line in the $(\r,\theta)$-, respectively the $(\c,h)$-plane. This 
tuning of initial conditions is commonly referred to as the Cosmic Coincidence Problem (see e.g. \cite{Weinberg:2000yb})\footnote{For a discussion of the fine-tuning issue in the particular case of a Quintessence field with an exponential potential, see \cite{LopesFranca:2002ek}.}. Our model does not alleviate this problem with respect to other quintessence models. In fact, if one allows for an additional dark energy component, present if $\b> 0$, the set of acceptable initial conditions extends to an infinite region. In that case, while the fine-tuning issue does not concern the initial conditions, the parameter $\beta$ has to be finely tuned. Hence, in either case some ``fine-tuning'' is needed. At this point, it should be recalled that although the Cosmic Coincidence Problem is an undesirable feature, it is not a consistency problem and therefore does not invalidate this and other models of dynamical dark energy.

Finally, we can briefly comment on the case of initial conditions lying in the region where $\Lambda_0$ can not be neglected. Initial conditions lie in this 
region ($\upsilon_2>1$) whenever $\theta_{in}$ is sufficiently close to $\pi/2$, respectively when $h_{in}/\c_{in}$ is sufficiently big. Note, however, that
as a consequence of condition \eqref{boundrho} and for typical parameter values, this only happens for extreme values  $\pi/2-\theta_{in}\lesssim 10^{-112}$, 
respectively $h_{in}/\c_{in}\gsim 10^{109}$. In that region $\rho$ is no longer constant. The E-frame potential \eqref{potentialE}  becomes dominated by
 the term proportional to $\Lambda_0$, i.e. 
$\tilde{U}(h,\chi)\simeq\tilde{V}_{\Lambda_0}(h,\chi)=\frac{M_P^4\Lambda_0}{\left(\xi_\chi \chi^2 
+\xi_h h^2\right)^2}$. The effect of this potential is to drive the scalar fields to larger values of $\c$ and $h$, respectively larger values of $\r$, before they enter into the scale-invariant region. Qualitatively, this means that in the non-scale-invariant region the line of successful initial conditions is no longer given by \eqref{boundrho} (respectively \eqref{boundchih}) but turns towards the origin. Still, the discussion related to the Cosmic Coincidence Problem equally applies to initial conditions in this region.

\section{Conclusions}\label{conclusions}
We have considered a minimal scale-invariant extension of the Standard Model, non-minimally
coupled to gravity, including a scalar dilaton. All mass scales 
at the classical level, including the Planck scale and the Electroweak scale, are induced by the spontaneous breaking of the scale invariance. The physical dilaton is almost massless but hardly affects particle physics phenomenology. Our findings rely on SI both at the classical
and the quantum level \cite{Shaposhnikov:2008xi,Englert:1976ep}. 
The replacement of standard General Relativity by Unimodular Gravity gives rise to an arbitrary
constant in the equations of motion, which in a minimally coupled theory would play the role of a cosmological constant. However, due to the non-minimal couplings between the scalar and the gravitational sectors, this constant gives rise to a non-trivial ``run-away'' potential for the dilaton. As a consequence, the dilaton can play the role of a quintessence field, responsible for a late dark energy dominated stage.
For appropriate values of the free parameters and initial conditions, the constructed model presents a rich cosmological phenomenology, providing mechanisms both for inflation and dark energy.

We find that the amplitude $P_\zeta$ of CMB anisotropies depends mainly on the ratio $\xi_h/\sqrt\lambda$ while the spectral tilt $n_s$, the associated running $\alpha_\zeta$ as well as the scalar-to-tensor ratio $r$ depend mainly on $\xi_\chi$. The observational limits on $P_\zeta$ and $n_s$ put bounds on $\xi_h/\sqrt\lambda$ and $\xi_\chi$, which in turn provide the bounds $\alpha_\zeta\lesssim-0.00015$ and $r\gsim 0.0009$. In addition, the model predicts the bounds $n_s<0.97$, $\alpha_\zeta>-0.0006$ and $r<0.0033$, which are obtained in the limit $\xi_\chi\rightarrow 0$ and correspond to the predictions of the Higgs-Inflation model of \cite{Shaposhnikov:2008xb}. The confrontation of these bounds with the results of the Plack satellite mission will constitute an important test of the Higgs-Dilaton model.

Neither SI nor Unimodular Gravity forbid the existence of a quartic term $\beta\chi^4$ in the Jordan frame, which would correspond to a proper cosmological constant in the Einstein frame. However, the parameter choice forbidding such a term $(\beta=0)$ appears to be specially interesting, both from the cosmological and the quantum theory point of view. For this choice, the dilaton is alone responsible for dark energy. The associated equation of state parameter $w_{DE}^0$ is found to practically depend on $\xi_\chi$ only. This has the interesting consequence that the spectral index $n_s$ can be expressed as $n_s=n_s(w_{DE}^0)$ thus relating an observable from the very early universe to an observable of the present universe. For a particular parameter region, this relation takes the simple form $-3(w^0_{DE}+1)\simeq(n_s-1)$. The observational bound on $n_s$ translates into a bound $0\leq1+w^0_{DE}\lesssim 0.02$, which might be tested by future experiments. Further, we were able to derive a relation between the running of the spectral index $\alpha_\zeta$ and the rate of change $3w^a_{DE}$ of the equation of state parameter. Notice that for the dilaton to provide the measured abundance of dark energy, initial conditions have to be finely tuned. Hence, as is the case for all quintessence models, the Cosmological Coincidence Problem remains unsolved.

\section*{Acknowledgements} 
We thank Julien Lesgourgues and Andrei Linde for helpful discussions and useful comments. JGB thanks the Institute of Theoretical Physics in Geneva for hospitality during his sabbatical year 2010, when this work initiated. JR thanks EPFL for hospitality during his stay in Lausanne. We also acknowledge financial support from the Madrid Regional Government (CAM) under the program HEPHACOS P-ESP-00346, and MICINN under grant  AYA2009-13936-C06-06. We also participate in the Consolider-Ingenio 2010 PAU (CSD2007-00060), as well as in the European Union Marie Curie Network ``UniverseNet" under contract MRTN-CT-2006-035863. JR would like to acknowledge financial support from UAM/CSIC. The work of M. S. and D. Z. was supported by the Swiss
National Science Foundation and by the Tomalla Foundation.


\appendix
\section{Higgs-Dilaton Inflation in the Jordan Frame}\label{appendix}


This appendix is devoted to the study of the inflationary trajectories in the Jordan frame.  We perform an analytical study of the trajectories 
of the scalar fields during slow-roll and compare it with the results of an exact numerical computation in Jordan and Einstein representations. The numerical
 computation in the Einstein frame is performed in the way described in \cite{Ringeval:2007am} and takes into account the non-minimal kinetic terms in \eqref{lagrangianE}. As the
 classical level different frames just correspond to different choice of
variables, and therefore the final physical results should not differ. Notice however the different units used in every frame. For homogeneous fields, $h=h(t)$ and $\chi=\chi(t)$, 
the conformal transformation of the metric from the Jordan to the Einstein frame (\ref{conftransf}) depends only on time $\Omega=\Omega(t)$  
and implies a redefinition of the cosmic time, $d\tilde t = \Omega(t) dt$ , as well as the scale factor, $\tilde a(\tilde t)= \Omega(t) a(t)$ , in the Einstein frame.  
The associated Hubble rate should be also redefined as
\begin{equation}\label{HE}
\tilde H \equiv \frac{1}{\tilde a}\frac{d\tilde a}{d\tilde t}= \frac{H}{\Omega}\left( 1+\frac{\Omega'}{\Omega}\right) \,,
\end{equation}
where the prime denotes derivatives wrt the number of e-folds $N$ in the Jordan frame.
Relation (\ref{HE}) allows us to easily obtain a useful relation between the number of e-folds computed in both frames,
\begin{equation}\label{Delta}
\Delta\equiv \frac{d N}{d \tilde N}= 1 - \frac{d \ln \Omega}{d \tilde N}\,.
 \end{equation}
Integrating this equation from the initial field configuration $\phi_0$ at the beginning of inflation we get
\begin{equation}
\tilde N-N=\ln\frac{\Omega(\phi)}{\Omega(\phi_{0})}\leq 0\,.
\end{equation}
As expected, the number of e-folds is not an invariant under conformal transformations. However, the difference between the two frames turns out to be 
practically irrelevant during the inflationary stage. To obtain an upper bound, we focus on 
 the value at the end of inflation,  $\ln \Omega_{end}/\Omega_{0}$, where the discrepancy between $\tilde N$ and $N$ is larger. As we saw in Section \ref{newvar}, the inflationary 
are well described by ellipses with constant radius, $r_0^2\equiv (1+6\xi_h)h_0^2 +(1+6\xi_\chi)\chi_0^2 $. Here $h_0$ and $\chi_0$ are the initial values for the Higgs and dilaton 
fields respectively. Let us assume that they are roughly equal. In this case, it is possible to relate the initial and final amplitude of the $h$ field to obtain
\begin{equation}\label{heh0}
\frac{h_{end}}{h_0}\simeq\sqrt{\frac{6\xi_\chi}{1+12\xi_\chi}}\,,
\end{equation}
where we have used $\xi_h \gg \xi_\chi$ as well as the approximate relation among the field amplitudes at the end of inflation $\chi\simeq \sqrt{\frac{\xi_h}{\xi_\chi}}h$. 
Taking into account (\ref{heh0}), we obtain $\Omega_{end}/\Omega_{0} \simeq \sqrt{\frac{12\xi_\chi}{1+12\xi_\chi}}$, which corresponds, for a typical value $\xi_\chi=0.005$, to 
\begin{equation}\label{Ndiff}
\Big|\frac{N-\tilde N}{N}\Big|\leq 2 \% \,.
\end{equation}
It can be shown numerically that during most of the inflationary stage the difference among the number of e-folds defined in both frames \eqref{Delta},
is indeed quite smaller than the previous bound. Given  the small difference between the number of e-folds defined in Jordan and Einstein frames, we will from now on
 identify $N=\tilde N$.
\begin{figure}
\begin{center}
\begin{tabular}{cc}
         \includegraphics[scale=0.7]{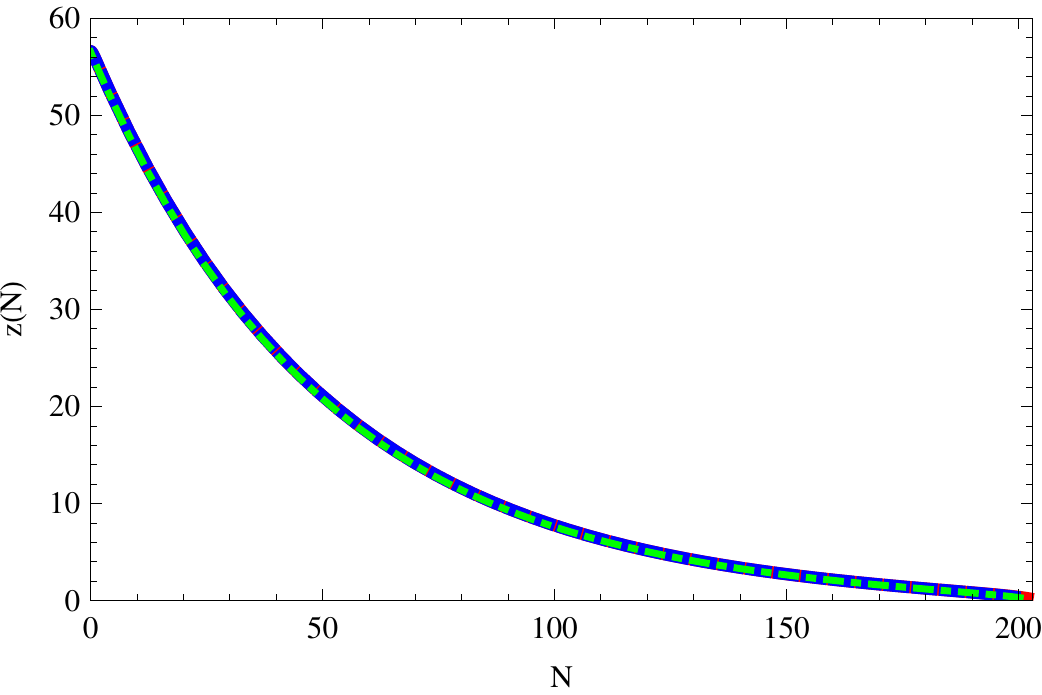} & \includegraphics[scale=0.7]{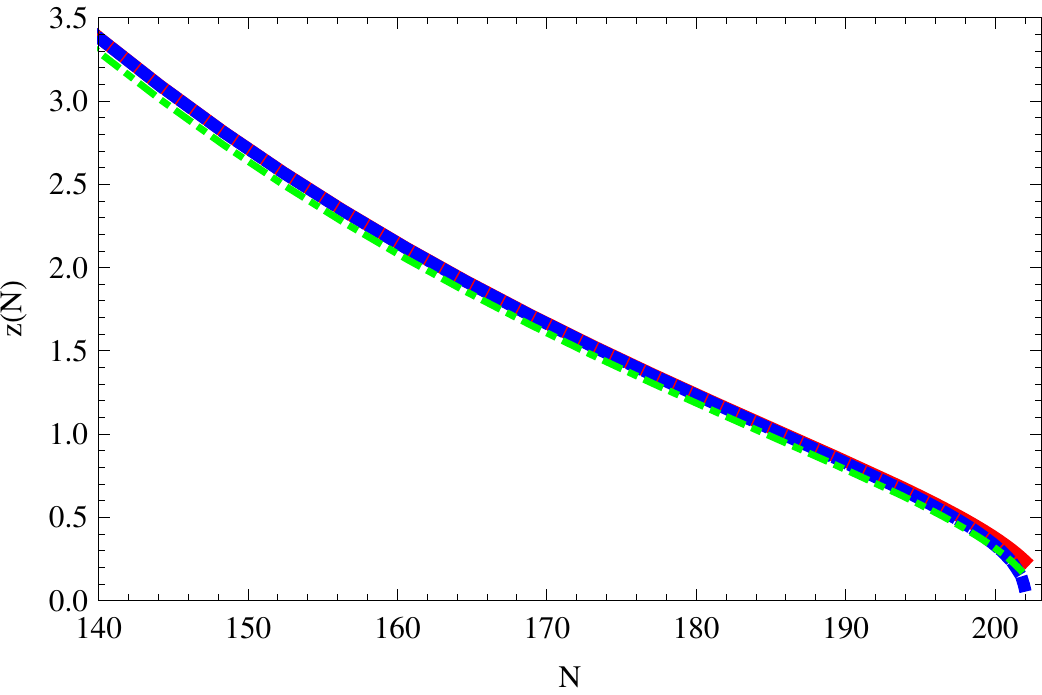}
 \end{tabular} 
\end{center}
\caption{Evolution of the angular variable $z$ as a function of the number of e-folds $N$ and detailed view of the last 60 e-folds. The green (dot-dashed) lines represent 
the approximate slow-roll solutions given by \eqref{rsolapprox}, while the red (solid) and blue (dashed) curves correspond to the result of an exact numerical computation performed in
the Jordan and Einstein frames respectively.}
\label{ztraject}
\end{figure}
\begin{figure}
\begin{center}
\begin{tabular}{cc}
         \includegraphics[scale=0.7]{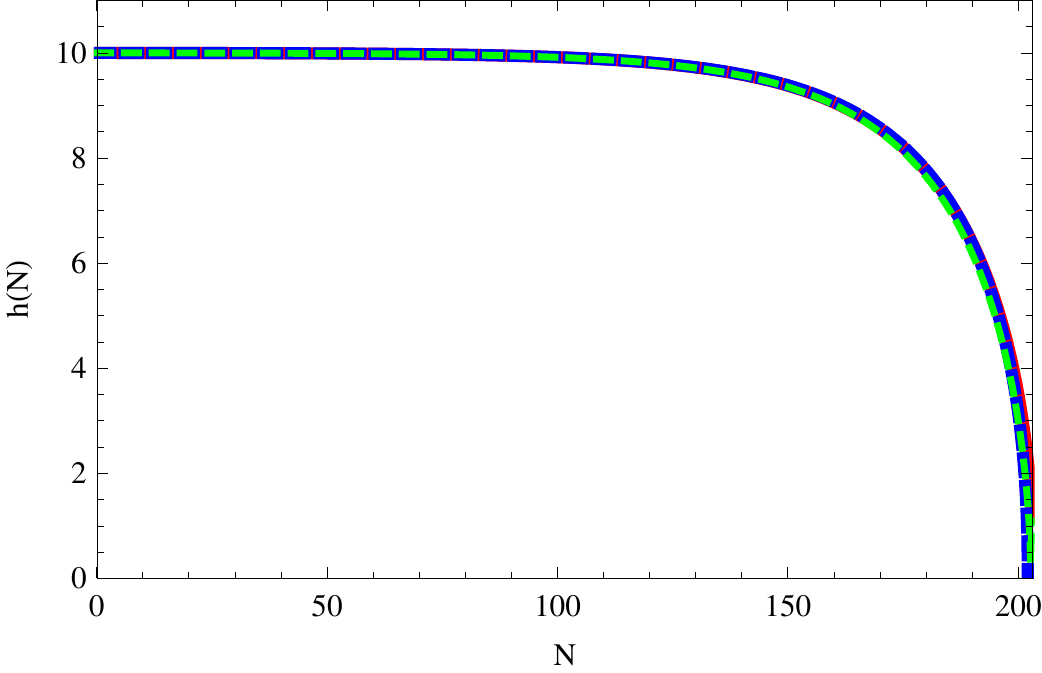} & \includegraphics[scale=0.71]{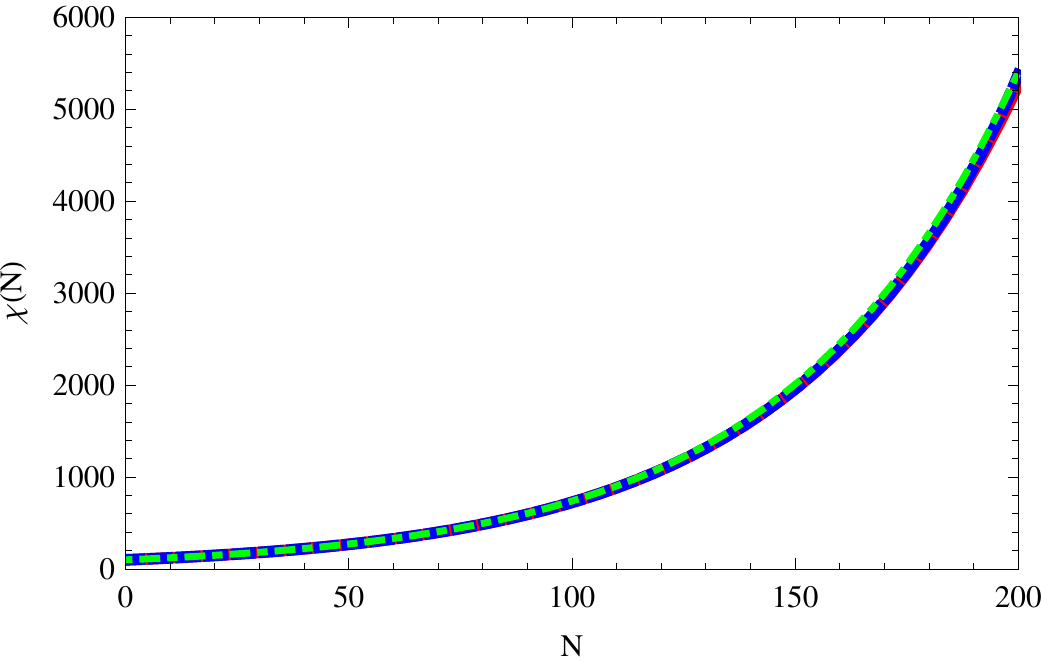}
 \end{tabular}
\end{center}
\caption{Evolution of the Higgs $h$ and dilaton $\chi$ fields as a function of the number of e-folds $N$. The green (dot-dashed) lines represent 
the approximate slow-roll solutions given by \eqref{hz}, while the red (solid) and blue (dashed) curves are exact numerical results in the Jordan and Einstein frames respectively.}
\label{hchiapprox}
\end{figure}
Let us now consider the Higgs-Dilaton Lagrangian density in the Jordan frame (\ref{lagrangianJ}). We will assume that the initial values of the fields are such that they evolve within the scale-invariant
 region, in which the $\Lambda_0$-term in (\ref{potentialJ}) can be neglected. Far away from the valleys of the potential the contribution of terms proportional to
 $\alpha\sim {\cal O}(10^{-30})$ can also be safely ignored. The Klein-Gordon equations of motion for homogeneous scalar fields are then given by
\begin{equation}\label{eomJ}
\ddot\phi^a+3H\dot\phi^a + V^{,a}-\frac{1}{2}f^{,a}R=0\,,
\end{equation}
where the Ricci scalar $R$ for a FLRW geometry and is given by $R=6(\dot H+2 H^2)$. In the Jordan frame field-space indices are raised and lowered with the Euclidean metric $\delta_{ij}$. The two Friedmann equations can be written as
\begin{equation}\label{FriedmannJ1}
3H^2f(\phi)= \frac{1}{2}\dot \phi^a\dot \phi_a+V(\phi)- 3H\partial_0 f(\phi)\,,
\end{equation}
\begin{equation}\label{FriedmannJ2}
f(\phi)R= -3(\partial_0^2+3H\partial_0)f(\phi)-\dot \phi^a\dot \phi_a + 4V(\phi)\,.
\end{equation}
If we assume the fields to be homogenous during inflation, together with the standard slow-roll approximation, $\dot\phi^a\dot\phi_a\ll V$,
$\ddot\phi^a\ll V^{,a}$ and $\ddot \phi^a \ll H\dot \phi^a$ 
the equations of motion for the scalar fields (\ref{eomJ}), expressed in terms of the number of e-folds $N$, become
\begin{equation}\label{eomJSR}
3H^2 \phi^a{}'  \simeq -V^{,a}+\frac{1}{2}f^{,a}R \,,
\end{equation}
while the Friedmann equations (\ref{FriedmannJ1}) and  (\ref{FriedmannJ2}) simplify respectively to (note that $\dot f= Hf'$)
\begin{equation}\label{FriedmannJSR}
V\simeq3H^2\left(f  + f'\right)\,,
\end{equation}
\begin{equation}\label{FriedmannJSR2}
f R\simeq   4V-9H^2 f' \,.
\end{equation}
 In the last equation, we have assumed extended slow-roll 
conditions, namely  $1+6\xi_a\dot\phi^a\ll V(\phi)$ and  $1+6\xi_a\dot\phi^a \ll H \dot f(\phi)$, which should 
be checked numerically \textit{a posteriori}. Equations \eqref{FriedmannJSR} and \eqref{FriedmannJSR2} imply that the Ricci scalar can be approximated
 as $R\simeq 12H^2\left(1+f'/(4f)\right)$, which does not correspond to the usual approximation $\dot H\ll H^2$. Although it can be checked numerically that the contribution of the
extra term $f'/(4f)$ is indeed very small, it must be explicitly maintained to preserve the conservation of the dilatational current in the slow-roll approximation. Indeed, combining 
equations \eqref{eomJSR}, \eqref{FriedmannJSR} and \eqref{FriedmannJSR2} we obtain the field space constraint
\begin{equation}
(1+6\xi_\c)\chi\chi' +(1+6\xi_h) h h' \simeq 0\,,
\end{equation}
which, as shown in Section \ref{newvar}, gives rise (in those cases in which the $\Lambda_0$ term can be neglected)to inflationary trajectories that can
 be properly described as ellipses in field space. Therefore, it will be useful to apply the same strategy of Section \ref{newvar}
and  rewrite the problem in terms of polar coordinates
 $(r,z)$, defined as
\begin{equation}\label{rz}
r^2\equiv (1+6\xi_h)h^2 +(1+6\xi_\chi)\chi^2\,,\hspace{10mm}
z\equiv \sqrt{\frac{(1+6\xi_h)}{(1+6\xi_\chi)}}\frac{h}{\chi}\,,
\end{equation}
where $z=\tan\theta$, cf. \eqref{thetadef}. The evolution equation for the previous variables can be computed making use of \eqref{eomJSR}, \eqref{FriedmannJSR} and \eqref{FriedmannJSR2} to
obtain 
\begin{equation}\label{revol}
 r' \simeq 0\,,\hspace{10mm}
\frac{z'}{z} \simeq -4 \xi_\chi \frac{z^2+\varsigma}{z^2+\varsigma+2\xi_\chi}\left(1+\frac{1}{z^2}\right)\,,
\end{equation}
where $\varsigma$ depends on the couplings $\xi_h,\xi_\chi$ and is given by \eqref{constants}. The previous equations can be easily solved to obtain the evolution of the radial
 and angular coordinates with the number of e-folds 
\begin{equation}\label{rsolapprox}
r=r_{0} 
\,,\hspace{10mm}
\frac{\left(1+z^2\right)^{1-2\xi_\chi}\left(z^2+\varsigma \right)^{2\xi_\chi}}{\left(1+z_0^2\right)^{1-2\xi_\chi}\left(z_0^2+\varsigma\right)^{2\xi_\chi}}=e^{-8\xi_\chi N}\,,
\end{equation}
where $r_{0}$ and $z_{0}$ stand for the initial values of the fields. The comparison between the slow-roll
 solution \eqref{rsolapprox} for the $z$ variable and  the
 exact solutions obtained numerically in Jordan and Einstein frames is shown in Fig. \ref{ztraject}.  Notice that, as pointed out above, we have
 identified the number  of e-folds computed in Jordan with that computed Einstein frame, $N\simeq \tilde N$, given the small difference between the 
two during the whole inflationary period. As expected, the evolution of the dimensionless quantity $z$ does not depend on the chosen frame. 
Making use of (\ref{rsolapprox}) it is also possible to compute the corresponding values of the original Higgs and dilaton fields, which 
in terms of the $z$ variable can be written as
\begin{equation}\label{hz}
h(N)=\frac{r(N)}{\sqrt{1+6\xi_h}} \left( 1+z^{-2}(N)\right)^{-1/2}\,,\hspace{10mm} \chi(N)=\frac{r(N)}{\sqrt{1+6\xi_\chi}} \left( 1+z^2(N)\right)^{-1/2}\,.
\end{equation}
The comparison with the numerical solutions is shown in Fig.~\ref{hchiapprox}.

\cleardoublepage

\end{document}